\documentclass[12pt,preprint]{aastex}

\newcommand{\mdot}{M$_{\odot}$ yr$^{-1}$}
\newcommand{\ldot}{L$_{\odot}$}
\newcommand{ \um}{$\mu$m~}
\newcommand{ \ums}{$\mu$m}
\def\kmsMpc{\ifmmode {\rm\,km\,s^{-1}\,Mpc^{-1}}\else
    ${\rm\,km\,s^{-1}\,Mpc^{-1}}$\fi}

\shorttitle{Dust and Luminosity for Starbursts}
\shortauthors{Sargsyan et al.}

\begin{document}

\title{Comparing Ultraviolet and Infrared-Selected Starburst Galaxies in Dust Obscuration and Luminosity}

\author{Lusine A. Sargsyan\altaffilmark{1}, Daniel W. Weedman\altaffilmark{2}, and James R. Houck\altaffilmark{2}}
\altaffiltext{1}{Byurakan Astrophysical Observatory, 378433, Byurakan, Aragatzotn Province, Armenia; sarl11@yahoo.com}
\altaffiltext {2}{Astronomy Department, Cornell University, Ithaca,
NY 14853; dweedman@isc.astro.cornell.edu}

\begin{abstract}
  
We present samples of starburst galaxies that represent the extremes discovered with infrared and ultraviolet observations, including 25 Markarian galaxies, 23 ultraviolet luminous galaxies discovered with GALEX, and the 50 starburst galaxies having the largest infrared/ultraviolet ratios.  These sources have z $<$ 0.5 and cover a luminosity range of $\sim$ 10$^{4}$. Comparisons between infrared luminosities determined with the 7.7 \um PAH feature and ultraviolet luminosities from the stellar continuum at 153 nm are used to determine obscuration in starbursts and dependence of this obscuration on infrared or ultraviolet luminosity.  A strong selection effect arises for the ultraviolet-selected samples: the brightest sources appear bright because they have the least obscuration.  Obscuration correction for the ultraviolet-selected Markarian+GALEX sample has the form log[UV(intrinsic)/UV(observed)] = 0.07($\pm$0.04)M(UV)+2.09$\pm$0.69 but for the full infrared-selected $Spitzer$ sample is log[UV(intrinsic)/UV(observed)] = 0.17($\pm$0.02)M(UV)+4.55($\pm$0.4). The relation of total bolometric luminosity $L_{ir}$ to M(UV) is also determined for infrared-selected and ultraviolet-selected samples.  For ultraviolet-selected galaxies, log $L_{ir}$  = -(0.33$\pm$0.04)M(UV)+4.52$\pm$0.69.  For the full infrared-selected sample, log $L_{ir}$ = -(0.23$\pm$0.02)M(UV)+6.99$\pm$0.41, all for $L_{ir}$ in \ldot~ and M(UV) the AB magnitude at rest frame 153 nm.  These results imply that obscuration corrections by factors of two to three determined from reddening of the ultraviolet continuum for Lyman Break Galaxies with z $>$ 2 are insufficient, and should be at least a factor of 10 for M(UV) $\sim$ -17, with decreasing correction for more luminous sources.  

\end{abstract}



\section{Introduction}

The formation and evolution of galaxies in the Universe is now being observed to redshifts z $\sim$ 8 \citep[e.g.][]{bou09,ouc09} using photometric imaging to wavelengths $\sim$ 1 \um in search of "Lyman Break Galaxies" (LBGs; Steidel et al. 1999).  These observations reveal the rest frame ultraviolet luminosity of star-forming galaxies and allow the measurement of changes in the star formation rate density of the Universe as a function of redshift \citep{mad98}.  (Because such star-forming galaxies are observationally characterised by indicators of young, massive, short-lived stars, we refer to them in the remainder of this paper as "starburst" galaxies.)  

Observations of LBGs indicate an increase in the characteristic luminosity of starburst galaxies for 0 $<$ z $\la$ 2, and then a decrease in luminosity for z $\ga$ 2 until the observational limit at z $\sim$ 8 \citep{red09}.  Is this evolution of starburst galaxies determined correctly based on LBGs?  The most uncertain factor in quantifying the luminosities is determining an accurate correction for obscuration by dust, because this obscuration can be extreme in the rest frame ultraviolet.  


The importance of dust in the early universe is emphasized by the result that the most luminous galaxies known have most of their luminosity emerging as reradiation by dust of intrinsic luminosity initially generated at much shorter wavelengths and absorbed by the dust.  Initially termed "Ultraluminous Infrared Galaxies" (ULIRGS; Soifer et al. 1987) when discovered with the Infrared Astronomical Satellite (IRAS), such galaxies are found in large numbers in surveys at mid-infrared wavelengths at all redshifts to z $\sim$ 3 \citep[e.g.][]{lef05,dey08} and millimeter wavelengths \citep{cha05,pop08,men09}. The evolution of starbursts among ULIRGs has been tracked to z $\sim$ 3 \citep{wee08} with the $Spitzer$ Space Telescope using redshifts measured with The Infrared Spectrograph on $Spitzer$ (IRS; Houck et al. 2004).  Because of the inability to measure redshifts of dusty starbursts with z $>$ 3 using the IRS, these have not yet been found at redshifts as high as LBGs.     

Combining our understanding of dusty galaxies dominated by infrared luminosity with galaxies discovered by ultraviolet observations is essential to a full understanding of the formation and evolution of galaxies (as well as the formation and evolution of dust).  The availability of IRS spectroscopy for the same galaxies which are detected in the rest frame ultraviolet by the Galaxy Evolution Explorer (GALEX, Martin et al. 2005) allows a direct comparison between the infrared luminosity from dust and the ultraviolet luminosity from hot stars.  

Our initial comparison in this way for 287 starburst galaxies having both $Spitzer$ spectra and GALEX ultraviolet measures found large differences in the star formation rates (SFRs) deduced from infrared compared to ultraviolet measures of SFR \citep{sar09}.  These differences are attributed to extreme obscuration of the ultraviolet luminosity. 

Our previous sample arose primarily from selection in the infrared, chosen as starbursts having IRS spectra available.  In the present paper, we use IRS spectra for samples of ultraviolet-selected starburst galaxies to study obscuration effects in sources with similar selection criteria as LBGs, in order to understand how conclusions depend on selection criteria.  Our objective is to use observations in the infrared and ultraviolet to make an independent test of dust corrections for starbursts detected as LBGs at high redshift, such as in \citet{red09}, and to determine reliable corrections for dust obscuration which are independent of selection technique.  We make measures only of the obscuration in order to avoid uncertain assumptions regarding transformation of luminosities to SFRs.  

In section 2, below, we describe the sample selection which includes 25 Markarian galaxies, 23 ultraviolet luminous galaxies discovered with GALEX, and the 50 starburst galaxies having the largest infrared/ultraviolet ratios.  In section 3, we present the infrared spectra of all 98 sources and compare the characteristics of infrared and ultraviolet spectra in section 4.  Results are used in section 5 to develop an empirical method for determining obscuration corrections as a function of M(UV), and in section 6 to compare bolometric luminosities $L_{ir}$ with M(UV).  

Our most important result (section 7) indicates that the bolometric luminosities deduced for starburst galaxies using LBGs at z $>$ 2 have been underestimated, because obscuration corrections   have been significantly underestimated for starbursts.  This means that SFRs have also been underestimated at high redshifts, and we propose a new observational test to verify these conclusions.

\section{Selection of Starburst Galaxies}

For the various comparisons which follow in this paper, we define three independent samples of starburst galaxies which have been selected using either ultraviolet or infrared criteria.  These are the "Markarian" sample, the "GALEX" sample, and the $Spitzer$ sample.  These samples are then compared using various spectroscopic parameters and conclusions are reached regarding how measures of dust obscuration and SFRs depend on sample selection criteria and source luminosities.

\subsection {Markarian Starburst Galaxies}

The first survey for galaxies selected on the basis of ultraviolet brightness was the Markarian survey \citep{mar67}, which detected "galaxies with an ultraviolet continuum" based on their apparent brightness in objective prism spectra extending to wavelengths $\ga$ 340 nm.  Although the Markarian survey was not homogeneous for flux limits or selection criteria, it produced large samples of galaxies which subsequently proved important for classification of ultraviolet-luminous galaxies.  For example, the prototype "starburst galactic nucleus" was defined by Markarian 538, also NGC 7714 \citep{wee81}.  The first set of starburst galaxies arising from this definition was the list of 50 Markarian galaxies in \citet{bal83}, selected using optical spectroscopy for measurement of emission line widths to distinguish starbursts from active galactic nuclei (AGN).  

Because the selection of these Markarian starburst galaxies was based only on optical criteria, these galaxies provide an important comparison sample to starbursts selected in the infrared (section 2.3).  (It was subsequently found that most Markarian galaxies would also be detected by the Infrared Astronomical Satellite (IRAS), but IRAS had not flown when the Markarian lists were compiled.)  For this reason, we present in this paper new IRS observations from our $Spitzer$ program 50834 of 25 Markarian galaxies selected from the Balzano (1983) list, and also include archival observations of NGC 7714.

\subsection {GALEX Starburst Galaxies}

The availability of the GALEX all sky survey enables the identification of nearby galaxies (z $<$ 0.3) which are similar in ultraviolet luminosity to the distant LBGs; these local galaxies have been called ultraviolet luminous galaxies, or UVLGs \citep{hec05}.  They represent the extremes in luminosity of ultraviolet-selected local starbursts, and they are sufficiently bright that various other data are available. 

A number of these UVLGs selected with GALEX have IRS observations available ($Spitzer$ archival program 40444, L. Armus, P.I.).  We have made new extractions of the IRS spectra for these sources and refer to them as the GALEX sample.

\subsection {Highly Obscured Spitzer Starburst Galaxies}

Our previous paper \citep{sar09} compared IRS spectra with GALEX data for 287 starburst galaxies with z $<$ 0.5, sources chosen only by having both IRS spectra and GALEX fluxes.  In that paper, the SFR measured from the luminosity of the 7.7 \um polycyclic aromatic hydrocarbon (PAH) feature, SFR(PAH), was compared to the SFR measured from the rest frame ultraviolet luminosity at 153 nm, SFR(UV).  The relation found was log [SFR(PAH)/SFR(UV)] = 0.53log [$\nu$L$_{\nu}$(7.7 $\mu$m)] - 21.5 for SFR in \mdot and $\nu$L$_{\nu}$(7.7 $\mu$m) in erg s$^{-1}$.  This relation indicates a much higher SFR as measured in the infrared, which indicates very severe extinction in the ultraviolet for these starbursts, implying the escape of $<$ 1\% of the ultraviolet luminosity for the most luminous starbursts.

To compare with the Markarian and GALEX ultraviolet-selected starbursts as one extreme of selection, we desire a comparably sized sample representing the other extreme of infrared selection.  For this sample, we utilize the infrared-selected starbursts tabulated in \citet{sar09} discovered in $Spitzer$ flux-limited surveys and having the most extreme values of SFR(PAH)/SFR(UV), those with log [SFR(PAH)/SFR(UV)] $>$ 2.   This gives a sample of 50 sources, which we call the $Spitzer$ sample.  Original IRS observations of this $Spitzer$ sample are in \citet{wee09}, taking sources from a flux limited survey to f$_{\nu}$(24\ums) $>$ 10 mJy with the Multiband Imaging Photometer for $Spitzer$ (MIPS, Rieke et al. 2004), and in archival program 40539 (G. Helou, P.I.), which derives sources from various flux limited surveys with MIPS to 5 mJy. 

This $Spitzer$ sample also provides a low redshift sample that approaches in infrared luminosity the high luminosity, optically faint, and very dusty starbursts discovered with $Spitzer$ at z $\sim$ 2 (source list and original references in \citet{wee08} plus new sources in \citet{des09}), which have been defined as "Dust Obscured Galaxies", or DOGs \citep{dey08}.  In the same fashion, therefore, that our ultraviolet selected samples represent local LBGs, the $Spitzer$ sample represents local DOGs. (Luminosity comparisons between DOGs and the local sample are shown in Figure 16.)

\section {Observations and Results}

\subsection{$Spitzer$ Infrared Spectra}

The infrared spectra used for our analysis were all obtained with the IRS \footnote{The IRS was a collaborative venture between Cornell
University and Ball Aerospace Corporation funded by NASA through the
Jet Propulsion Laboratory and the Ames Research Center.} using the Short Low module in
orders 1 and 2 (SL1 and SL2) and the Long Low module in orders 1 and 2 (LL1 and
LL2). Combining these orders gives total low resolution spectral
coverage from $\sim$5\,\um to $\sim$35\,\ums.  All spectra which are used in this paper are presented below, although spectra of 7 sources were originally given in \citet{wee09}.   


An IRS observation in one order consists of two independent spectra obtained at two different nod positions on the slit.  Each spectrum is derived from several different images taken at the same nod position.  All images with the source in one of the two nod positions were coadded to obtain the image of the source spectrum.  The background was determined from coadded background images that added both nod positions with the source in the other slit (i.e., both nods on the LL1 slit provide a background image when the source is in the LL2 slit). The difference between coadded source images minus coadded background images was used for the spectral extraction, giving two independent extractions of the spectrum for each order.  These independent spectra were compared to reject any highly outlying pixels in either
spectrum, and a final mean spectrum was produced.  

The extraction of one dimensional spectra from the two dimensional images was done with the SMART analysis package \citep{hig04,leb09}, beginning with the Basic Calibrated Data products, version 18, of the $Spitzer$ flux calibration pipeline.  Sources in the $Spitzer$ sample and GALEX sample are faint and always unresolved, so that use of the SMART Optimal Extraction in \citet{leb09} improves signal to noise (S/N).  The brighter Markarian sources, sometimes slightly resolved, were extracted with the "tapered column" extraction which uses an extraction window that is wider in the direction perpendicular to dispersion. 

That some Markarian starburst galaxies are slightly resolved at the scale of the IRS slits is noticed when the overlapping portions of the SL1 and LL2 spectra show a larger flux in the wider LL2 slit (the SL1 slit is 3.7 \arcsec wide, and the LL2 slit is 10.5 \arcsec wide).  When the SL1 and LL2 flux densities are different in the overlapping wavelengths, the spectra are matched, or "stitched", by raising the SL1 by the factor needed to match LL2.  These "stitching" factors are given in Table 1. 

Figure 1 shows spectra of all starbursts in the Markarian sample, Figure 2 all starbursts in the GALEX sample, and Figure 3 all starbursts in the $Spitzer$ sample.  Twelve AGN found among the Markarian and GALEX samples (section 4.2) are shown in Figure 4.  In all Figures, spectra are truncated at 25\,\um in the rest frame because of the absence of any significant features in the low resolution spectra beyond that wavelength.  Illustrated spectra are not smoothed.  For convenience of illustration, spectra are arranged by spectral slope; i.e by the ratio f$_{\nu}$(24 \ums)/f$_{\nu}$(7.7 \ums).  Spectra in Figures are normalized to 1 mJy at 7.7 \um and offset for illustration.  Observed spectroscopic data for all sources are given in Tables 1 and 2 (Markarian sample), 3 and 4 (GALEX sample), and 5 and 6 ($Spitzer$ sample).

Average spectra for all starbursts in the 3 samples are shown in Figure 5.  These averages were produced by averaging the individual normalized spectra of starburst galaxies in Figures 1, 2, and 3.  After averaging, the average spectra were boxcar-smoothed to 0.2 \um for illustration in Figure 5.

\subsection{GALEX Ultraviolet and SDSS Optical Data}

GALEX \citep{mar05} obtains images in the ultraviolet at wavelengths 134-179 nm (FUV) and 177-283 nm (NUV)\footnote{GALEX data used in this paper were those available in August, 2009, at the MultiMission Archive at Space Telescope Science Institute [http://archive.stsci.edu/]}.  As discussed in \citet{sar09}, GALEX ultraviolet-selected starbursts and $Spitzer$ infrared-selected starbursts generally agree well in position, to accuracies much better than the observing apertures, which are of similar sizes.  Point source GALEX images have full width at half maximum (FWHM) of 4.2\arcsec~ for FUV and 5.3\arcsec~ for NUV (Morrissey et al. 2005).  For the $Spitzer$ IRS, spatial resolution is wavelength dependent, so the FWHM of an unresolved source at the mid-infrared wavelengths used for our spectroscopic analysis is 3\arcsec $\la$ FWHM $\la$ 6\arcsec.  

The observed flux density at a rest frame FUV wavelength of 153 nm is determined by using the GALEX observed flux densities at FUV (effective wavelength 153 nm) and NUV (effective wavelength 227 nm) to fit a power law continuum between these wavelengths, and then interpolate to the flux density at observed wavelength (1+z)153 nm using the resulting power law index.  This f$_{\nu}$[(1+z)153 nm] is the value of f$_{\nu}$(FUV) listed in Tables 1, 3, and 5.  When we refer in this paper to ultraviolet luminosities or absolute magnitudes M(UV), we mean M(153 nm).  

For comparison with the IRS spectra, we also use optical spectra from the Sloan Digital Sky Survey (SDSS, Gunn et al. 1998). We classified all such spectra to determine if the Balmer lines show broad wings characteristic of type 1 AGN, or have unusually strong [OIII]/H$\beta$ ratios, which may indicate type 2 AGN.  Sources with H$\beta$ comparable in strength to [OIII] $\lambda$ 4959 without broad H$\beta$ wings are classified as starbursts.  These SDSS classifications result in some of the AGN discussed in section 4.2.



\section{Infrared Spectroscopic Characteristics of Starbursts}

\subsection{Infrared Luminosities and PAH Strengths}

Starburst galaxies show very similar spectra in the mid-infrared, characterised by strong PAH emission features \citep[e.g.][]{rig00,gen98,bra06}.  The exceptional similarity among such spectra is well illustrated in Figures 1, 2, and 3.  Starburst spectra can be separated from spectra containing a contribution to dust emission from an AGN by using the equivalent widths (EW) of the PAH features.  The best PAH feature for such quantitative use in starburst spectra observed with the IRS over all redshift ranges 0 $<$ z $<$ 3 is the 6.2 \um feature, because this feature is spectrally isolated and does not require deconvolution from adjacent features. 

Because of the strength of the PAH features, luminosities of these features provide a measure of starburst luminosity which relates to the star formation rate \citep[e.g.][]{for04,bra06}.  The PAH luminosity parameter which we have defined and used previously is $\nu$L$_{\nu}$(7.7$\mu$m), where L$_{\nu}$(7.7$\mu$m) is determined from the flux density f$_{\nu}$(7.7 \ums) at the peak of the 7.7$\mu$m feature \citep{hou07}. This parameter was chosen to enable comparisons with starbursts at redshifts z $\sim$ 2, for which the 7.7 \um feature is the strongest spectral feature and whose peak flux density is the simplest parameter to measure in sources of poor S/N \citep{wee08}.  Measurement of a peak f$_{\nu}$(7.7 \ums) also avoids the uncertainties of deconvolution and definition of the underlying continuum for the broad PAH features, which can lead to different measures of PAH flux even from the same spectrum.  Although such deconvolutions are valuable for sources of sufficiently high S/N \citep{smi07}, they are difficult to implement in sources at high redshift with poor S/N and few visible PAH features. 

An EW(6.2 \ums) $>$ 0.4 \um has been found empirically to be a good discriminant for "pure" starbursts; sources with a PAH feature of this strength invariably have optical spectra classified as starbursts.  We examine this criterion for the present samples in Figure 6, where EW(6.2 \ums) is compared with source luminosity determined from $\nu$L$_{\nu}$(7.7$\mu$m).  In this Figure, the starbursts in Figures 1, 2, and 3 are distinguished from the AGN in Figure 4 (classification of the AGN sources is discussed below in section 4.2).  The 3 GALEX sources with starburst classification but EW(6.2 \ums) $<$ 0.4 are especially interesting and are discussed in section 4.2 regarding whether they are really starbursts or have an AGN contribution.

Our new results provide an independent confirmation of the EW criterion adopted previously to define pure starbursts.  It is expected that the $Spitzer$ sample should have EW(6.2 \ums) $>$ 0.4 \um, because this was a criterion for inclusion in the tabulation of \citet{sar09}, from which we drew this sample.  It is notable, however, that the Markarian starbursts all satisfy the criterion of EW(6.2 \ums) $>$ 0.4 \um.  This is independent confirmation of the EW criterion, because the \citet{bal83} sample of Markarian starbursts was defined using only optical criteria.  The four Markarian sources with EW(6.2 \ums) $<$ 0.4 \um now appear with SDSS spectra to be AGN, so the original classification was incorrect, overlooking weak, broad wings on the hydrogen lines.

We note, however, that two GALEX sources with a starburst classification from both SDSS and IRS spectra have EW(6.2 \um) $<$ 0.4 \um.  These are sources 3 and 16 in Figure 2B (the low EW  measurement for source 13 is not reliable because the feature is so close to the end of the spectrum).   In previous studies, such weak PAH strengths in sources without AGN were observed only in blue compact dwarfs (BCDs; Wu et al. 2006).   These two sources have luminosities greatly exceeding BCDs, however, so they are potentially an important class of luminous starburst with weak PAH features.  As discussed in efforts to understand BCD spectra, such sources could arise if the ionization radiation is unusually hard, such as in a very young starburst, or if the HII region is density bounded and not surrounded by molecular clouds, or in circumstances of low metallicity.  

No such sources with low EW(6.2 \ums) are encountered among the Markarian sample, and the presence of several other AGN within the GALEX sample (section 4.2) raises the question of whether these objects also have a strong dust continuum and weak PAH because of the presence of an unrecognized AGN.  Because of this possibility, we can at present only call attention to the unusual nature of these sources but cannot confirm their nature. 

Figure 6 also shows the luminosity range of our samples, a factor of 10$^{4}$, and illustrates that all three samples overlap in luminosity.  Especially notable is that the ultraviolet-selected starbursts provide a continuous range of luminosities, typically lower luminosities from the Markarian sample and higher from the GALEX sample, but they overlap in both EW and luminosity parameters.  That these samples also overlap in $\nu$L$_{\nu}$(7.7$\mu$m) with the $Spitzer$ sample of highly obscured starbursts is evidence that these very different selection techniques have nevertheless returned samples of similarly luminous starbursts and with similar spectroscopic nature regarding strength of the PAH star formation indicators. 

\subsection{Sources Classified as AGN}

When comparing ultraviolet and infrared luminosities in order to study starbursts, it is important that AGN are not included in the samples.  An AGN can raise the ultraviolet luminosity and also raise the level of infrared dust continuum.  We are not considering details of AGN in the present paper, but after definition of the samples described in section 2, it was found that spectra showed AGN in a few cases.  Because we have IRS spectra of those sources, we identify them and show the spectra, but we do not use them subsequently in our analysis of starbursts.

In Tables 1 and 3, sources in the Markarian and GALEX samples which are classified by us as AGN are noted.  These spectra are shown in Figure 4, and these sources are not included as starbursts in the various Figures.  For the five Markarian AGN (Mkn 93, 149, 193, 764, and 863), the classification as AGN arises from SDSS spectra which show weak, broad wings to the Balmer lines.  These were overlooked in the initial classification of \citet{bal83} from which we drew the sample.  All IRS spectra for these AGN except for Mkn 193 (source 16 in Table 1) show strong PAH features.  This means that the infrared spectrum is characterized by a circumnuclear starburst, and the PAH features could be used to measure the luminosity of this starburst.  The ultraviolet luminosity may be contaminated by the AGN, however, so we do not use these sources to compare ultraviolet and infrared luminosities.  Mkn 193 also shows strong [SIV] 10.5 \um emission, which is an indicator of the hard ionizing radiation from an AGN. 

For the GALEX sources in Table 3, we classify 7 as AGN.  This classification includes any source which could be an AGN based on either optical or infrared spectra.  Such conservative criteria are designed to assure that we do not include sources with ultraviolet luminosity arising in part from an AGN.  In most cases, the AGN classification arises because of weak PAH 6.2 \um together with absorption in the silicate feature at $\sim$ 10 \um.  This combination is interpreted as evidence of an AGN with hot dust close to the AGN whose emission raises the level of continuum beneath the PAH and is also absorbed in the silicate feature by intervening cooler dust (Imanishi 2007). 

Five GALEX AGN are classified in this way with IRS spectra (sources 1, 2, 4, 10, and 23 in Table 3 and Figure 4).  There is some ambiguity in this classification, however.  Sources 3 and 16 in Table 3 are included in Figure 2 as starbursts despite having weak EW(6.2 \ums), because they show no evidence of silicate absorption.  Sources 3 and 16 have high [OIII]/H$\beta$ ratios from SDSS spectra but no evidence that the lines are broad as in type 2 AGN.  Such high ratios also arise in BCDs, for example, because of harder ionizing radiation, so we do not exclude these as starbursts.  The question of emission line ratios and ionizing radiation is considered more in section 4.3, below.  GALEX sources 12 and 15 from Table 3 are classified as AGN both because of SDSS spectra and because of strong [SIV] 10.5 \um in the IRS spectra. 

These Markarian and GALEX AGN are not included in the remaining discussion, because we do not want to confuse conclusions regarding starbursts by including sources whose spectroscopic parameters may be influenced by an AGN.

\subsection{Dust and Ionizing Radiation in Starbursts}

Mid-infrared spectra of starbursts such as those in Figure 5 contain data for several independent indicators for the luminosity of the starburst. These are the luminosity of the PAH emission features discussed above, the luminosity of the dust continuum underlying the PAH features, and the luminosity of the emission lines.  These three indicators measure different characteristics of the starburst.  The PAH emission is excited by photons of various energies penetrating the photodissociation region (PDR) at the boundary between the HII region and the surrounding molecular cloud \citep{pee04}.  Such photons are not the most energetic photons which have ionized the HII region, because such energetic photons would destroy the PAH molecules.  The dust continuum arises from dust intermixed in the HII region and heated primarily by the hot stars, and the emission lines arise within the HII region from ionizing photons arising in the hot, young stars.  

Differences among these three parameters (PAH, dust, emission lines) give information about differences in the nature of different starbursts.  Some notable differences are seen when comparing the three samples of starbursts considered herein.   These differences are first illustrated in Figure 5, which shows average spectra for each sample. The well established dominance of PAH features in starburst spectra is again confirmed in the three samples, as shown for individual sources in Figures 1, 2, and 3.  There are no measurable differences among relative PAH strengths for the three samples.  There are, however, differences in the relative strengths of the emission lines and the PAH features.

The relative strengths of the emission lines compared to PAH gives a measure of relative luminosities from the HII region compared to the surrounding PDR where the PAH features arise.  Relative intensities of the emission lines themselves provide a diagnostic for the hardness of ionizing radiation in the HII region.  The most important emission lines observed in the low resolution IRS spectra are labeled in Figure 5.

The [NeII] 12.8 \um emission line is blended with an adjacent PAH feature so the actual measurement of its flux is uncertain in these low resolution spectra, but the shape and relative flux of this blended feature does not change among the three samples.  The total flux of the blended feature is dominated by [NeII], so we can conclude that the HII luminosity relative to the PDR luminosity is similar among the samples. Compared to the fluxes of the PAH features and the [NeII] line, there is a clear progression in relative strength of the higher ionization lines [SIV] 10.5 \um and [NeIII] 15.6 \um, from strongest in the GALEX sources, intermediate in the Markarian sources, to weakest in the $Spitzer$ sources.  The [SIII] 18.7 \um line is also strong in the Markarian sources compared to the $Spitzer$ sources, although S/N is too low for measurement in the GALEX average spectrum.  The average spectra indicate, therefore, that the ultraviolet-selected samples contain hotter ionizing stars than does the extreme infrared-selected $Spitzer$ sample.

The average spectra also show increased dust continuum for the GALEX sample compared to the other samples, easily seen by the higher continuum for wavelengths $>$ 15 \um but also measurable by an increased continuum at $\sim$ 10 \ums.  That dust emission increases while the hardness of the ionization also increases is consistent with having warmer dust within the HII region in the sources with harder ionization.  Comparisons are plotted for individual sources in Figures 7 and 8.

In Figure 7, the ratio of total flux in [NeIII] to PAH 11.3 \um is compared with EW of PAH 11.3 \um (in \ums).  This plot includes all three samples plus the nearby starburst galaxies in the \citet{bra06} sample as measured with the high resolution spectroscopy in \citet{ber09}. [NeIII]/PAH is a measure of hardness of the ionizing radiation, and EW(11.3 \ums) shows the strength of the dust continuum relative to the strength of PAH.  The median values of log ([NeIII]/PAH), including limits, are -0.29, -1.01, -1.13, and -1.25 for the GALEX, Markarian, $Spitzer$, and nearby starbursts.  Both Figure 5 and these results show decreasing EW(11.3 \ums) PAH with increasing  [NeIII]/PAH.  The average EW(11.3 \ums) for the $Spitzer$ sample is 0.65 \ums, for the Markarian sample is 0.55 \ums, and for the GALEX sample is 0.50 \ums.

The values for EW(11.3 \ums) indicate that the Markarian starbursts have 18\% more dust luminosity in the continuum at $\sim$ 11 \um relative to PAH luminosity than has the $Spitzer$ sample,  and the GALEX sample has 30\% more dust luminosity than the $Spitzer$ sample.  A similar result is found by measuring dust continuum directly.  For example, the value of continuum f$_{\nu}$(10 \ums) in the normalized average spectra is 0.16/0.29/0.40 for $Spitzer$/Markarian/GALEX.  The ratio of dust continuuum at 15 \um to PAH peak at 7.7 \um is 0.48/0.68/0.91 for $Spitzer$/Markarian/GALEX.  These results are all consistent with the simple interpretation that the harder ionizing radiation in the ultraviolet-selected starbursts results in stronger dust emission at mid-infrared wavelengths, compared to the extreme $Spitzer$ sample. 

Figure 8 compares the [NeIII]/PAH ratio with luminosity of the PAH features, measured by $\nu$L$_{\nu}$(7.7 $\mu$m) in erg s$^{-1}$. There is no trend with luminosity for this ratio, but the result shows as in Figure 7 the systematically stronger [NeIII] compared to PAH in the UV-selected GALEX and Markarian samples compared to the $Spitzer$ infrared-selected sample.  


\section{Obscuration in Starbursts}

Because the starbursts at highest redshift have been discovered in the rest frame ultraviolet \citep{bou09,red09}, understanding their intrinsic luminosities requires corrections for dust obscuration.  The measure of obscuration is the greatest uncertainty in measuring SFRs.   \citet{cal07} and \citet{cal08} have summarized the various methods for determining such corrections and their application to measurement of SFRs.  

When observing the rest frame ultraviolet, the obscuration is estimated by comparing the observed spectral slope (or "reddening") with an assumed intrinsic slope \citep[e.g.][]{bur05}.  This is the source of the correction applied to the LBGs discovered at the highest redshifts.  This has led to the result that the dust correction remains small for lower luminosity galaxies, which yields the conclusion that low luminosity galaxies progressively dominate the luminosity density for z $>$ 3 \citep{bou09}.  

Such obscuration corrections apply only for the portion of the ultraviolet luminosity that is able to emerge from the obscuration.  Completely hidden sources cannot be measured in this way.  This is the primary reason why infrared measures of starbursts are so important, because infrared luminosities are much less subject to extinction.  In our discussion below, we present a new, direct comparison between the parameters most directly measuring the starburst luminosity as observed in the infrared (PAH luminosity) and in the ultraviolet (continuum luminosity).  These parameters are compared to determine empirically the relation for unobscured starbursts, and then to use this relation to determine obscuration corrections for obscured starbursts. 

The PAH emission and the ultraviolet monochromatic continuum provide two spectral "features" which are intrinsic to the same starburst.  The f$_{\nu}$(7.7 $\mu$m) arises from the photodissociation region immediately surrounding the ongoing starburst and directly excited by photons from the starburst. This PAH emission is intrinsic to the photodissociation region and is present regardless of whether any dust obscures the starburst.  It is just as intrinsic to the starburst as radiation from the hot stars or emission lines from the HII region.  Both f$_{\nu}$(7.7 $\mu$m) and f$_{\nu}$(153 nm) features are affected by the same obscuring dust in the surrounding cold molecular cloud, but the ultraviolet suffers heavy extinction whereas the PAH feature suffers negligible extinction.  

If we can determine empirically the intrinsic ratio between the PAH emission and ultraviolet continuum in unobscured sources, we can then simply use the observed ratio in obscured sources to determine how much extinction applies to the ultraviolet.  This empirical determination is made using Figures 11, 12, and 14, discussed below, which illustrate the most extreme, limiting values of f$_{\nu}$(7.7 $\mu$m)/f$_{\nu}$(153 nm) found in any of the starbursts.  In all three figures, this limit is log[f$_{\nu}$(7.7 $\mu$m)/f$_{\nu}$(153 nm)] = 1.  We will adopt this as the intrinsic value for sources with negligible obscuration of f$_{\nu}$(153 nm).  If these extreme sources also have some ultraviolet obscuration, then the final values of obscuration which we derive are lower limits, and sources are even more obscured in the ultraviolet than our results indicate.


\subsection{Comparison of Star Formation Rates}

The ultimate objective of studying the luminosities of starbursts is to determine the SFR in starbursts and to follow the resulting cosmic evolution of the SFR density.  Various assumptions enter the transformation from the luminosity of a starburst into a quantitative measure of SFR in \mdot.  For infrared measures based on luminosities of the dust continuum, the SFR is determined by equating the bolometric luminosity of the starburst to the luminosity absorbed and reradiated by dust.  For ultraviolet luminosities, the SFR derives from the total luminosity of the hot stars in the starburst measured at a rest frame ultraviolet wavelength.

For sources with a mid-infrared spectrum arising only from a starburst with no AGN contamination, the luminosity $\nu$L$_{\nu}$(7.7 $\mu$m) relates empirically to the total infrared luminosity of the starburst, $L_{ir}$, according to log $L_{ir}$ = log [$\nu$L$_{\nu}$(7.7 $\mu$m)] + 0.78 \citep{hou07}.  This transformation has no dependence on luminosity, as is seen by combining the IRAS-derived $L_{ir}$ for the local, low luminosity starburst sample of \citet{bra06} with the high luminosity, high redshift submillimeter galaxies in \citet{pop08}.  Using the relation from \citet{ken98} to convert $L_{ir}$ into a SFR , this result for $L_{ir}$ yields that log [SFR(PAH)] = log [$\nu$L$_{\nu}$(7.7 $\mu$m)] - 42.57$\pm$0.2, for $\nu$L$_{\nu}$(7.7 $\mu$m) in ergs s$^{-1}$ and SFR in \mdot.  This conversion was confirmed in \citet{sar09} who found no offset between this measure of SFR and that derived from radio continuum observations of the same starbursts \citep{con92}.

SFR(UV) is determined using the relation SFR(UV)= 1.08x10$^{-28}$L$_{\nu}$(153 nm), for L$_{\nu}$(153 nm) in ergs s$^{-1}$ Hz$^{-1}$ \citep{sam07}.  For our comparisons, SFR(UV) is the observed value without applying any extinction corrections to the ultraviolet.  For comparison to SFR(PAH), we convert this to log [SFR(UV)] = log $\nu$L$_{\nu}$(FUV) - 43.26, for $\nu$L$_{\nu}$(FUV) in ergs s$^{-1}$ and SFR in \mdot, using f$_{\nu}$(153 nm) from Tables 1, 3, and 5.  The ratio SFR(PAH)/SFR(UV) is a measure of the fraction of luminosity absorbed by dust and seen in the infrared compared to the fraction which escapes absorption and can be seen in the ultraviolet.

Figure 9 shows SFRs determined from 7.7 \um PAH luminosity compared to SFRs determined from observed GALEX UV continuum for the three samples of starbursts.  The line is reproduced form the full sample of 287 infrared-selected starbursts in \citet{sar09}.  The GALEX and Markarian UV-selected samples show similar values of SFR(PAH)/SFR(UV), with median log SFR(PAH)/SFR(UV) = 0.8.  If the discrepancy between SFR(PAH) and SFR(UV) is caused only by the obscuration of the UV, then this result indicates that only $\sim$ 15\% of the intrinsic UV luminosity is observed, even for these ultraviolet-selected samples.  Much greater obscuration is found for the $Spitzer$-selected sample, with median log SFR(PAH)/SFR(UV) = 2.4, indicating that only 0.4\% of the intrinsic ultraviolet luminosity emerges.  The GALEX sample represents the most extreme UV-selected starbursts, and these are systematically more luminous than the Markarian sources.  


In Figure 10, we transform this comparison to use M(UV) as a measure of luminosity. M(UV) is an AB magnitude, determined with the fundamental definition m$_{AB}$ = -2.5log f$_{\nu}$  - 48.60, for f$_{\nu}$ in erg cm$^{-2}$ s$^{-1}$ Hz$^{-1}$ \citep{oke74}.  M(UV) is determined as the m(UV) which arises using L$_{\nu}$(153 nm) to determine f$_{\nu}$(153 nm) at a distance of 10 pc.  The line shown is the linear least squares fit to all of the points.  It is notable that this line has the opposite slope of the line in Figure 9.  This result indicates that objects selected using ultraviolet observations of luminosity select in favor of sources having less obscuration.  As a result, the dependence of obscuration on luminosity has the opposite sign of the dependence found when starbursts are selected in the infrared.  Figure 10 shows the similarity of GALEX and Markarian sources in terms of obscuration as measured by SFR(PAH)/SFR(UV), but shows that the GALEX sample is systematically more luminous.

\subsection{Comparing Infrared and Ultraviolet Starburst Luminosities to Determine Obscuration}

While the ultimate use of infrared or ultraviolet spectra of starbursts is to determine the star formation rates, the dependence of this result on the amount of obscuration means that it is useful to discuss only those parameters which can measure the obscuration.  In this way, we hope to determine a quantitative measure of obscuration in starbursts which is independent of the many assumptions required to relate observed luminosity to SFR from either ultraviolet or infrared spectra. 

Therefore, we introduce a new comparison which is only between the observed parameters used to determine PAH or UV luminosities.  This avoids any of the uncertainties in converting to SFRs.  These parameters are the observed f$_{\nu}$(7.7 $\mu$m) and f$_{\nu}$(153 nm) (Tables 1, 3, and 5).  We assume neglible extinction of the PAH, so that the PAH luminosity represents the intrinsic luminosity of the starburst.  This is justified by the extinction curve of \citet{dra89}, which shows small PAH extinction compared to optical or UV extinction.  For example, from Draine, A(H$\alpha$)/E(J-K) = 6.5, but A(7.7$\mu$m)/E(J-K) = 0.07, so 5 magnitudes of H$\alpha$ extinction (1\% emerges) corresponds to only 0.05 mag of 7.7 \um extinction (95\% emerges). 

Figure 11 shows the comparison of infrared PAH flux density to ultraviolet flux density, f$_{\nu}$(7.7 $\mu$m)/f$_{\nu}$(153 nm), as a function of luminosity $\nu$L$_{\nu}$(7.7 $\mu$m).  The distribution of points among the three samples illustrates an empirical result that the smallest value of log[f$_{\nu}$(7.7 $\mu$m)/f$_{\nu}$(153 nm)] = 1.   We take this as the empirical value which relates PAH luminosity to ultraviolet luminosity for a starburst with negligible obscuration of the UV. This is the most important result of the comparisons shown in Figure 11.  It is an empirical result which can then be used to compare ultraviolet and infrared spectra for a quantitative determination of obscuration in the ultraviolet spectra. 

Figure 12 shows the same comparison but using M(UV) as the luminosity measure.  This Figure also shows the empirical result that the limiting value of log[f$_{\nu}$(7.7 $\mu$m)/f$_{\nu}$(153 nm)] = 1, so we take this to be the value for an unobscured source. The UV-selected sources in both Markarian and GALEX samples have similar values of log[f$_{\nu}$(7.7 $\mu$m)/f$_{\nu}$(153 nm)], with a median of log[f$_{\nu}$(7.7 $\mu$m)/f$_{\nu}$(153 nm)] = 1.8, compared to the median for the $Spitzer$ sample of 3.4. 

Figures 11 and 12 show the strong offset in the samples resulting from the selection criteria.  The infrared-selected starbursts have median values of f$_{\nu}$(7.7 $\mu$m)/f$_{\nu}$(153 nm) larger by about a factor of 100 compared to the ultraviolet-selected starbursts.  This is a direct indication that the infrared-selected sample shows extreme obscuration in the ultraviolet. Demonstrating obscuration in this way avoids the uncertainties and assumptions required when comparing SFRs determined from the different samples. 

Using the limit for f$_{\nu}$(7.7 $\mu$m)/f$_{\nu}$(153 nm) as representing sources without obscuration, such that log[f$_{\nu}$(7.7 $\mu$m)/f$_{\nu}$(153 nm)] = 1 means no obscuration, Figures 11 and 12 then give the correction factor which can be used to determine the obscuration correction for a source.  The fraction of UV luminosity intrinsically emitted compared to the luminosity observed, UV(intrinsic)/UV(observed), is given by log[UV(intrinsic)/UV(observed)] = log[f$_{\nu}$(7.7 $\mu$m)/f$_{\nu}$(153 nm)] - 1.  

\subsection{Obscuration Corrections to M(UV)}

The correction for dust obscuration determined above in section 5.2 is compared with luminosity M(UV) in Figure 13.  Figure 13 illustrates our most important results regarding the selection effects arising from obscuration by dust for discovering starbursts .  The luminosities of infrared-selected and ultraviolet-selected samples overlap, but the most luminous ultraviolet sources (the GALEX sample, diamonds in Figure 13) appear luminous because they have low obscuration.  There is a very large range between the miminally obscured ultraviolet sources and the most obscured infrared sources, with the latter having dust correction factors exceeding 100 at M(UV) = -18, compared to a correction of less than a factor of 10 at this M(UV) for the ultraviolet-selected sources. 

We show in Figure 13 the linear fit for the most obscured $Spitzer$ starbursts (upper line), the sample in Table 5, but we also plot in Figure 13 the full sample of 287 infrared-selected starbursts (center line) having IRS and GALEX measures from \citet{sar09}.  The distribution of these points shows that the ultraviolet-selected starbursts from the Markarian and GALEX samples (fit by the lower line) generally lie at the extremes of the full infrared-selected sample.
 
The fits to the sources in Figure 13 give measures of obscuration which apply for starbursts so obscured that most of the ultraviolet luminosity is hidden.  These fits are independent of measures derived only from reddening of ultraviolet observations, as in \citet{bou09} and \citet{bur05}.  Even in the ultraviolet-selected samples, the obscuration is significant, being a factor of 8 to 10 with little dependence on luminosity.  The linear least squares fit to the ultraviolet selected samples is log[UV(intrinsic)/UV(observed)] = 0.07($\pm$0.04)M(UV)+2.09$\pm$0.69.  This obscuration is about twice that used by \citet{red09} to correct for LBGs with z $>$ 3, although the M(UV) in Figure 13 overlap with the M(UV) of these high redshift sources.

For the full infrared-selected $Spitzer$ sample (central line in Figure 13), The linear least squares fit is log[UV(intrinsic)/UV(observed)] = 0.17($\pm$0.02)M(UV)+4.55($\pm$0.4); the fit to the extreme infrared-selected $Spitzer$ sample in Table 5 (upper line) is log [UV(intrinsic)/UV(observed)] = 0.14($\pm$0.03)M(UV)+4.68$\pm$0.50.

We make another estimate of effects of dust obscuration in sections 6 and 7, by determining the relation between M(UV) and total bolometric luminosities $L_{ir}$.  This comparison is possible because we have an empirical scaling between $L_{ir}$ and $\nu$L$_{\nu}$(7.7 $\mu$m).

\subsection{Ratio H$\alpha$/H$\beta$ as Measure of Obscuration compared to Infrared/Ultraviolet}

A method often used to determine obscuration corrections for starbursts is the measure of reddening from the  H$\alpha$/H$\beta$ ratio \citep{cal07}.  This method works only for the emission lines which are able to emerge from the starbursts and so cannot account for totally obscured starbursts.  It is useful, therefore, to attempt a comparison between the obscuration which would be deduced from H$\alpha$/H$\beta$ and that deduced from f$_{\nu}$(7.7 $\mu$m)/f$_{\nu}$(153 nm).  Our primary objective of this comparison is to obtain another measure of the value f$_{\nu}$(7.7 $\mu$m)/f$_{\nu}$(153 nm) for an unobscured starburst.

Figure 14 compares H$\alpha$/H$\beta$ determined from SDSS spectra of the starbursts (Tables 1, 3, 5) with the ratio f$_{\nu}$(7.7 $\mu$m)/f$_{\nu}$(153 nm) discussed in section 5.2.  The relative H$\alpha$/H$\beta$ line fluxes were measured using $\lambda$f$_{\lambda}$ taking f$_{\lambda}$ at the peak of the line in the SDSS spectra. Values given in Tables 1, 3, and 5 are for all sources with SDSS spectra available.  The intrinsic H$\alpha$/H$\beta$ ratio in an unobscured source should be 2.87, which indeed agrees with lowest value in Figure 14. 

The comparisons in Figure 14 confirm the determination of log[f$_{\nu}$(7.7 $\mu$m)/f$_{\nu}$(153 nm)] = 1 for an unobscured source, found above in section 5.2.  The line is a linear least squares fit to all of the points.  The result that there is a trend line shows that there is indeed some relation between reddening observed for the Balmer lines and the obscuration shown by comparing PAH and UV luminosities. 

To quantify this relation, we show in Figure 15 a measure of obscuration for H$\alpha$ compared with the obscuration deduced from comparing f$_{\nu}$(7.7 $\mu$m)/f$_{\nu}$(153 nm) discussed in section 5.2.  From \citet{cal07},  k(H$\alpha$) = 2.47 and K(H$\beta$)-K(H$\alpha$)=1.16.  Knowing the intrinsic value H$\alpha$/H$\beta$ = 2.87,  this differential extinction means that 0.47log H$\alpha$(intrinsic) = 1.47log H$\alpha$(observed)-log H$\beta$(observed)-0.46.  From this, we derive a value H$\alpha$(intrinsic)/H$\alpha$(observed). 

This fraction of escaping H$\alpha$ is compared to the fraction of escaping UV by using the empirical result in section 5.2 that log[UV(intrinsic)/UV(observed)] = log[f$_{\nu}$(7.7 $\mu$m)/f$_{\nu}$(153 nm)]-1.  The comparison is shown in Figure 15. 

Figure 15 shows the expected result for highly obscured regions that the obscuration of the intrinsic UV is much greater than the obscuration which would be deduced from the Balmer lines.  This is because the Balmer lines which can be observed emerge from outside of the most obscured regions, whereas our measure of ultraviolet obscuration applies to the total obscuration of the most hidden regions.

\subsection{Comparing Starburst and Stellar Luminosities for Starburst Galaxies}

Understanding the formation of galaxies requires, in addition to a measure of the SFR, a determination of the mass of stars already present.   Such masses can be estimated from observations at various rest-frame wavelengths if stellar models and mixtures are assumed, as explained, for example, in \citet{red09}, \citet{lon09} and \citet{bus09}.  For the starburst samples in the present paper, we compare the luminosities of starbursts with the luminosities of the underlying stellar component for the galaxy.  

This comparison is done by showing in Figure 16 the ratio of f$_{\nu}$(7.7 \ums) at the peak of the PAH feature to the rest frame f$_{\nu}$(1.6 \ums), which arises from the stellar continuum.   For the Markarian, GALEX, and $Spitzer$ samples, the stellar luminosity is measured from the Two Micron All Sky Survey (2MASS; Skrutskie et al. 2006) $H$ band flux densities; no corrections for redshift are made because there are only small differences between the $H$ flux density and $K$ (2.2 \ums) flux density (Tables 1, 3, and 5.)  Figure 16 is also useful for comparison of PAH luminosities among the various samples, including the DOGs at z $\sim$ 2.

In Figure 16, we include numerous starburst DOGs from the optically faint, high redshift samples selected at f$_{\nu}$(24 \ums).  All have published IRS spectra that allow the measure of f$_{\nu}$(7.7 \ums), and have photometry with the $Spitzer$ Infrared Array Camera (IRAC; Fazio et al. 2004) that allows measurement of rest frame 1.6 \um luminosity.  These sources are found in \citet{wee06,yan07,far08} and \citet{des09}. 

These sources represent the most luminous sources known as measured with $\nu$L$_{\nu}$(7.7 $\mu$m) so they show the most luminous, most obscured, starburst ULIRGs \citep{wee08}.  Most of these sources were selected as "bump" sources, for which an excess continuum is observed at 4.6 \um to 5.8 \um from giant and supergiant stars having lower opacity at rest frame $\sim$ 1.6 \um \citep{sim99}.  The comparison in Figure 16 is with IRAC 5.8 \um fluxes to estimate rest frame f$_{\nu}$(1.6 \ums) because fluxes at 4.6 \um and 5.8 \um are so similar that no redshift corrections are made. 

Because the high redshift starbursts were selected based both on f$_{\nu}$(24 \ums) and f$_{\nu}$(5.8 \ums), it is expected that selection effects will restrict the range of f$_{\nu}$(7.7 \ums)/f$_{\nu}$(1.6 \ums), the measure of starburst luminosity/stellar luminosity.  No such selection effect can apply for the lower redshift samples from the present paper, however, because 2MASS magnitudes played no role in choosing these sources.

The similarity in the starburst to stellar ratio is striking among all of the samples.  The range of 10$^{4}$ in PAH luminosity compares to a range of only 4 relative to stellar luminosity.  We do not attempt to calculate stellar masses because of uncertainties in converting a 1.6 \um stellar luminosity into a stellar mass.  More importantly, the homogeneity of the f$_{\nu}$(7.7 \ums)/f$_{\nu}$(1.6 \ums) ratio among the samples in Figure 16 raises a fundamental question of what such a stellar mass measure at 1.6 \um means.   We wonder if the stellar luminosity at 1.6 \um arises from supergiant stars very recently evolved from the starburst itself and does not reflect the underlying stellar mass of old stars. 

There is some evidence for this interpretation if we compare the ratios f$_{\nu}$(7.7 \ums)/f$_{\nu}$(1.6 \ums) among the different samples.  This evidence is that the median value of the ratio f$_{\nu}$(7.7 \ums)/f$_{\nu}$(1.6 \ums) scales with obscuration determined from f$_{\nu}$(7.7 \ums)/f$_{\nu}$(153 nm), as measured in section 5.2.   The ultraviolet-selected GALEX and Markarian sample combined have a median f$_{\nu}$(7.7 \ums)/f$_{\nu}$(1.6 \ums) = 8 (counting limits), the low redshift infrared-selected extreme $Spitzer$ sample in Table 5 (local DOGs) has median f$_{\nu}$(7.7 \ums)/f$_{\nu}$(1.6 \ums) = 13, and the high redshift DOGs sample of the most luminous, most dusty starbursts have median f$_{\nu}$(7.7 \ums)/f$_{\nu}$(1.6 \ums) = 20.   

While extinction at 1.6 \um is much less than in the ultraviolet, some extinction can also be expected at 1.6 \um for the most obscured sources.  Because extinction at 7.7 \um is less than at 1.6 \um \citep{dra89}, the ratio f$_{\nu}$(7.7 \ums)/f$_{\nu}$(1.6 \ums) should scale with the amount of obscuration, more obscured sources having a higher ratio, if the 1.6 \um stellar luminosity arises within the starburst and is also obscured by dust.   This is the trend which is observed among the samples, from most obscured $Spitzer$ sources to least obscured ultraviolet sources.  By contrast, we would not expect such a scaling if the 1.6 \um luminosity arises from an underlying stellar component which is outside of the obscured starbursts.


\section{Comparing Bolometric Luminosities $L_{ir}$ with M(UV)}

The fundamental question for measurement of SFRs is which parameter best represents the integrated SFR over the life of the starburst \citep[e.g.][]{cal08, ros02}. This is the estimate required to determine the rate at which galaxies form.  Different parameters are sensitive to different stars in the starbursts and to starbursts of different ages.  For example, the H$\alpha$ emission depends on the luminosity of the most massive, short lived stars with the greatest ionizing radiation shortward of the Lyman limit, but L$_{\nu}$(153 nm) measures luminosity from hot stars longward of the Lyman limit, so can be influenced by lower mass, longer lived stars.   Comparing these measures would not give the same result for starbursts with different mass distributions for the ionizing stars or different ages for the starburst.

The preferred parameter is the total luminosity of all stars in the starburst, which does not depend on what fraction of the stellar luminosity ionizes hydrogen.  This parameter is measured by $L_{ir}$, the total bolometric luminosity over all wavelengths, and it is dominated by infrared luminosity because most starburst luminosity is absorbed by obscuring dust.  This is the basis of  measures of SFR using infrared luminosities \citep{ken98}.   

The measure of SFR that we have used from $\nu$L$_{\nu}$(7.7 $\mu$m) is related empirically to $L_{ir}$ by using the starburst galaxies in \citet{bra06}, which all have IRAS flux measures allowing the determination of $L_{ir}$.  The transformation adopted is log $L_{ir}$ = log[$\nu$L$_{\nu}$(7.7$\mu$m)] + 0.78 \citep{hou07}, and this is confirmed to extend to starbursts at z $\ga$ 2 using submillimeter determinations of $L_{ir}$ \citep{pop08}.  (The $L_{ir}$ in Brandl et al. is a total 8-1000 \um infrared luminosity as estimated from IRAS fluxes according to \citet{san96}.)  Using this transformation, we determine log $L_{ir}$ compared to M(UV) for all sources in our samples.

Results are shown in Figures 17 and 18, comparing the ultraviolet-selected samples to the infrared-selected samples.  In Figure 17, the lines which are fit compare the extreme $Spitzer$ sample of Table 5 to the ultraviolet Markarian+GALEX samples.  the extremes of infrared selection and ultraviolet selection for starbursts.  This Figure demonstrates in alternative fashion to the discussion in section 5.2 that there exists a population of very obscured starbursts which have larger luminosities $L_{ir}$ than found among ultraviolet-selected starbursts and which have faint M(UV) because of this obscuration. 

The offset between the fits in Figure 17 indicates that ultraviolet-selected starbursts of a given M(UV) can underestimate bolometric luminosities by about a factor of 30 over all luminosities when compared with the most obscured starbursts. Another way to state the result is that starbursts of the same $L_{ir}$ can be found at all -16 $>$ M(UV) $>$ -21, and those which appear faintest in M(UV) are so because of obscuration.  If there are large populations of starbursts at z $>$ 3 which are as dusty and obscured as these extreme $Spitzer$ sources, the conclusion that only low luminosity galaxies exist at such redshifts is incorrect.  

In Figure 18, we show the samples compared with the fit found for the full infrared-selected sample of 287 starbursts from \citet{sar09}.  In this case, differences between lines for the ultraviolet-selected sample (lower line) and the infrared-selected sample (upper line) are not as large as in Figure 17.  The offset gives another estimate of obscuration corrections for ultraviolet-selected galaxies when compared to infrared-selected samples for comparison with the obscuration corrections in section 5.2. 

For the samples and fits in Figures 17 and 18, we conclude the following.  For ultraviolet-selected galaxies, log $L_{ir}$  = -(0.33$\pm$0.04)M(UV)+4.52$\pm$0.69.  For the full $Spitzer$ infrared-selected starburst sample, log $L_{ir}$ = -(0.23$\pm$0.02)M(UV)+6.99$\pm$0.41. For the extreme infrared-selected $Spitzer$ sample in Tables 5 and 6, log $L_{ir}$ = -(0.25$\pm$0.03)M(UV)+7.2$\pm$0.48, for $L_{ir}$ in \ldot~and M(UV) the AB magnitude at rest frame 153 nm.


All of these interpretations depend on an accurate determination of $L_{ir}$.  A crucial requirement is to improve the calibration of  $L_{ir}$ to $\nu$L$_{\nu}$(7.7$\mu$m).  We can make one test of the adopted calibration using our new IRS spectra of Markarian galaxies, because most have IRAS fluxes.  We show in Figure 19 the result of this ratio for both the Brandl starbursts and for our new observations of Markarian galaxies in Table 2.


Results in Figure 19 show that, within the uncertainties, the ultraviolet-selected Markarian galaxies have the same transformation from $L_{ir}$ to $\nu$L$_{\nu}$(7.7$\mu$m) as we have explained and adopted in previous analyses for $Spitzer$-selected sources \citep{hou07,wee08}.  This is an important confirmation that this transformation and the subsequent use of $L_{ir}$ to determine SFRs is not modified by starbursts with the hotter ionizing continuum seen in the Markarian starbursts.  Nevertheless, improvements for this calibration using far infrared and millimeter observations of starbursts with a variety of luminosities and redshifts are essential.

\section{Obscuration Corrections and Evolution for Starburst Galaxies at High Redshift}

Our motivation for comparing ultraviolet-selected and infrared-selected starbursts is to determine an independent measure of obscuration by dust, because the value of such a dust correction is  crucial for measuring the SFR density.  Present conclusions (summarized in Bouwens et al. 2009), which track the SFR density to z $\sim$ 7, indicate that obscuration for luminous galaxies, M(UV) $\la$ -20, is substantial, requiring a correction factor to ultraviolet luminosities of 6$\pm$2.5.  The overall dust correction factor to total SFR density is taken as only a factor of $\sim$ 2, however, because the correction is smaller for low luminosity galaxies which dominate the observed SFR density. As concluded by Bouwens et al., "Because these same lower luminosity galaxies dominate the luminosity density in the UV continuum, the overall dust extinction correction remains modest at all redshifts and the evolution of this correction with redshift is only modest." 

Having a correct measurement of dust obscuration is crucial to determine the true SFR density as a function of redshift because the form of evolution for SFR density depends on this correction.  For example, the observed decrease in the SFR density from z $\sim$ 2 to z $\sim$ 7 in \citet{red09} is comparable to the factors applied for dust correction for the most luminous sources. Many assumptions and explanations enter in the determination of such dust corrections, as thoroughly discussed in \citet{bou09} and \citet{red09}.  Fundamentally, the adopted corrections depend on interpreting as reddening by dust the observed differences in the slopes of the ultraviolet continuum for LBGs of different luminosities.

The ultimate measure of SFR derives from the corrected values of $L_{ir}$ which are finally adopted after applying the corrections.  The consequences of different methods for determining corrections can be compared, therefore, simply by comparing the derived $L_{ir}$ as a function of ultraviolet luminosity, M(UV).  These are the comparisons we now make using our methodology from section 6 for determining $L_{ir}$ based on infrared PAH luminosities.

Results are shown in Figure 20.  We consider that Figure 20 summarizes the most important results of our present paper.  The relation between $L_{ir}$ and M(UV) for the various samples which we consider is compared to that adopted in \citet{bou09} for LBGs at high redshift used to determine the evolution of SFR density for z $>$ 2.  In Figure 20, the line with asterisks is taken from Figure 10 of Bouwens et al. for z = 2.5.  The comparison of this line with the points in Figure 20 indicates that the result adopted for high redshift LBGs really applies only to the lower envelope for all starbursts.  

Differences between this line and the other lines in Figure 20 show differences that would also arise for resulting measures of SFR density.  The adopted results for LBGs agree with the full infrared-selected sample at M(UV) $\sim$ -21, but differ by a factor $>$ 10 at M(UV) = -17.  Differences are greater at fainter M(UV) for reasons discussed in section 5.2, that the faintest sources in the ultraviolet are those with the greatest obscuration by dust.

The crucial question that results from Figure 20 is:  which line represents the true population of starburst galaxies at high redshift?  Are the dust properties determined for our local samples (z $<$ 0.5) similar to properties in high redshift starbursts of similar M(UV)?  Answering this question for z $>$ 2 is not feasible at present, because there are no significant samples of sources discovered at such redshifts using their dust emission.  

A direct comparison could be made at z $\sim$ 2 because of the large population of DOGs discovered by $Spitzer$ at about this redshift.  A strong selection effect leads to a median redshift z = 1.8$\pm$0.2 for the starburst DOGs with f$_{\nu}$(24 \ums) $>$ 0.5 mJy, using the high redshift, dusty starbursts shown in Figure 16 (asterisks).  This selection effect arises when the strong PAH 7.7 \um feature falls within the MIPS 24 \um band used for the surveys that find these DOGs.  When the 24 \um selection is also coupled to a criterion of having a rest-frame 1.6 \um excess from stellar luminosity ("bump" sources; see section 5.5), there is nearly complete success in finding a dusty starburst with z $\sim$ 2  \citep {wee06,far08,des09}.  

This selection using "bump" sources means that existing $Spitzer$ surveys covering tens of deg$^{2}$, such as the First Look Survey \citep{fad06}, the Bootes survey \citep{dey08}, and the SWIRE surveys \citep{lon04}, can yield reliable lists of dusty starbursts in a well defined redshift interval near z $\sim$ 2 , even if spectroscopic redshifts are not available for all of the sources.  There is also a straightforward empirical transformation in this redshift interval that f$_{\nu}$(7.7 \ums) = (1.5$\pm$0.3)f$_{\nu}$(24 \ums), for DOGs like those in Figure 16.  The space density and luminosity function of such starbursts could be determined to f$_{\nu}$(24 \ums) $>$ 0.3 mJy, for which the f$_{\nu}$(7.7 \ums) $\sim$ 0.5 mJy, or log $\nu$L$_{\nu}$(7.7 $\mu$m) = 45.2 at z = 1.8.  

From Figures 17 and 18, this value of $\nu$L$_{\nu}$(7.7 $\mu$m) corresponds to -16 $>$ M(UV) $>$ -22 , depending on whether the source is highly obscured or has lower obscuration such as in the ultraviolet-selected samples.  This range of M(UV) is easily accessible in surveys for UVLGs at z $\sim$ 2,  so such surveys in the same regions where are available the $Spitzer$ surveys would make possible a direct comparison between infrared and ultraviolet selection of the starburst population at z $\sim$ 2. This would provide a definitive answer regarding obscuration corrections and measures of true SFR densities as a function of M(UV).

Even though existing wide-area infrared surveys reach only the top end of the bolometric luminosity function at high redshifts, it is encouraging that tracking the most luminous sources gives results for the form of evolution over all redshifts that are consistent with results found from total SFR densities which reach deeper into luminosity functions.  We reach this conclusion by comparing the evolution of the SFR density as illustrated in \citet{red09}, their Figure 10, with the luminosity evolution factor of (1+z)$^{2.5}$ found from $Spitzer$ surveys for the most luminous starbursts \citep{wee08}.  This scaling for luminosity evolution matches, within the uncertainties, the scaling of total SFR density for 0 $<$ z $<$ 2 summarized in \citet{red09}.

\section{Summary and Conclusions}

1. We have selected starburst galaxies discovered using both infrared and ultraviolet selection for comparison of spectral parameters measured with the $Spitzer$ Infrared Spectrograph to ultraviolet photometry from GALEX.  Our primary objective was to determine how the luminosity and dust obscuration of starbursts depend on selection technique.  Sources have z $<$ 0.5 and cover a luminosity range of $\sim$ 10$^{4}$; the highest luminosities are comparable to starburst luminosities found at z $\sim$ 3.

2. We measure and illustrate spectra of 26 Markarian galaxies, 23 ultraviolet luminous galaxies discovered with GALEX, and 50 galaxies with the most extreme infrared/ultraviolet ratios discovered with $Spitzer$.  We find (section 4.3) that these samples differ in the nature of the ionizing stars, with hotter stars in the ultraviolet-selected sources as evidenced by stronger [NeIII] emission and stronger mid-infrared dust continuum (section 4, Figures 6 - 8.)

3. It is found that a strong selection effect arises for the ultraviolet-selected samples: the most luminous UV sources appear luminous because they have the least obscuration, not because they have the largest bolometric luminosities (section 5.3).  Comparisons between rest-frame infrared fluxes f$_{\nu}$(7.7 \ums) arising from PAH emission and ultraviolet fluxes f$_{\nu}$(153 nm) arising from the stellar continuum allow an empirical method for determining obscuration in starbursts and dependence of this obscuration on infrared or ultraviolet luminosity.    Even sources selected in the ultraviolet and which resemble Lyman Break Galaxies require obscuration corrections of about a factor of 10.  These results are contrary to those adopted in studies of LBGs, which find that the most luminous galaxies in M(UV) are also the most obscured. Obscuration correction for the ultraviolet-selected Markarian+GALEX sample has the form log[UV(intrinsic)/UV(observed)] = 0.07($\pm$0.04)M(UV)+2.09$\pm$0.69; the full infrared-selected $Spitzer$ sample is log[UV(intrinsic)/UV(observed)] = 0.17($\pm$0.02)M(UV)+4.55($\pm$0.4); and is log [UV(intrinsic)/UV(observed)] = 0.14($\pm$0.03)M(UV)+4.68$\pm$0.50 for the extreme infrared-selected $Spitzer$ sample (Figure 13).

4. Analysis of obscuration determined from the H$\alpha$/H$\beta$ ratio in SDSS spectra confirms the trends observed by comparing infrared/ultraviolet ratios, although obscuration measured by H$\alpha$/H$\beta$ is much less than measured by infrared/ultraviolet (section 5.4 and Figure 15).  

5. Stellar luminosities measured at rest frame 1.6 \um are exceptionally uniform compared to starburst luminosity over a redshift range 0 $<$ z $<$ 2 and luminosity range of 10$^{4}$, but the ratio from different samples indicates that the 1.6 \um stellar luminosity is also obscured by dust in proportion to the ultraviolet obscuration.  This leads to our suggesting that stellar luminosities measured at 1.6 \um reveal the recently evolved stars from the starburst rather than an underlying old population (section 5.5 and Figure 16).

6. We examine the relation of total bolometric luminosity $L_{ir}$ to M(UV) for infrared-selected and ultraviolet-selected samples.  Spectroscopy of Markarian galaxies confirms the transformation assumed for $\nu$L$_{\nu}$(7.7 $\mu$m) to $L_{ir}$ and implies that the hotter ionizing stars in the uv-selected samples do not affect this calibration.  For ultraviolet-selected galaxies, log $L_{ir}$  = -(0.33$\pm$0.04)M(UV)+4.52$\pm$0.69.  For the full infrared-selected sample, log $L_{ir}$ = -(0.23$\pm$0.02)M(UV)+6.99$\pm$0.41, and log $L_{ir}$ = -(0.25$\pm$0.03)M(UV)+7.2$\pm$0.48 for the extreme infrared-selected $Spitzer$ sample, all for $L_{ir}$ in \ldot and M(UV) the AB magnitude at rest frame 153 nm (section 6 and Figure 18). 

7. Our most important summary result (section 7 and Figure 20) indicates that the bolometric luminosities deduced for starburst galaxies using LBGs at z $>$ 2 are insufficient, by a factor of at least 10 for M(UV) $\sim$ -17.  This also implies that obscuration corrections by factors of two to three applied for 3 $<$ z $<$ 7 to track the formation and evolution of the earliest galaxies are insufficient, if dusty galaxies exist at such epochs comparable to those with z $<$ 3.  We propose an observational test which can determine more accurate dust corrections and $L_{ir}$ for large samples of infrared-selected starbursts and LBGs at z $\la$ 2.  This test is possible because existing $Spitzer$ surveys allow reliable discovery of dusty starbursts with z = 1.8 $\pm$ 0.2 based only on photometric properties.

\acknowledgments
We thank Don Barry for continued technical assistance and V. Lebouteiller and J. Bernard-Salas for providing the SMART-Optimal Extraction software. This work is based on observations made with the
$Spitzer$ Space Telescope, which is operated by the Jet Propulsion
Laboratory, California Institute of Technology, under NASA contract
1407.  Support for this work by the IRS GTO team at Cornell University was provided by NASA through Contract
Number 1257184 issued by JPL/Caltech. Support was also provided by the US Civilian Research and Development Foundation under grant ARP1-2849-YE-06.  LAS and DWW thank D. Engels and the Hamburger Sternwarte for hospitality during preparation of this paper, and LAS acknowledges the Deutsche Forschungsgemeinschaft for support through grant En 176/36-1.

\clearpage


\begin{deluxetable}{lccccccccccccc} 
\rotate
\tablecolumns{14}
\tabletypesize{\footnotesize}
\tablewidth{0pc}
\tablecaption{Spectroscopic Data for Markarian Sample}
\tablehead{
\colhead{No.} & \colhead{Name} & \colhead{J2000 coordinates} & \colhead{z\tablenotemark{a}}& \colhead{EW(6.2)\tablenotemark{b}} & \colhead{f$_{\nu}$(7.7)\tablenotemark{c}} & \colhead{f(11.3)\tablenotemark{d}} & \colhead{EW(11.3)} & \colhead{f(15.5)\tablenotemark{e}} & \colhead{f$_{\nu}$(24)\tablenotemark{f}} & \colhead{f$_{\nu}$(153)\tablenotemark{g}} &  \colhead{H$\alpha$/H$\beta$)\tablenotemark{h}} & \colhead{f($H$)\tablenotemark{i}} & \colhead{f($K$)\tablenotemark{j}} \\
\colhead{} & \colhead{} & \colhead{} & \colhead{} & \colhead{\ums} & \colhead{mJy} & \colhead{} & \colhead{\ums} & \colhead{} & \colhead{mJy} & \colhead{$\mu$Jy} & \colhead{}  &  \colhead{mJy} & \colhead{mJy}
}

\startdata
1 & Mkn 353 & 010316.58+222032.6 & 0.014 & 0.71 & 176(2.04) & 23.18 & 0.78 & 1.47 & 422 & 266 & \nodata & 18.1 & 17.0  \\%
2 & Mkn 1158 & 013459.36+350222.1 & 0.015 & 0.69 & 47.3(1.19) & 3.43 & 0.50 & 1.27 & 160 & 403 & \nodata & 5.6 & 5.3  \\%
3 & Mkn 363 & 015058.56+215948.0 & 0.010 & 0.65 & 64.5(1.41) & 6.64 & 0.70 & 1.16 & 145 & 772 & \nodata & 9.4 & 8.0  \\%
4 & Mkn 1012 & 015657.73-052411.7 & 0.020 & 0.48 & 90.8(1.41) & 9.35 & 0.54 & $<$0.52 & 466 & 218 & \nodata & 14.6 & 14.7  \\%
5 & Mkn 603 & 030856.85-025718.4 & 0.007 & 0.61 & 258(1.63) & 26.67 & 0.48 & 13.88 & 1790 & 1140 & \nodata & 17.5 & 18.9 \\%
6 & Mkn 14 & 081059.31+724740.3 & 0.008 & 0.59 & 28.1(1.47) & 2.62 & 0.51 & 0.51 & 120 & 884 & \nodata & 5.9 & 4.7  \\%
7 & Mkn 87 & 082140.17+735917.5 & 0.009 & 0.52 & 48.7(1.28) & 4.68 & 0.74 & 0.07 & 71.3 & 187 & \nodata & 14.6 & 12.3  \\%
8 & Mkn 90 & 083000.10+524148.9 & 0.014 & 0.52 & 30.3(1.32) & 2.88 & 0.65 & 0.16 & 80.4 & 194 & 5.5 & 4.8 & 5.0  \\%
9 & Mkn 93\tablenotemark{k} & 083642.35+661358.3 & 0.017 & 0.36 & 48.7\nodata & 2.77 & 0.19 & 3.02 & 367 & 147 & \nodata & 3.5 & 3.4  \\%
10 & Mkn 105 & 092026.34+712416.0 & 0.012 & 0.62 & 13.9\nodata & 0.88 & 0.50 & 0.17 & 45.3 & 120 & \nodata & 2.9 & 2.2  \\%
11 & Mkn 21 & 094830.73+575814.4 & 0.027 & 0.45 & 24.4(1.54) & 2.66 & 0.70 & 0.1 & 40.3 & 305 & 7.2 & 7.3 & 6.1  \\%
12 & Mkn 140 & 101628.23+451919.4 & 0.004 & 0.51 & 13.9\nodata & 1.02 & 0.58 & 0.60 & 43.1 & 738 & 3.6 & 2.4 & 1.9  \\%
13 & Mkn 31 & 101942.83+572524.9 & 0.026 & 0.50 & 22.8\nodata & 1.74 & 0.66 & 0.15 & 58.0 & 135 & 4.6 & 4.3 & 3.8  \\%
14 & Mkn 149\tablenotemark{k} & 103801.78+641559.0 & 0.005 & 0.24 & 17.0\nodata & 1.53 & 0.23 & $<$0.08 & 165 & 163 & 5.3 & 7.3 & 7.4  \\%
15 & Mkn 165 & 111835.60+631641.0 & 0.011 & 0.61 & 16.8(1.57) & 1.95 & 0.78 & 0.26 & 23.0 & 785 & 4.9 & 8.1 & 6.6  \\%
16 & Mkn 193\tablenotemark{k} & 115528.50+573944.0 & 0.021 & $<$0.04 & $<$2.9\nodata & 0.07 & 0.03 & 0.4 & 60.4 & 332 &  \nodata & $<$0.3 & \nodata  \\%
17 & Mkn 764\tablenotemark{k} & 121600.04+124114.0 & 0.066 & 0.47 & 32.9(1.49) & 3.23 & 0.64 & 0.07 & 42.9 & 242 & 7.2 & 3.6 & 3.6  \\%
18 & Mkn 439 & 122436.30+392259.0 & 0.005 & 0.61 & 132.0(1.80) & 17.86 & 0.72 & 0.71 & 409.0 & 3760 & \nodata & 22.6 & 18.8  \\%
19 & Mkn 769 & 122525.54+162811.8 & 0.006 & 0.59 & 194(1.10) & 14.48 & 0.61 & 5.62 & 638 & 5770 & \nodata & 29.1 & 21.7  \\%
20 & Mkn 799 & 140045.70+591942.0 & 0.009 & 0.51 & 207(1.67) & 22.54 & 0.51 & 0.46 & 857 & \nodata & 6.9 & 31.6 & 43.6  \\%
21 & Mkn 286 & 141926.97+713517.4 & 0.026 & 0.64 & 187(1.46) & 16.16 & 0.62 & 1.14 & 597 & 1220 & \nodata & 14.5 & 13.0  \\%
22 & Mkn 480 & 150633.73+423829.2 & 0.019 & 0.66 & 101(1.45) & 8.91 & 0.72 & 0.29 & 170 & 523 & 5.2 & 8.5 & 9.2  \\%
23 & Mkn 489 & 154421.68+410509.8 & 0.032 & 0.62 & 77.0(1.57) & 7.7 & 0.58 & 1.14 & 217 & 972 & 4.3 & 3.6 & 3.2  \\%
24 & Mkn 691 & 154658.90+175308.0 & 0.011 & 0.65 & 106(1.34) & 8.13 & 0.52 & 0.88 & 355 & \nodata & 4.4 & 12.7 & 11.2  \\%
25 & Mkn 863\tablenotemark{k} & 155625.90+090318.8 & 0.043 & 0.30 & 47.6(1.16) & 3.02 & 0.38 & 0.14 & 103 & 341 & 5.3 & 4.8 & 6.5  \\%
26 & NGC 7714\tablenotemark{l} & 233614.10+020919.0 & 0.010 & 0.59 & 354(1.17) & 25.35 & 0.36 & 7.36 & 2150 & 4190 & \nodata & 21.8 & 20.4  \\%

\enddata
\tablenotetext{a}{Redshift from PAH features in IRS spectra.}
\tablenotetext{b}{Rest frame equivalent width of 6.2 \um PAH feature fit with single gaussian on a linear continuum within rest wavelength range 5.5 \um to 6.9 \um.}
\tablenotetext{c}{Flux density in mJy at peak of 7.7 \um PAH feature.  The value in parenthesis is the "stitching" factor by which the SL1 spectrum has been multiplied to match the LL2 spectrum to account for a slightly resolved source that gives larger LL2 flux because the SL1 slit is 3.7 \arcsec~ wide, and the LL2 slit is 10.5 \arcsec~ wide.}
\tablenotetext{d}{Total flux of PAH 11.3 \um in units of 10$^{-20}$W cm$^{-2}$ s$^{-1}$.}
\tablenotetext{e}{Total flux of [NeIII] 15.5 \um in units of 10$^{-20}$W cm$^{-2}$ s$^{-1}$.}
\tablenotetext{f}{Continuum flux density at rest frame 24 \ums.}
\tablenotetext{g}{Flux density in $\mu$Jy observed at wavelength (1+z)153 nm interpolated with a power law between flux densities in the GALEX catalog measured with the FUV filter (observed 134-179 nm) and the NUV filter (observed 177-283 nm).}
\tablenotetext{h}{Observed emission line ratio H$\alpha$/H$\beta$ measured by us from SDSS spectra using ratios of $\lambda$f$_{\lambda}$ for f$_{\lambda}$ at the line peak shown in the SDSS spectra.}
\tablenotetext{i}{Observed flux density in the $H$ band of 2MASS using conversion that zero magnitude corresponds to 1024 Jy.}
\tablenotetext{j}{Observed flux density in the $K$ band of 2MASS using conversion that zero magnitude corresponds to 667 Jy.}
\tablenotetext{k}{SDSS spectrum shows classification as AGN, as discussed in section 4.2.}
\tablenotetext{l}{All spectra from our new $Spitzer$ program 50834, except NGC 7714 from archival program 14.}

\end{deluxetable}

\clearpage

\begin{deluxetable}{lcccccccc} 
\rotate
\tablecolumns{9}
\tabletypesize{\footnotesize}
\tablewidth{0pc}
\tablecaption{Luminosities and Star Formation Rates for Markarian Sample}
\tablehead{
\colhead{No.} & \colhead{Name} & \colhead{J2000 coordinates} &\colhead{logL(7.7\ums)\tablenotemark{a}} & \colhead{logL(UV)\tablenotemark{b}} & \colhead{M(UV)\tablenotemark{c}} & \colhead{log SFR(PAH)\tablenotemark{d}} & \colhead{log SFR(UV)\tablenotemark{e}} & \colhead{$L_ir$\tablenotemark{f}} \\
\colhead{} & \colhead{} & \colhead{} &  \colhead{log erg s$^{-1}$}  & \colhead{log erg s$^{-1}$} & \colhead{} & \colhead{log\mdot} & \colhead{log\mdot} & \colhead{log erg s$^{-1}$}
}

\startdata
1 & Mkn 353 & 010316.58+222032.6 & 43.46 & 42.34 & -16.03 & 0.89 & -0.92 & 44.07,(0.70) \\%
2 & Mkn 1158 & 013459.36+350222.1 & 42.95 & 42.58 & -16.63 & 0.38 & -0.68 & $<$43.65,(0.70) \\%
3 & Mkn 363 & 015058.56+215948.0 & 42.73 & 42.51 & -16.45 & 0.16 & -0.75 & $<$43.24,(1.20) \\%
4 & Mkn 1012 & 015657.73-052411.7 & 43.48 & 42.57 & -16.61 & 0.91 & -0.69 & 44.41,(0.83) \\%
5 & Mkn 603 & 030856.85-025718.4 & 43.02 & 42.37 & -16.09 & 0.45 & -0.89 & 43.99,(0.79) \\%
6 & Mkn 14 & 081059.31+724740.3 & 42.17 & 42.37 & -16.11 & -0.40 & -0.89 & $<$43.04,(0.77) \\%
7 & Mkn 87 & 082140.17+735917.5 & 42.52 & 41.80 & -14.68 & -0.05 & -1.46 & $<$43.23,(0.61)\\%
8 & Mkn 90 & 083000.10+524148.9 & 42.70 & 42.20 & -15.69 & 0.13 & -1.06 & $<$43.41,(0.49) \\%
9 & Mkn 93\tablenotemark{g} & 083642.35+661358.3 & 43.07 & 42.25 & -15.80 & \nodata & \nodata & 43.85,(0.75) \\%
10 & Mkn 105 & 092026.34+712416.0 & 42.22 & 41.86 & -14.83 & -0.35 & -1.40 & $<$43.01,(0.53) \\%
11 & Mkn 21 & 094830.73+575814.4 & 43.17 & 42.97 & -17.61 & 0.60 & -0.29 & $<$43.67,(0.36) \\%
12 & Mkn 140 & 101628.23+451919.4 & 41.26 & 41.69 & -14.40 & -1.31 & -1.57 & $<$41.86,(0.39) \\%
13 & Mkn 31 & 101942.83+572524.9 & 43.11 & 42.59 & -16.64 & 0.54 & -0.67 & \nodata,\nodata \\%
14 & Mkn 149\tablenotemark{g} & 103801.78+641559.0 & 41.55 & 41.23 & -13.25 & \nodata & \nodata & $<$42.46,(0.82) \\%
15 & Mkn 165 & 111835.60+631641.0 & 42.23 & 42.60 & -16.68 & -0.34 & -0.66 & $<$42.57,(0.29) \\%
16 & Mkn 193\tablenotemark{g} & 115528.50+573944.0 & $<$42.03 & 42.79 & -16.73 & \nodata & \nodata & 43.50,(0.57) \\%
17 & Mkn 764\tablenotemark{g} & 121600.04+124114.0 & 44.09 & 43.66 & -19.32 & \nodata & \nodata & $<$44.30,(0.23) \\%
18 & Mkn 439 & 122436.30+392259.0 & 42.44 & 42.60 & -16.66 & -0.13 & -0.66 & 43.28,(0.58) \\%
19 & Mkn 769 & 122525.54+162811.8 & 42.76 & 42.90 & -17.43 & 0.19 & -0.36 & 43.54,(0.60) \\%
20 & Mkn 799 & 140045.70+591942.0 & 43.15 & \nodata & \nodata & 0.58 & \nodata & 43.99,(0.50) \\%
21 & Mkn 286 & 141926.97+713517.4 & 44.03 & 43.54 & -19.04 & 1.46 & 0.28 & 44.71,(0.77) \\%
22 & Mkn 480 & 150633.73+423829.2 & 43.49 & 42.90 & -17.43 & 0.92 & -0.36 & 44.07,(0.83) \\%
23 & Mkn 489 & 154421.68+410509.8 & 43.82 & 43.63 & -19.24 & 1.25 & 0.37 & $<$44.65,(0.91) \\%
24 & Mkn 691 & 154658.90+175308.0 & 43.03 & \nodata & \nodata & 0.46 & \nodata & 43.85,(0.64) \\%
25 & Mkn 863\tablenotemark{g} & 155625.90+090318.8 & 43.87 & 43.43 & -18.75 & \nodata & \nodata  & 44.26,(0.04)\\%
26 & NGC 7714 & 233614.10+020919.0 & 43.44 & 43.22 & -18.22 & 0.87 & -0.04 & 44.19,(0.71) \\%

\enddata

\tablenotetext{a}{Rest frame luminosity of 7.7 \um PAH feature in erg s$^{-1}$ determined as L(7.7 \ums) = $\nu$L$_{\nu}$(7.7 $\mu$m) =  4$\pi$D$_{L}$$^{2}$[$\nu$/(1+z)]$f_{\nu}$(7.7 $\mu$m), for $\nu$ corresponding to 7.7 \um and using f$_{\nu}$(7.7 $\mu$m) at observed wavelength of the 7.7 \um feature from Table 1.  Luminosity distances D$_{L}$ from \citet{wri06}:  http://www.astro.ucla.edu/~wright/CosmoCalc.html, for H$_0$ = 71 \kmsMpc, $\Omega_{M}$=0.27 and $\Omega_{\Lambda}$=0.73.  (Log [$\nu$L$_{\nu}$(\ldot)] = log [$\nu$L$_{\nu}$(erg s$^{-1}$)] - 33.59.)}
\tablenotetext{b}{Rest frame ultraviolet continuum luminosity at 153 nm in erg s$^{-1}$ determined as L(UV) = $\nu$L$_{\nu}$(153 nm), using f$_{\nu}$(153 nm) from Table  at observed wavelength (1+z)153 nm.}
\tablenotetext{c}{M(UV) is an absolute AB magnitude at continuum wavelength 153 nm, determined with the fundamental definition m$_{AB}$ = -2.5logf$_{\nu}$  - 48.60, for f$_{\nu}$ in erg cm$^{-2}$ s$^{-1}$ Hz$^{-1}$ \citep{oke74}.  M(UV) is determined as the m(UV) which arises using L$_{\nu}$(153 $\mu$m) to determine f$_{\nu}$(153 nm) at a distance of 10 pc.}
\tablenotetext{d}{Star formation rates in units of \mdot determined as log [SFR(PAH)] = log L(7.7 $\mu$m) - 42.57.}
\tablenotetext{e}{Star formation rates in units of \mdot determined as log [SFR(UV)] = log L(UV) - 43.26.} 
\tablenotetext{f}{Total infrared luminosity determined from IRAS flux densities by taking total infrared flux from  8-1000 \um to be 1.8[13.48S(12\ums) + 5.16S(25\ums) + 2.58S(60\ums) + S(100\ums)] from \citet{san96}, where S is IRAS flux density in Jy.  Value in parenthesis is the ratio f$_{\nu}$(IRS 25 \ums)/f$_{\nu}$(IRAS 25 \ums) determined by using synthetic photometry with SMART for the IRAS 25 \um band to determine (IRS 25 \ums) from the IRS spectrum in Figure 1.}
\tablenotetext{g}{SDSS spectrum shows classification as AGN, as discussed in section 4.2.}

\end{deluxetable}

\clearpage

\begin{deluxetable}{lccccccccccccc} 
\rotate
\tablecolumns{14}
\tabletypesize{\footnotesize}
\tablewidth{0pc}
\tablecaption{Spectroscopic Data for GALEX Sample}
\tablehead{
\colhead{No.} & \colhead{Name} & \colhead{J2000 coordinates} & \colhead{z\tablenotemark{a}}& \colhead{EW(6.2)\tablenotemark{b}} & \colhead{f$_{\nu}$(7.7)\tablenotemark{c}} & \colhead{f(11.3)\tablenotemark{d}} & \colhead{EW(11.3)} & \colhead{f(15.5)\tablenotemark{e}} & \colhead{f$_{\nu}$(24)\tablenotemark{f}} & \colhead{f$_{\nu}$(153)\tablenotemark{g}} &  \colhead{H$\alpha$/H$\beta$)\tablenotemark{h}} & \colhead{f($H$)\tablenotemark{i}} & \colhead{f($K$)\tablenotemark{j}} \\
\colhead{} & \colhead{} & \colhead{} & \colhead{} & \colhead{\ums} & \colhead{mJy} & \colhead{} & \colhead{\ums} & \colhead{} & \colhead{mJy} & \colhead{$\mu$Jy} & \colhead{}  &  \colhead{mJy} & \colhead{mJy}
}

\startdata
1 & SDSS\tablenotemark{k} & 001054.84+001451.3 & 0.244 & 0.15 & 6.3 & 0.34 & 0.42 & $<$0.01 & 13.5 & 26.5 & 8.21 & 0.8 & 1.1 \\%
2 & SDSS & 005439.79+155446.9 & 0.239 & 0.27 & 3.3 & 0.15 & 0.33 & $<$0.01 & 7.1 & 13.2 & 5.68 & $<$0.3 & \nodata \\%
3 & SDSS\tablenotemark{k} & 005527.46-002148.8 & 0.168 & 0.08 & 6.6 & 0.20 & 0.10 & 0.25 & 50.3 & 77.9 & 3.94 & 0.3 & 0.4 \\%
4 & SDSS\tablenotemark{k} & 015028.41+130858.4 & 0.147 & 0.16 & 18.1 & 0.34 & 0.26 & 0.31 & 93.4 & 82.0 & 4.01 & 0.5 & 0.5 \\%
5 & SDSS & 032846.00+011150.8 & 0.143 & 0.46 & 1.2 & 0.09 & 0.65 & $<$0.03 & 6.1 & 31.3 & 3.76 & $<$0.3 & \nodata \\%
6 & SDSS & 035733.99-053719.7 & 0.204 & 0.42 & 1.9 & 0.07 & 0.37 & $<$0.06 & 7.2 & 12.6 & 4.14 & $<$0.3 & \nodata \\%
7 & SDSS & 040208.87-050642.1 & 0.138 & 0.74 & 0.5 & 0.03 & 0.35 & $<$0.02 & 1.4 & 45.2 & 3.88 & $<$0.3 & \nodata \\%
8 & SDSS & 080232.34+391552.6 & 0.267 & 0.66 & 4.1 & 0.16 & 0.43 & 0.01 & 12.3 & 34.8 & 5.84 & $<$0.3 & \nodata \\%
9 & SDSS & 080844.27+394852.3 & 0.093 & 0.53 & 9.2 & 0.52 & 0.29 & $<$0.02 & 51 & 160.5 & 4.69 & 0.9 & 0.6 \\%
10 & SDSS\tablenotemark{k} & 082001.72+505039.1 & 0.219 & 0.04 & 3.6 & 0.03 & 0.06 & 0.05 & 23.9 & 30.5 & 3.70 & $<$0.3 & \nodata \\%
11 & SDSS & 082550.95+411710.2 & 0.156 & 0.46 & 1.7 & 0.09 & 0.50 & 0.05 & 4.7 & 42.2 & 4.06 & 0.2 & 0.4 \\%
12 & SDSS\tablenotemark{k} & 083803.72+445900.3 & 0.145 & $<$0.09 & 0.4 & 0.03 & 0.40 & 0.03 & 3.6 & $<$77.0 & 3.09 & $<$0.3 & \nodata \\%
13 & SDSS & 092159.38+450912.3 & 0.236 & 0.23 & 11.3 & 0.42 & 0.51 & 0.06 & 44.7 & 78.3 & 5.37 & 0.4 & 0.6 \\%
14 & SDSS & 092336.45+544839.2 & 0.227 & 0.58 & 1.1 & 0.08 & 0.64 & 0.03 & 4.1 & 44.5 & 4.19 & $<$0.3 & \nodata \\%
15 & SDSS\tablenotemark{k} & 092600.41+442736.1 & 0.182 & $<$0.17 & $<$0.3 & $<$0.002 & $<$0.01 & 0.06 & 5.9 & $<$94.1 & 3.24 & $<$0.3 & \nodata \\%
16 & SDSS & 093813.50+542825.1 & 0.101 & 0.18 & 2.3 & 0.11 & 0.16 & 0.29 & 28.3 & 249.8 & 3.38 & 0.3 & 0.5 \\%
17 & SDSS & 102613.97+484458.9 & 0.157 & 0.91 & 1.3 & 0.09 & 0.57 & 0.07 & 5.9 & 85.8 & 3.34 & $<$0.3 & \nodata \\%
18 & SDSS & 124819.75+662142.7 & 0.260 & 0.67 & 1.4 & 0.08 & 0.52 & 0.04 & 5.5 & 33.8 & 4.08 & $<$0.3 & \nodata \\%
19 & SDSS & 135355.90+664800.6 & 0.200 & 0.90 & 3.0 & 0.20 & 0.81 & 0.05 & 8.2 & 100.0 & 3.80 & 0.4 & 0.6 \\%
20 & SDSS & 143417.15+020742.5 & 0.181 & 0.61 & 3.7 & 0.24 & 0.68 & 0.06 & 9 & 50.7 & 4.47 & $<$0.3 & \nodata \\%
21 & SDSS & 210358.75-072802.5 & 0.136 & 0.38 & 21.1 & 0.11 & 0.19 & 0.17 & 128.9 & 84.8 & 5.20 & 1.7 & 1.9 \\%
22 & SDSS & 214500.25+011157.5 & 0.205 & 0.66 & 3.9 & 0.22 & 0.77 & 0.08 & 8.5 & 32.7 & 4.13 & $<$0.3 & \nodata \\%
23 & SDSS\tablenotemark{k} & 231812.99-004126.1 & 0.253 & $<$0.03 & $<$6.4 & 0.17 & 0.31 & 0.17 & 31.4 & $<$67.6 & 3.68 & $<$0.3 & \nodata \\%

\enddata

\tablenotetext{a}{Redshift from PAH features in IRS spectra.}
\tablenotetext{b}{Rest frame equivalent width of 6.2 \um PAH feature fit with single gaussian on a linear continuum within rest wavelength range 5.5 \um to 6.9 \um.}
\tablenotetext{c}{Flux density in mJy at peak of 7.7 \um PAH feature.}
\tablenotetext{d}{Total flux of PAH 11.3 \um in units of 10$^{-20}$W cm$^{-2}$ s$^{-1}$.}
\tablenotetext{e}{Total flux of [NeIII] 15.5 \um in units of 10$^{-20}$W cm$^{-2}$ s$^{-1}$.}
\tablenotetext{f}{Continuum flux density at rest frame 24 \ums.}
\tablenotetext{g}{Flux density in $\mu$Jy observed at wavelength (1+z)153 nm interpolated with a power law between flux densities in the GALEX catalog measured with the FUV filter (observed 134-179 nm) and the NUV filter (observed 177-283 nm).}
\tablenotetext{h}{Observed emission line ratio H$\alpha$/H$\beta$ measured by us from SDSS spectra using ratios of $\lambda$f$_{\lambda}$ for f$_{\lambda}$ at the line peak shown in the SDSS spectra.}
\tablenotetext{i}{Observed flux density in the $H$ band of 2MASS using conversion that zero magnitude corresponds to 1024 Jy.}
\tablenotetext{j}{Observed flux density in the $K$ band of 2MASS using conversion that zero magnitude corresponds to 667 Jy.}
\tablenotetext{k}{SDSS spectrum or IRS spectrum shows classification as AGN, as discussed in section 4.2.}

\end{deluxetable}

\clearpage

\begin{deluxetable}{lccccccc} 
\rotate
\tablecolumns{8}
\tabletypesize{\footnotesize}
\tablewidth{0pc}
\tablecaption{Luminosities and Star Formation Rates for GALEX Sample}
\tablehead{
\colhead{No.} & \colhead{Name} & \colhead{J2000 coordinates} &\colhead{logL(7.7\ums)\tablenotemark{a}} & \colhead{logL(UV)\tablenotemark{b}} & \colhead{M(UV)\tablenotemark{c}} & \colhead{log SFR(PAH)\tablenotemark{d}} & \colhead{log SFR(UV)\tablenotemark{e}} \\
\colhead{} & \colhead{} & \colhead{} &  \colhead{log erg s$^{-1}$}  & \colhead{log erg s$^{-1}$} & \colhead{} & \colhead{log\mdot} & \colhead{log\mdot}
}

\startdata
1 & SDSS\tablenotemark{f} & 001054.84+001451.3 & 44.54 & 43.87 & -19.84 & \nodata & \nodata \\%
2 & SDSS & 005439.79+155446.9 & 44.24 & 43.55 & -19.04 & \nodata & \nodata \\%
3 & SDSS\tablenotemark{f} & 005527.46-002148.8 & 44.22 & 44.00 & -20.17 & 1.65 & 0.74 \\%
4 & SDSS\tablenotemark{f} & 015028.41+130858.4 & 44.54 & 43.90 & -19.93 & \nodata & \nodata \\%
5 & SDSS & 032846.00+011150.8 & 43.35 & 43.46 & -18.82 & 0.78 & 0.20 \\%
6 & SDSS & 035733.99-053719.7 & 43.86 & 43.38 & -18.64 & 1.29 & 0.12 \\%
7 & SDSS & 040208.87-050642.1 & 42.92 & 43.59 & -19.14 & 0.35 & 0.33 \\%
8 & SDSS & 080232.34+391552.6 & 44.43 & 44.07 & -20.35 & 1.86 & 0.81 \\%
9 & SDSS & 080844.27+394852.3 & 43.84 & 43.78 & -19.64 & 1.27 & 0.52 \\%
10 & SDSS\tablenotemark{f} & 082001.72+505039.1 & 44.20 & 43.83 & -19.75 & \nodata & \nodata \\%
11 & SDSS & 082550.95+411710.2 & 43.57 & 43.67 & -19.34 & 1.00 & 0.41 \\%
12 & SDSS\tablenotemark{f} & 083803.72+445900.3 & 42.87 & 43.86 & -19.83 & \nodata & \nodata \\%
13 & SDSS & 092159.38+450912.3 & 44.76 & 44.31 & -20.94 & 2.19 & 1.05 \\%
14 & SDSS & 092336.45+544839.2 & 43.72 & 44.03 & -20.24 & 1.15 & 0.77 \\%
15 & SDSS\tablenotemark{f} & 092600.41+442736.1 & $<$43.0 & 44.15 & -20.56 & \nodata & \nodata \\%
16 & SDSS & 093813.50+542825.1 & 43.31 & 44.05 & -20.30 & 0.74 & 0.79 \\%
17 & SDSS & 102613.97+484458.9 & 43.47 & 43.98 & -20.13 & 0.90 & 0.72 \\%
18 & SDSS & 124819.75+662142.7 & 43.93 & 44.03 & -20.25 & 1.36 & 0.77 \\%
19 & SDSS & 135355.90+664800.6 & 44.04 & 44.26 & -20.84 & 1.47 & 1.00 \\%
20 & SDSS & 143417.15+020742.5 & 44.04 & 43.88 & -19.87 & 1.47 & 0.62 \\%
21 & SDSS & 210358.75-072802.5 & 44.54 & 43.85 & -19.79 & 1.97 & 0.59 \\%
22 & SDSS & 214500.25+011157.5 & 44.18 & 43.80 & -19.68 & 1.61 & 0.54 \\%
23 & SDSS\tablenotemark{f} & 231812.99-004126.1 & $<$44.58 & 44.31 & -20.94 & \nodata & \nodata \\%

\enddata

\tablenotetext{a}{Rest frame luminosity of 7.7 \um PAH feature in erg s$^{-1}$ determined as L(7.7 \ums) = $\nu$L$_{\nu}$(7.7 $\mu$m) =  4$\pi$D$_{L}$$^{2}$[$\nu$/(1+z)]$f_{\nu}$(7.7 $\mu$m), for $\nu$ corresponding to 7.7 \um and using f$_{\nu}$(7.7 $\mu$m) at observed wavelength of the 7.7 \um feature from Table 1.  Luminosity distances D$_{L}$ from \citet{wri06}:  http://www.astro.ucla.edu/~wright/CosmoCalc.html, for H$_0$ = 71 \kmsMpc, $\Omega_{M}$=0.27 and $\Omega_{\Lambda}$=0.73.  (Log [$\nu$L$_{\nu}$(\ldot)] = log [$\nu$L$_{\nu}$(erg s$^{-1}$)] - 33.59.)}
\tablenotetext{b}{Rest frame ultraviolet continuum luminosity at 153 nm in erg s$^{-1}$ determined as L(153 nm) = $\nu$L$_{\nu}$(153 nm), using f$_{\nu}$(153 nm) from Table 3 at observed wavelength (1+z)153 nm.}
\tablenotetext{c}{M(UV) is an absolute AB magnitude at continuum wavelength 153 nm, determined with the fundamental definition m$_{AB}$ = -2.5logf$_{\nu}$  - 48.60, for f$_{\nu}$ in erg cm$^{-2}$ s$^{-1}$ Hz$^{-1}$ \citep{oke74}.  M(UV) is determined as the m(UV) which arises using L$_{\nu}$(153 $\mu$m) to determine f$_{\nu}$(153 nm) at a distance of 10 pc.}
\tablenotetext{d}{Star formation rates in units of \mdot determined as log [SFR(PAH)] = log L(7.7 $\mu$m) - 42.57.}
\tablenotetext{e}{Star formation rates in units of \mdot determined as log [SFR(UV)] = log L(UV) - 43.26.} 
\tablenotetext{f}{SDSS spectrum or IRS spectrum shows classification as AGN, as discussed in section 4.2.}

\end{deluxetable}

\clearpage

\begin{deluxetable}{lccccccccccccc} 
\rotate
\tablecolumns{14}
\tabletypesize{\footnotesize}
\tablewidth{0pc}
\tablecaption{Spectroscopic Data for $Spitzer$ Sample}
\tablehead{
\colhead{No.} & \colhead{Name} & \colhead{J2000 coordinates} & \colhead{z\tablenotemark{a}}& \colhead{EW(6.2)\tablenotemark{b}} & \colhead{f$_{\nu}$(7.7)\tablenotemark{c}} & \colhead{f(11.3)\tablenotemark{d}} & \colhead{EW(11.3)} & \colhead{f(15.5)\tablenotemark{e}} & \colhead{f$_{\nu}$(24)\tablenotemark{f}} & \colhead{f$_{\nu}$(153)\tablenotemark{g}} &  \colhead{H$\alpha$/H$\beta$)\tablenotemark{h}} & \colhead{f($H$)\tablenotemark{i}} & \colhead{f($K$)\tablenotemark{j}} \\
\colhead{} & \colhead{} & \colhead{} & \colhead{} & \colhead{\ums} & \colhead{mJy} & \colhead{} & \colhead{\ums} & \colhead{} & \colhead{mJy} & \colhead{$\mu$Jy} & \colhead{}  &  \colhead{mJy} & \colhead{mJy}
}

\startdata
1  & SST24 & 021849.78-052158.4 & 0.292 & 0.71 & 6.2 & 0.24 & 0.53 & $<$0.01 & 11.6 & 3.5 & \nodata & $<$0.3 & \nodata \\%
2  & SST24 & 022205.05-050536.7 & 0.259 & 0.52 & 7.7 & 0.35 & 0.81 & $<$0.02 & 12.7 & 1.9 & \nodata & 0.5 & 0.6 \\%
3  & SST24 & 022345.09-054234.4 & 0.142 & 0.55 & 10.7 & 0.6 & 0.76 & $<$0.04 & 13.7 & 8.2 & \nodata & 0.8 & 0.6 \\%
4  & SST24 & 022447.00-040851.4 & 0.096 & 0.36 & 12.5 & 0.6 & 0.74 & $<$0.02 & 8.7 & 3.4 & \nodata & 1.8 & 1.9 \\%
5   & SST24 & 022548.25-050051.8 & 0.145 & 0.42 & 4.2 & 0.2 & 0.55 & $<$0.01 & 14.3 & 0.5 & \nodata & 0.4 & 0.5 \\%
6  & SST24 & 022549.79-040025.1 & 0.043 & 0.36 & 23.8 & 1.41 & 0.50 & 0.27 & 69.6 & 15.1 & \nodata & 2.4 & 2.3 \\%
7  & SST24 & 022600.02-050145.1 & 0.203 & 0.55 & 6.4 & 0.32 & 0.72 & $<$0.06 & 9.6 & 3.0 & \nodata & 0.6 & 0.6 \\%
8  & SST24 & 022637.83-035841.8 & 0.073 & 0.38 & 5.1 & 0.28 & 0.52 & $<$0.02 & 18.5 & 3.2 & \nodata & 0.7 & 0.7 \\%
9  & SST24 & 022738.55-044702.9 & 0.171 & 0.51 & 5.8 & 0.33 & 0.68 & $<$0.03 & 11.9 & 3.8 & \nodata & 0.4 & 0.5 \\%
10  & SST24 & 104016.34+570845.9 & 0.115 & 0.64 & 7.3 & 0.35 & 0.70 & $<$0.02 & 7.2 & 2.3 & \nodata & 0.5 & 0.6 \\%
11  & SST24 & 104729.91+572843.3 & 0.230 & 0.64 & 11.2 & 0.59 & 0.79 & $<$0.02 & 10.5 & 6.0 & 7.71 & 0.6 & 0.8 \\%
12   & SST24 & 104907.19+565715.1 & 0.072 & 0.75 & 16.7 & 0.83 & 0.76 & $<$0.03 & 11.7 & 14.1 & 6.19 & 1.0 & 1.0 \\%
13  & SST24 & 105336.93+580350.7 & 0.455 & 0.39 & 4 & 0.13 & 0.22 & $<$0.03 & 6.2 & 3.0 & \nodata & $<$0.3 & \nodata \\%
14  & SST24 & 110133.82+575205.8 & 0.282 & 0.54 & 8.2 & 0.33 & 0.58 & $<$0.01 & 14.2 & 1.3 & \nodata & 0.5 & 0.5 \\%
15  & SST24 & 142504.04+345013.7 & 0.078 & 0.59 & 16.6 & 0.9 & 0.78 & 0.04 & 14.5 & 10.8 & 6.44 & 1.6 & 1.4 \\%
16   & SST24 & 142554.57+344603.2 & 0.035 & 0.56 & 38.6 & 2.19 & 0.72 & 0.11 & 60.2 & 27.5 & 6.03 & 3.5 & 3.3 \\%
17  & SST24 & 143039.27+352351.0 & 0.089 & 0.47 & 14.4 & 0.71 & 0.53 & $<$0.06 & 49.1 & 6.1 & 10.58 & 1.2 & 1.4 \\%
18  & SST24 & 143119.79+353418.1 & 0.033 & 0.68 & 39 & 2.39 & 0.78 & 0.18 & 39.7 & 42.3 & 5.82 & 3.4 & 3.3 \\%
19  & SST24 & 143126.81+344517.9 & 0.084 & 0.54 & 13.7 & 0.76 & 0.72 & $<$0.01 & 14.1 & 3.3 & 7.58 & 1.2 & 1.1 \\%
20  & SST24 & 160038.84+551018.4 & 0.143 & 0.42 & 9.7 & 0.49 & 0.43 & $<$0.04 & 30.6 & 2.2 & \nodata & 0.6 & 0.8 \\%
21  & SST24 & 160341.36+552612.5 & 0.145 & 0.59 & 8.9 & 0.52 & 0.75 & 0.03 & 8.0 & 4.3 & \nodata & 0.9 & 1.2 \\%
22  & SST24 & 160358.21+555504.9 & 0.333 & 0.39 & 3.9 & 0.04 & 0.08 & 0.03 & 11.5 & 0.3 & \nodata & $<$0.3 & \nodata \\%
23   & SST24 & 160408.22+542531.2 & 0.264 & 0.54 & 7 & 0.3 & 0.75 & 0.03 & 9.4 & 0.6 & \nodata & $<$0.3 & \nodata \\%
24  & SST24 & 160408.36+545812.7 & 0.064 & 0.52 & 20.1 & 1.13 & 0.67 & $<$0.07 & 32.8 & 5.4 & 9.99 & 2.6 & 2.6 \\%
25  & SST24 & 160832.64+552927.0 & 0.065 & 0.59 & 8.7 & 0.43 & 0.69 & $<$0.01 & 7.4 & 1.9 & \nodata & 0.7 & 0.9 \\%
26  & SST24 & 160907.60+552428.3 & 0.064 & 0.53 & 9.6 & 0.46 & 0.76 & $<$0.06 & 9.6 & 8.6 & \nodata & 0.7 & 0.8 \\%
27  & SST24 & 160908.34+552241.2 & 0.086 & 0.57 & 11.6 & 0.72 & 0.92 & 0.11 & 8.4 & 6.4 & \nodata & 1.0 & 1.1 \\%
28   & SST24 & 160937.59+541259.6 & 0.085 & 0.65 & 12 & 0.69 & 0.81 & $<$0.02 & 6.4 & 7.1 & \nodata & 1.0 & 1.5 \\%
29   & SST24 & 161823.14+552721.3 & 0.083 & 0.64 & 36.3 & 1.91 & 0.77 & 0.08 & 36.2 & 31.4 & \nodata & 1.7 & 1.8 \\%
30  & SST24 & 161827.74+552209.2 & 0.082 & 0.53 & 8.5 & 0.39 & 0.73 & $<$0.05 & 14.6 & 0.8 & \nodata & 0.7 & 0.8 \\%
31  & SST24 & 161950.56+543715.4 & 0.144 & 0.54 & 10.3 & 0.51 & 0.68 & $<$0.03 & 10.2 & 5.2 & \nodata & 0.7 & 0.9 \\%
32   & SST24 & 162038.14+553521.6 & 0.190 & 0.57 & 9.4 & 0.47 & 0.86 & $<$0.02 & 14.7 & 6.8 & \nodata & 0.5 & 0.6 \\%
33 & SST24 & 162210.98+550253.5 & 0.030 & 0.42 & 31.4 & 1.42 & 0.62 & 0.08 & 55.7 & 3.7 & \nodata & 4.7 & 4.4 \\%
34  & SST24 & 163001.55+410953.3 & 0.123 & 0.58 & 7.5 & 0.39 & 0.73 & $<$0.04 & 11.0 & 1.1 & \nodata & $<$0.3 & \nodata \\%
35  & SST24 & 163359.13+405305.0 & 0.031 & 0.58 & 15 & 0.82 & 0.79 & 0.07 & 12.0 & 13.7 & 5.86 & 1.0 & 1.2 \\%
36   & SST24 & 163506.08+411038.2 & 0.078 & 0.52 & 14.8 & 0.59 & 0.63 & $<$0.03 & 18.7 & 1.1 & 9.84 & 0.9 & 1.0 \\%
37  & SST24 & 163608.15+410507.8 & 0.167 & 0.41 & 14.6 & 0.51 & 0.59 & 0.1 & 26.5 & 10.5 & 9.53 & 0.8 & 1.0 \\%
38   & SST24 & 163715.60+414934.2 & 0.122 & 0.52 & 5.8 & 0.33 & 0.64 & $<$0.05 & 12.2 & 4.6 & 6.99 & 0.8 & 0.9 \\%
39  & SST24 & 163751.95+401503.6 & 0.072 & 0.69 & 12.4 & 0.63 & 0.83 & $<$0.03 & 10.8 & 5.9 & \nodata & 0.7 & 0.8 \\%
40   & SST24 & 163808.55+403214.2 & 0.225 & 0.46 & 9.2 & 0.37 & 0.57 & 0.04 & 26.0 & 3.4 & \nodata & 0.3 & 0.5 \\%
41   & SST24 & 164211.93+410816.4 & 0.143 & 0.51 & 9 & 0.43 & 0.62 & 0.06 & 20.9 & 4.8 & \nodata & 0.5 & 0.6 \\%
42  & SST24 & 171316.55+583235.5 & 0.080 & 0.52 & 9.5 & 0.5 & 0.74 & 0.02 & 9.4 & 9.7 & 9.20 & 1.6 & 1.6 \\%
43  & SST24 & 171414.89+585221.9 & 0.166 & 0.59 & 7.1 & 0.32 & 0.65 & 0.04 & 14.9 & 1.4 & \nodata & 0.4 & 0.6 \\%
44   & SST24 & 171446.54+593400.3 & 0.130 & 0.62 & 14.7 & 0.7 & 0.79 & 0.05 & 10.4 & 3.3 & \nodata & 0.8 & 1.1 \\%
45   & SST24 & 171542.00+591657.4 & 0.119 & 0.54 & 12 & 0.5 & 0.57 & 0.1 & 43.9 & 6.6 & \nodata & 1.0 & 1.3 \\%
46 & SST24 & 171614.52+595423.4 & 0.153 & 0.58 & 10.4 & 0.57 & 0.72 & $<$0.02 & 13.2 & 9.4 & 6.76 & 0.7 & 0.6 \\%
47  & SST24 & 171711.17+602709.4 & 0.110 & 0.53 & 6.8 & 0.36 & 0.66 & $<$0.04 & 12.6 & 1.3 & \nodata & 0.6 & 0.7 \\%
48  & SST24 & 171933.40+592743.2 & 0.142 & 0.45 & 9.1 & 0.37 & 0.61 & 0.03 & 13.3 & 2.4 & \nodata & 0.6 & 0.7 \\%
49   & SST24 & 172043.33+584026.5 & 0.125 & 0.54 & 9.7 & 0.57 & 0.73 & $<$0.04 & 15.0 & 5.7 & 7.51 & 1.1 & 1.1 \\%
50   & SST24 & 172400.61+590228.2 & 0.179 & 0.44 & 6.2 & 0.29 & 0.45 & $<$0.03 & 23.7 & 6.7 & \nodata & 0.5 & 0.6 \\%

\enddata

\tablenotetext{a}{Redshift from PAH features in IRS spectra.}
\tablenotetext{b}{Rest frame equivalent width of 6.2 \um PAH feature fit with single gaussian on a linear continuum within rest wavelength range 5.5 \um to 6.9 \um.}
\tablenotetext{c}{Flux density in mJy at peak of 7.7 \um PAH feature.}
\tablenotetext{d}{Total flux of PAH 11.3 \um in units of 10$^{-20}$W cm$^{-2}$ s$^{-1}$.}
\tablenotetext{e}{Total flux of [NeIII] 15.5 \um in units of 10$^{-20}$W cm$^{-2}$ s$^{-1}$.}
\tablenotetext{f}{Continuum flux density at rest frame 24 \ums.}
\tablenotetext{g}{Flux density in $\mu$Jy observed at wavelength (1+z)153 nm interpolated with a power law between flux densities in the GALEX catalog measured with the FUV filter (observed 134-179 nm) and the NUV filter (observed 177-283 nm).}
\tablenotetext{h}{Observed emission line ratio H$\alpha$/H$\beta$ measured by us from SDSS spectra using ratios of $\lambda$f$_{\lambda}$ for f$_{\lambda}$ at the line peak shown in the SDSS spectra.}
\tablenotetext{i}{Observed flux density in the $H$ band of 2MASS using conversion that zero magnitude corresponds to 1024 Jy.}
\tablenotetext{j}{Observed flux density in the $K$ band of 2MASS using conversion that zero magnitude corresponds to 667 Jy.}

\end{deluxetable}

\clearpage

\begin{deluxetable}{lccccccc} 
\rotate
\tablecolumns{8}
\tabletypesize{\footnotesize}
\tablewidth{0pc}
\tablecaption{Luminosities and Star Formation Rates for $Spitzer$ Sample}
\tablehead{
\colhead{No.} & \colhead{Name} & \colhead{J2000 coordinates} &\colhead{logL(7.7\ums)\tablenotemark{a}} & \colhead{logL(UV)\tablenotemark{b}} & \colhead{M(UV)\tablenotemark{c}} & \colhead{log SFR(PAH)\tablenotemark{d}} & \colhead{log SFR(UV)\tablenotemark{e}} \\
\colhead{} & \colhead{} & \colhead{} &  \colhead{log erg s$^{-1}$}  & \colhead{log erg s$^{-1}$} & \colhead{} & \colhead{log\mdot} & \colhead{log\mdot}
}

\startdata
1 &  SST24 & 021849.78-052158.4 & 44.70 & 43.16 & -18.08 & 2.13 & -0.10 \\%
2 &  SST24 & 022205.05-050536.7 & 44.68 & 42.78 & -17.13 & 2.11 & -0.48 \\%
3 &  SST24 & 022345.09-054234.4 & 44.29 & 42.87 & -17.35 & 1.72 & -0.39 \\%
4 &  SST24 & 022447.00-040851.4 & 44.00 & 42.13 & -15.50 & 1.43 & -1.13 \\%
5 & SST24 & 022548.25-050051.8 & 43.90 & 41.70 & -14.43 & 1.33 & -1.56 \\%
6 &  SST24 & 022549.79-040025.1 & 43.57 & 42.08 & -15.38 & 1.00 & -1.18 \\%
7 &   SST24 & 022600.02-050145.1 & 44.38 & 42.76 & -17.08 & 1.81 & -0.50 \\%
8 &  SST24 & 022637.83-035841.8 & 43.37 & 41.87 & -14.85 & 0.80 & -1.39 \\%
9 & SST24 & 022738.55-044702.9 & 44.19 & 42.70 & -16.93 & 1.62 & -0.56 \\%
10 &  SST24 & 104016.34+570845.9 & 43.93 & 42.13 & -15.50 & 1.36 & -1.13 \\%
11 &   SST24 & 104729.91+572843.3 & 44.74 & 43.17 & -18.10 & 2.17 & -0.09 \\%
12 &   SST24 & 104907.19+565715.1 & 43.87 & 42.50 & -16.43 & 1.30 & -0.76 \\%
13 &   SST24 & 105336.93+580350.7 & 44.91 & 43.48 & -18.88 & 2.34 & 0.22 \\%
14 &  SST24 & 110133.82+575205.8 & 44.78 & 42.69 & -16.90 & 2.21 & -0.57 \\%
15 &   SST24 & 142504.04+345013.7 & 43.96 & 42.46 & -16.33 & 1.39 & -0.80 \\%
16 &   SST24 & 142554.57+344603.2 & 43.62 & 42.16 & -15.58 & 1.05 & -1.10 \\%
17 &   SST24 & 143039.27+352351.0 & 44.01 & 42.32 & -15.98 & 1.44 & -0.94 \\%
18 &   SST24 & 143119.79+353418.1 & 43.58 & 42.30 & -15.93 & 1.01 & -0.96 \\%
19 &   SST24 & 143126.81+344517.9 & 43.94 & 42.00 & -15.18 & 1.37 & -1.26 \\%
20 &   SST24 & 160038.84+551018.4 & 44.24 & 42.31 & -15.95 & 1.67 & -0.95 \\%
21 &   SST24 & 160341.36+552612.5 & 44.22 & 42.61 & -16.70 & 1.65 & -0.65 \\%
22 &   SST24 & 160358.21+555504.9 & 44.61 & 42.27 & -15.85 & 2.04 & -0.99 \\%
23 &   SST24 & 160408.22+542531.2 & 44.66 & 42.29 & -15.90 & 2.09 & -0.97 \\%
24 &   SST24 & 160408.36+545812.7 & 43.85 & 41.98 & -15.13 & 1.28 & -1.28 \\%
25 &   SST24 & 160832.64+552927.0 & 43.50 & 41.54 & -14.03 & 0.93 & -1.72 \\%
26 &   SST24 & 160907.60+552428.3 & 43.53 & 41.82 & -14.73 & 0.96 & -1.44 \\%
27 &   SST24 & 160908.34+552241.2 & 43.87 & 42.31 & -15.95 & 1.30 & -0.95 \\%
28 &   SST24 & 160937.59+541259.6 & 43.88 & 42.35 & -16.05 & 1.31 & -0.91 \\%
29 &   SST24 & 161823.14+552721.3 & 44.34 & 42.98 & -17.63 & 1.77 & -0.28 \\%
30 &   SST24 & 161827.74+552209.2 & 43.70 & 41.37 & -13.60 & 1.13 & -1.89 \\%
31 &   SST24 & 161950.56+543715.4 & 44.28 & 42.69 & -16.90 & 1.71 & -0.57 \\%
32 &   SST24 & 162038.14+553521.6 & 44.49 & 43.05 & -17.80 & 1.92 & -0.21 \\%
33 &   SST24 & 162210.98+550253.5 & 43.38 & 41.15 & -13.05 & 0.81 & -2.11 \\%
34 &   SST24 & 163001.55+410953.3 & 44.00 & 41.87 & -14.85 & 1.43 & -1.39 \\%
35 &   SST24 & 163359.13+405305.0 & 43.08 & 41.75 & -14.55 & 0.51 & -1.51 \\%
36 &   SST24 & 163506.08+411038.2 & 43.89 & 41.46 & -13.83 & 1.32 & -1.80 \\%
37 &   SST24 & 163608.15+410507.8 & 44.56 & 43.12 & -17.98 & 1.99 & -0.14 \\%
38 &   SST24 & 163715.60+414934.2 & 43.88 & 42.49 & -16.40 & 1.31 & -0.77 \\%
39 &   SST24 & 163751.95+401503.6 & 43.74 & 42.12 & -15.48 & 1.17 & -1.14 \\%
40 &  SST24 & 163808.55+403214.2 & 44.63 & 42.91 & -17.45 & 2.06 & -0.35 \\%
41 &   SST24 & 164211.93+410816.4 & 44.22 & 42.65 & -16.80 & 1.65 & -0.61 \\%
42 &   SST24 & 171316.55+583235.5 & 43.72 & 42.43 & -16.25 & 1.15 & -0.83 \\%
43 &   SST24 & 171414.89+585221.9 & 44.24 & 42.24 & -15.78 & 1.67 & -1.02 \\%
44 &   SST24 & 171446.54+593400.3 & 44.34 & 42.40 & -16.18 & 1.77 & -0.86 \\%
45 &  SST24 & 171542.00+591657.4 & 44.17 & 42.62 & -16.73 & 1.60 & -0.64 \\%
46 &  SST24 & 171614.52+595423.4 & 44.34 & 43.00 & -17.68 & 1.77 & -0.26 \\%
47 &   SST24 & 171711.17+602709.4 & 43.86 & 41.86 & -14.83 & 1.29 & -1.40 \\%
48 &   SST24 & 171933.40+592743.2 & 44.21 & 42.34 & -16.03 & 1.64 & -0.92 \\%
49 &   SST24 & 172043.33+584026.5 & 44.13 & 42.60 & -16.68 & 1.56 & -0.66 \\%
50 &   SST24 & 172400.61+590228.2 & 44.25 & 42.99 & -17.65 & 1.68 & -0.27 \\%

\enddata

\tablenotetext{a}{Rest frame luminosity of 7.7 \um PAH feature in erg s$^{-1}$ determined as L(7.7 \ums) = $\nu$L$_{\nu}$(7.7 $\mu$m) =  4$\pi$D$_{L}$$^{2}$[$\nu$/(1+z)]$f_{\nu}$(7.7 $\mu$m), for $\nu$ corresponding to 7.7 \um and using f$_{\nu}$(7.7 $\mu$m) at observed wavelength of the 7.7 \um feature from Table 1.  Luminosity distances D$_{L}$ from \citet{wri06}:  http://www.astro.ucla.edu/~wright/CosmoCalc.html, for H$_0$ = 71 \kmsMpc, $\Omega_{M}$=0.27 and $\Omega_{\Lambda}$=0.73.  (Log [$\nu$L$_{\nu}$(\ldot)] = log [$\nu$L$_{\nu}$(erg s$^{-1}$)] - 33.59.)}
\tablenotetext{b}{Rest frame ultraviolet continuum luminosity at 153 nm in erg s$^{-1}$ determined as L(153 nm) = $\nu$L$_{\nu}$(153 nm), using f$_{\nu}$(153 nm) from Table 3 at observed wavelength (1+z)153 nm.}
\tablenotetext{c}{M(UV) is an absolute AB magnitude at continuum wavelength 153 nm, determined with the fundamental definition m$_{AB}$ = -2.5logf$_{\nu}$  - 48.60, for f$_{\nu}$ in erg cm$^{-2}$ s$^{-1}$ Hz$^{-1}$ \citep{oke74}.  M(UV) is determined as the m(UV) which arises using L$_{\nu}$(153 $\mu$m) to determine f$_{\nu}$(153 nm) at a distance of 10 pc.}
\tablenotetext{d}{Star formation rates in units of \mdot determined as log [SFR(PAH)] = log L(7.7 $\mu$m) - 42.57.}
\tablenotetext{e}{Star formation rates in units of \mdot determined as log [SFR(UV)] = log L(UV) - 43.26.}

\end{deluxetable}

\clearpage

\begin{figure}
\figurenum{1A}
\includegraphics[scale=0.9]{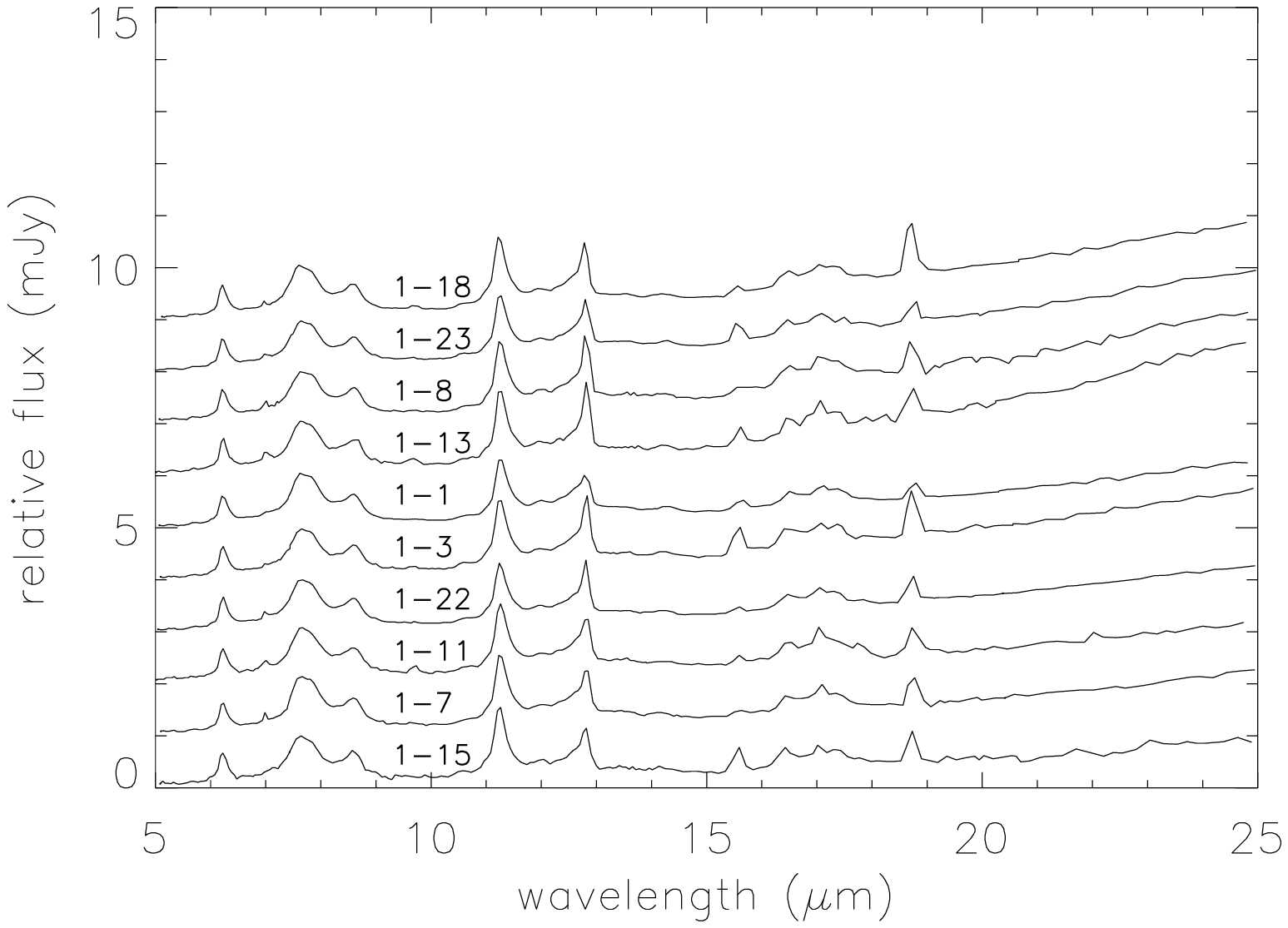}
\caption{Spectra of Markarian starbursts from Table 1.  Markarian AGN are in Figure 4.  All spectra are normalized to f$_{\nu}$(7.7 \ums) = 1.0 mJy, but zero points are displaced by 1 mJy for illustration.  The zero flux level for each spectrum is located 1 mJy below the value at 7.7 \ums. Spectra are labeled by running numbers in Table 1.  Spectra are arranged in order of relative continuum strength at 24 \ums, i.e. in order of ratio f$_{\nu}$(24 \ums)/f$_{\nu}$(7.7 \ums) in Table 1.}

\end{figure}

\clearpage

\begin{figure}
\figurenum{1B}
\includegraphics[scale=0.9]{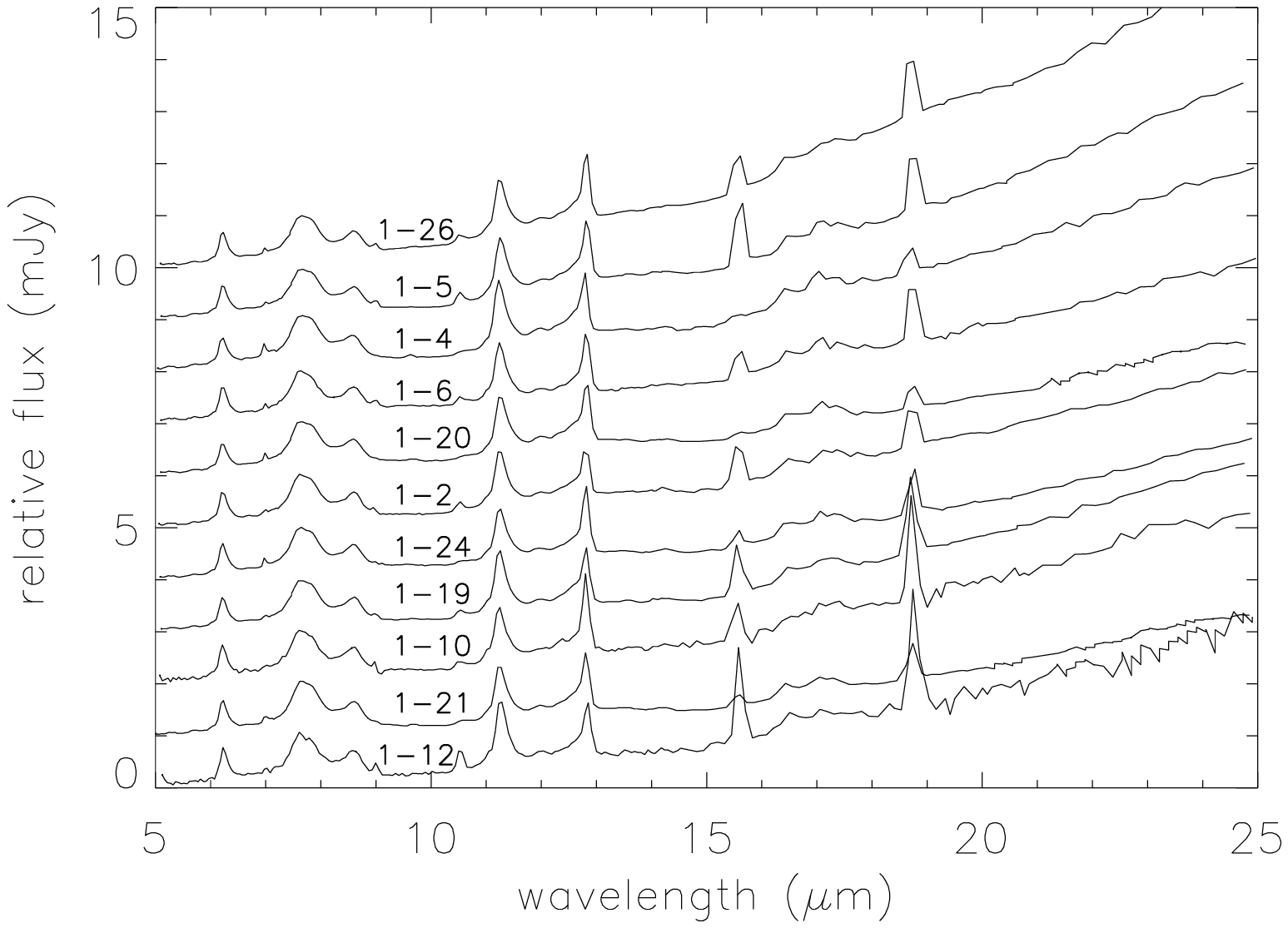}
\caption{Spectra of Markarian starbursts from Table 1.  Markarian AGN are in Figure 4.  All spectra are normalized to f$_{\nu}$(7.7 \ums) = 1.0 mJy, but zero points are displaced by 1 mJy for illustration.  The zero flux level for each spectrum is located 1 mJy below the value at 7.7 \ums. Spectra are labeled by running numbers in Table 1.  Spectra are arranged in order of relative continuum strength at 24 \ums, i.e. in order of ratio f$_{\nu}$(24 \ums)/f$_{\nu}$(7.7 \ums) in Table 1.}

\end{figure}

\clearpage
\begin{figure}
\figurenum{2A}
\includegraphics[scale=0.9]{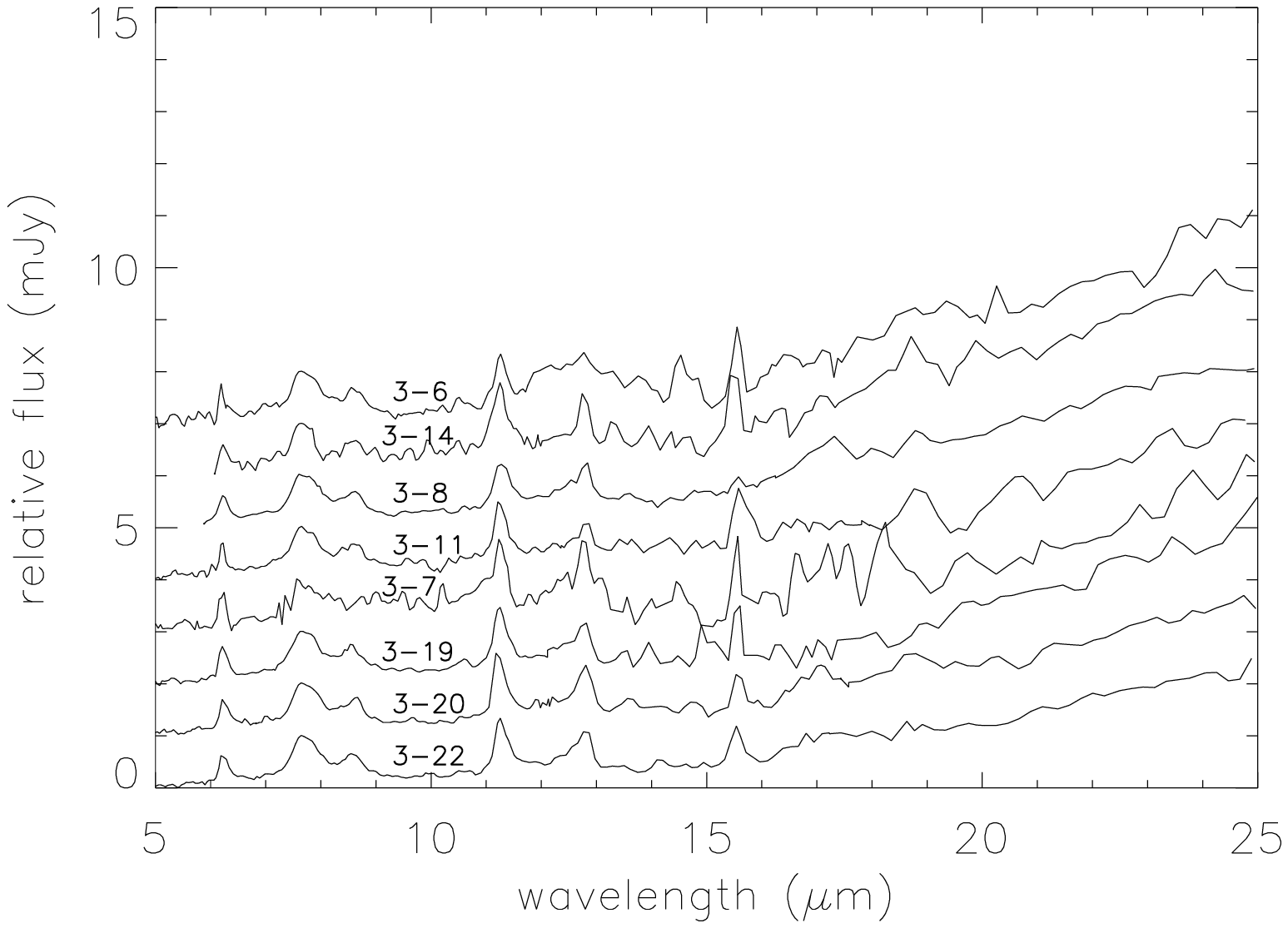}
\caption{Spectra of GALEX starbursts from Table 3.  GALEX AGN are in Figure 4.  All spectra are normalized to f$_{\nu}$(7.7 \ums) = 1.0 mJy, but zero points are displaced by 1 mJy for illustration.  The zero flux level for each spectrum is located 1 mJy below the value at 7.7 \ums. Spectra are labeled by running numbers in Table 3. Spectra are arranged in order of relative continuum strength at 24 \ums, i.e. in order of ratio f$_{\nu}$(24 \ums)/f$_{\nu}$(7.7 \ums) in Table 3.}

\end{figure}

\clearpage
\begin{figure}
\figurenum{2B}
\includegraphics[scale=0.9]{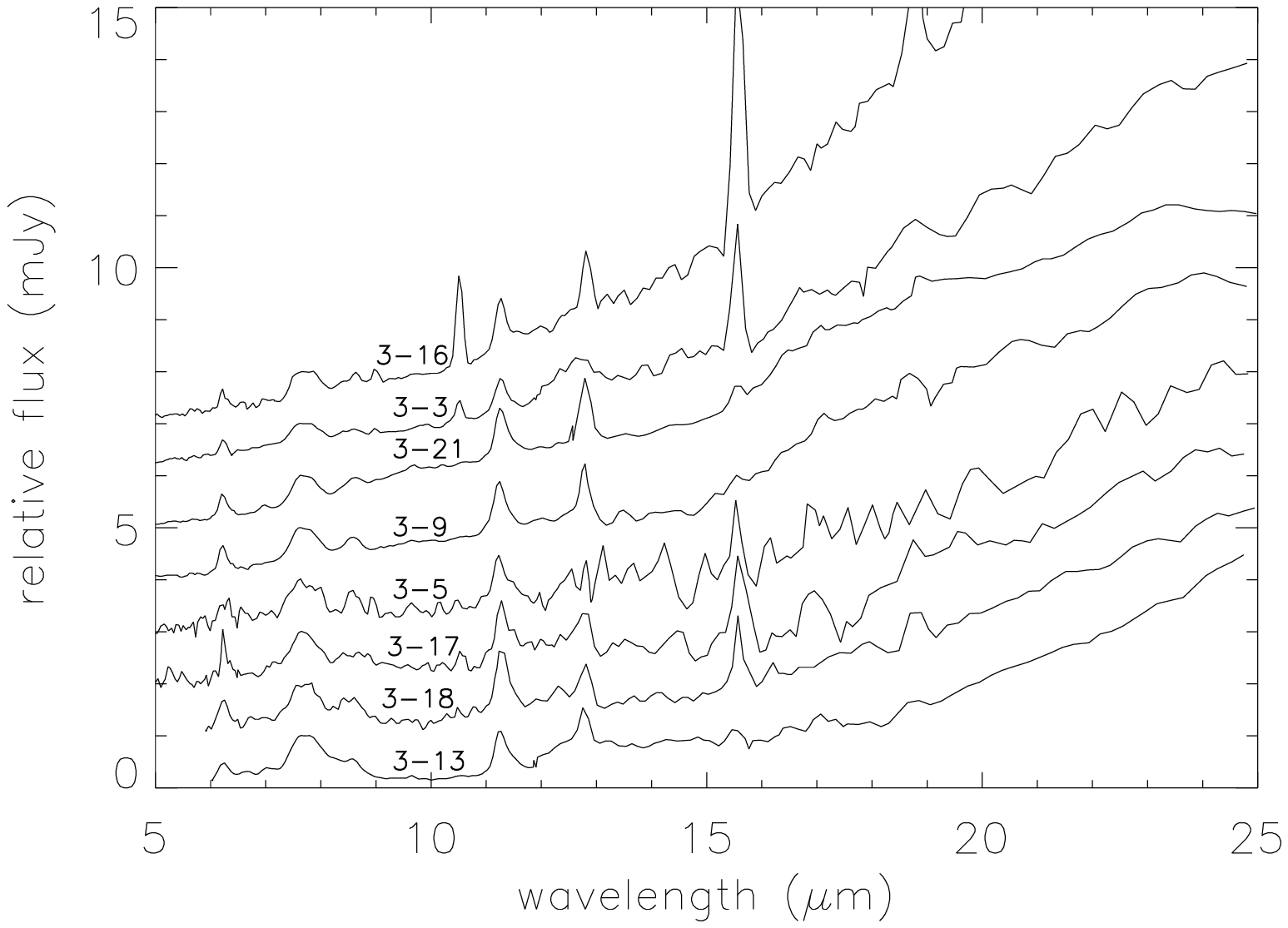}
\caption{Spectra of GALEX starbursts from Table 3. GALEX AGN are in Figure 4.  All spectra are normalized to f$_{\nu}$(7.7 \ums) = 1.0 mJy, but zero points are displaced by 1 mJy for illustration.  The zero flux level for each spectrum is located 1 mJy below the value at 7.7 \ums. Spectra are labeled by running numbers in Table 3. Spectra are arranged in order of relative continuum strength at 24 \ums, i.e. in order of ratio f$_{\nu}$(24 \ums)/f$_{\nu}$(7.7 \ums) in Table 3.}

\end{figure}

\clearpage

\begin{figure}
\figurenum{3A}
\includegraphics[scale=0.9]{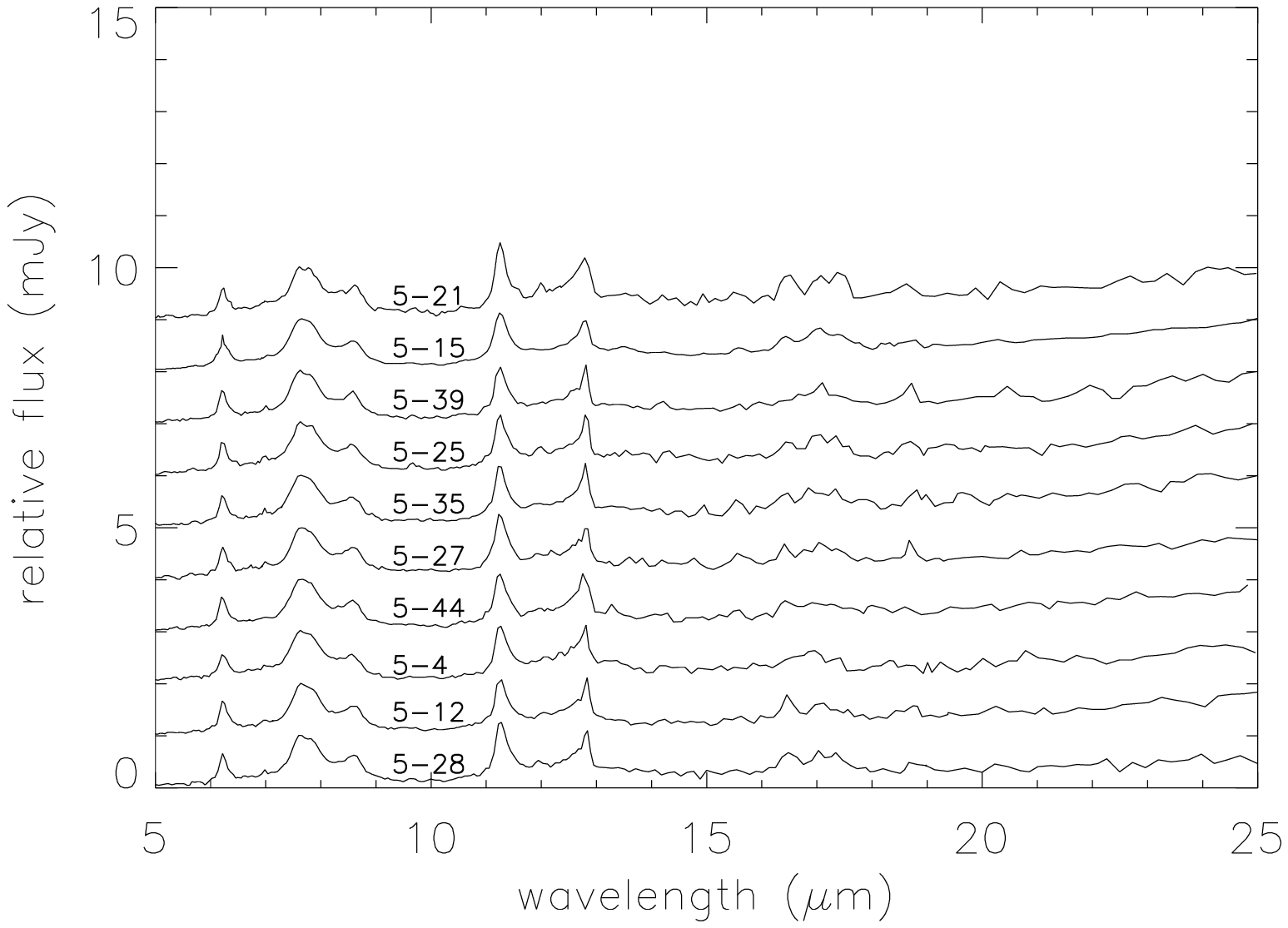}
\caption{Spectra of $Spitzer$ starbursts from Table 5.  All spectra are normalized to f$_{\nu}$(7.7 \ums) = 1.0 mJy, but zero points are displaced by 1 mJy for illustration.  The zero flux level for each spectrum is located 1 mJy below the value at 7.7 \ums. Spectra are labeled by running numbers in Table 5. Spectra are arranged in order of relative continuum strength at 24 \ums, i.e. in order of ratio f$_{\nu}$(24 \ums)/f$_{\nu}$(7.7 \ums) in Table 5.}

\end{figure}

\clearpage

\begin{figure}
\figurenum{3B}
\includegraphics[scale=0.9]{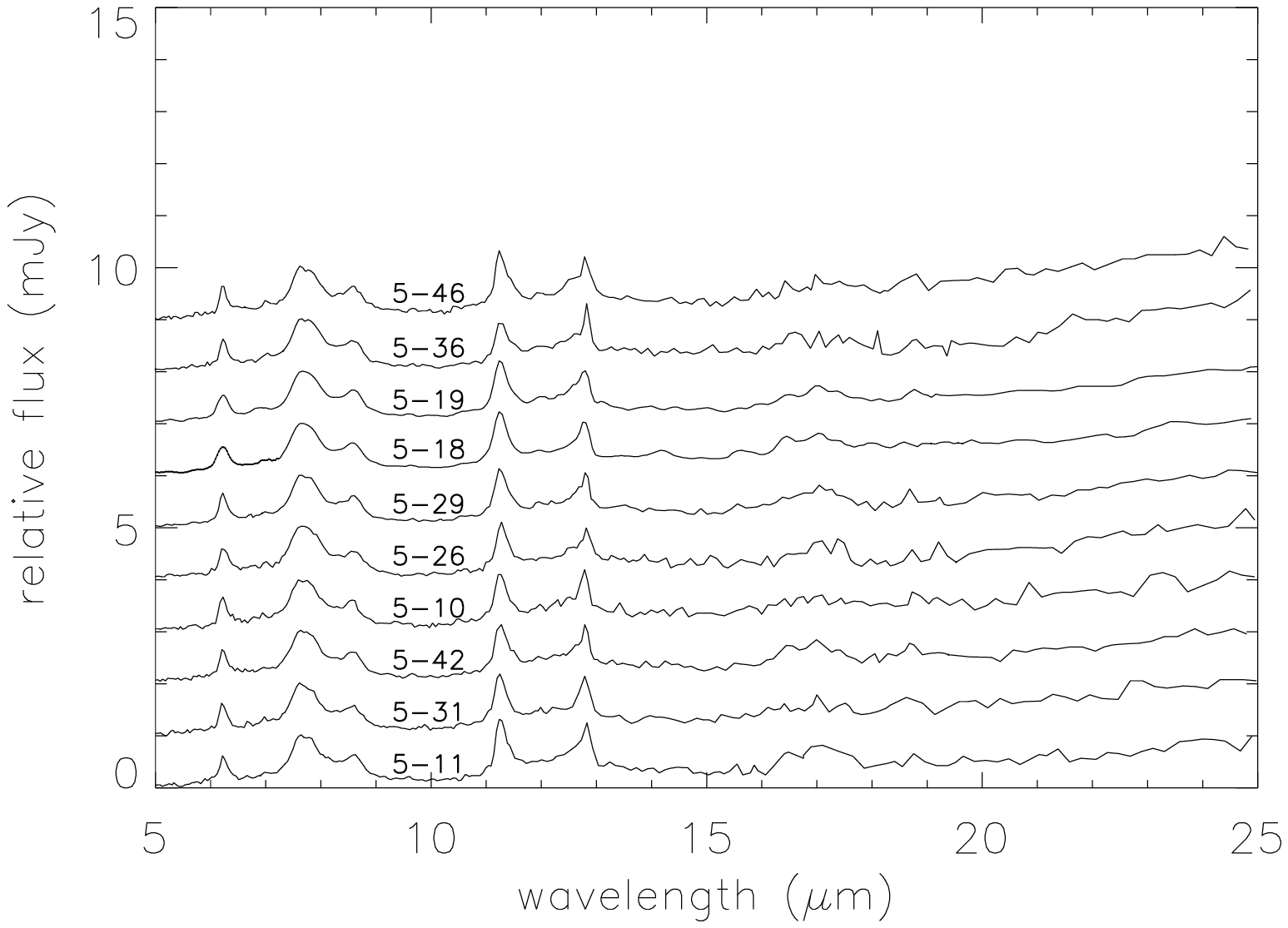}
\caption{Spectra of $Spitzer$ starbursts from Table 5.  All spectra are normalized to f$_{\nu}$(7.7 \ums) = 1.0 mJy, but zero points are displaced by 1 mJy for illustration.  The zero flux level for each spectrum is located 1 mJy below the value at 7.7 \ums. Spectra are labeled by running numbers in Tables 5. Spectra are arranged in order of relative continuum strength at 24 \ums, i.e. in order of ratio f$_{\nu}$(24 \ums)/f$_{\nu}$(7.7 \ums) in Table 5. }

\end{figure}

\clearpage

\begin{figure}
\figurenum{3C}
\includegraphics[scale=0.9]{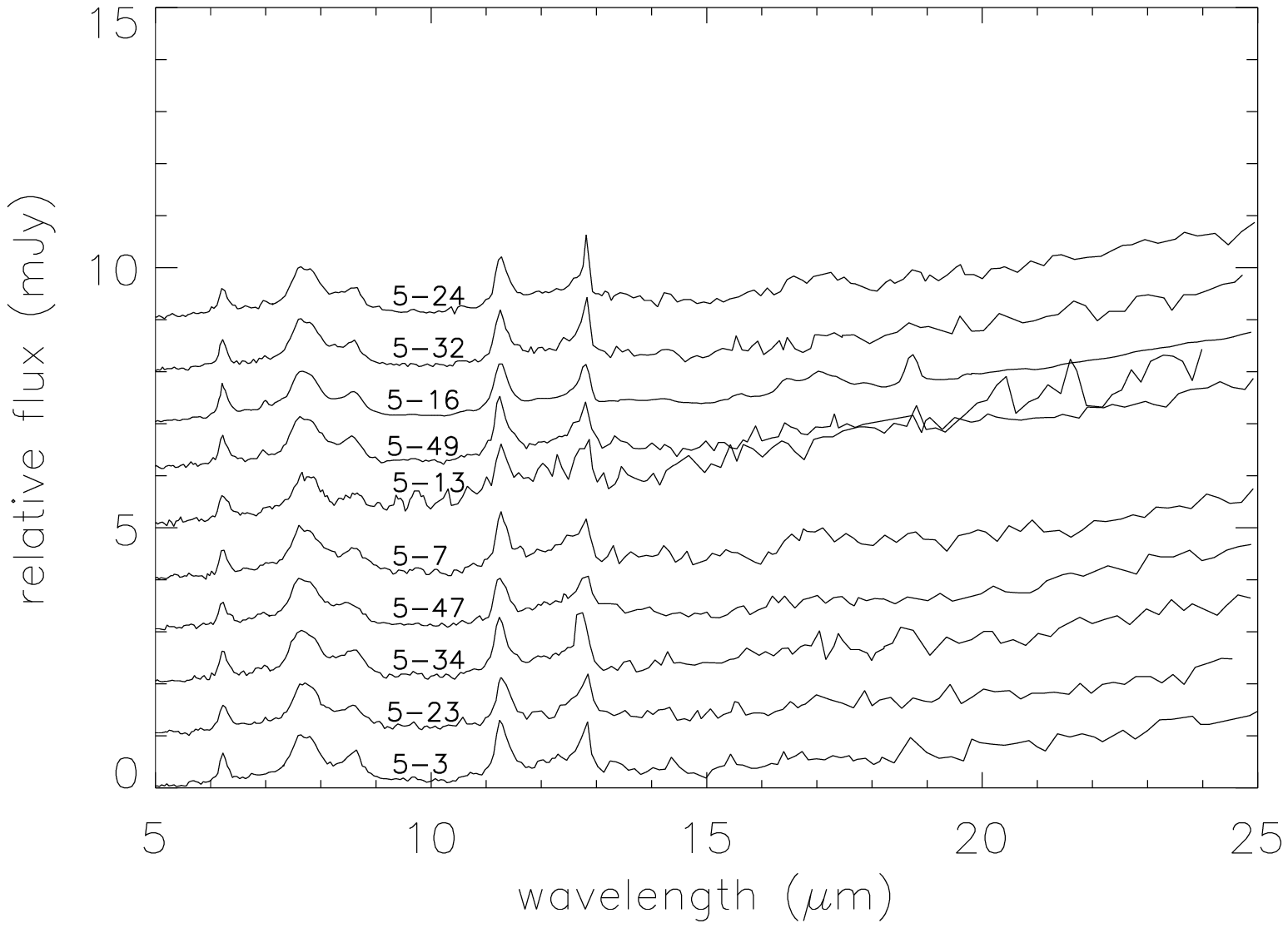}
\caption{Spectra of $Spitzer$ starbursts from Table 5.  All spectra are normalized to f$_{\nu}$(7.7 \ums) = 1.0 mJy, but zero points are displaced by 1 mJy for illustration.  The zero flux level for each spectrum is located 1 mJy below the value at 7.7 \ums. Spectra are labeled by running numbers in Table 5. Spectra are arranged in order of relative continuum strength at 24 \ums, i.e. in order of ratio f$_{\nu}$(24 \ums)/f$_{\nu}$(7.7 \ums) in Table 5. }

\end{figure}

\clearpage

\begin{figure}
\figurenum{3D}
\includegraphics[scale=0.9]{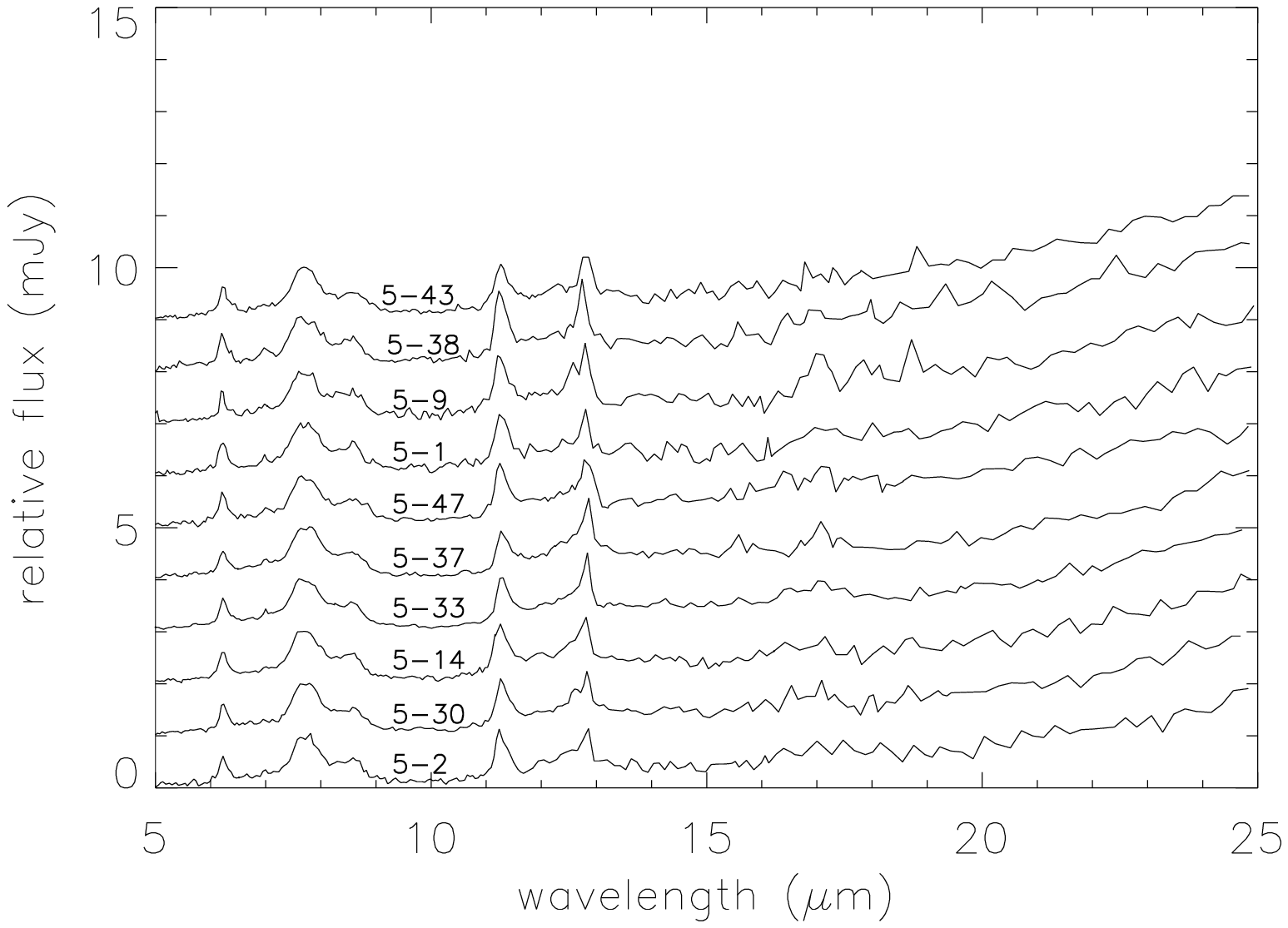}
\caption{Spectra of $Spitzer$ starbursts from Table 5.  All spectra are normalized to f$_{\nu}$(7.7 \ums) = 1.0 mJy, but zero points are displaced by 1 mJy for illustration.  The zero flux level for each spectrum is located 1 mJy below the value at 7.7 \ums. Spectra are labeled by running numbers in Table 5. Spectra are arranged in order of relative continuum strength at 24 \ums, i.e. in order of ratio f$_{\nu}$(24 \ums)/f$_{\nu}$(7.7 \ums) in Table 5. }

\end{figure}

\clearpage

\begin{figure}
\figurenum{3E}
\includegraphics[scale=0.9]{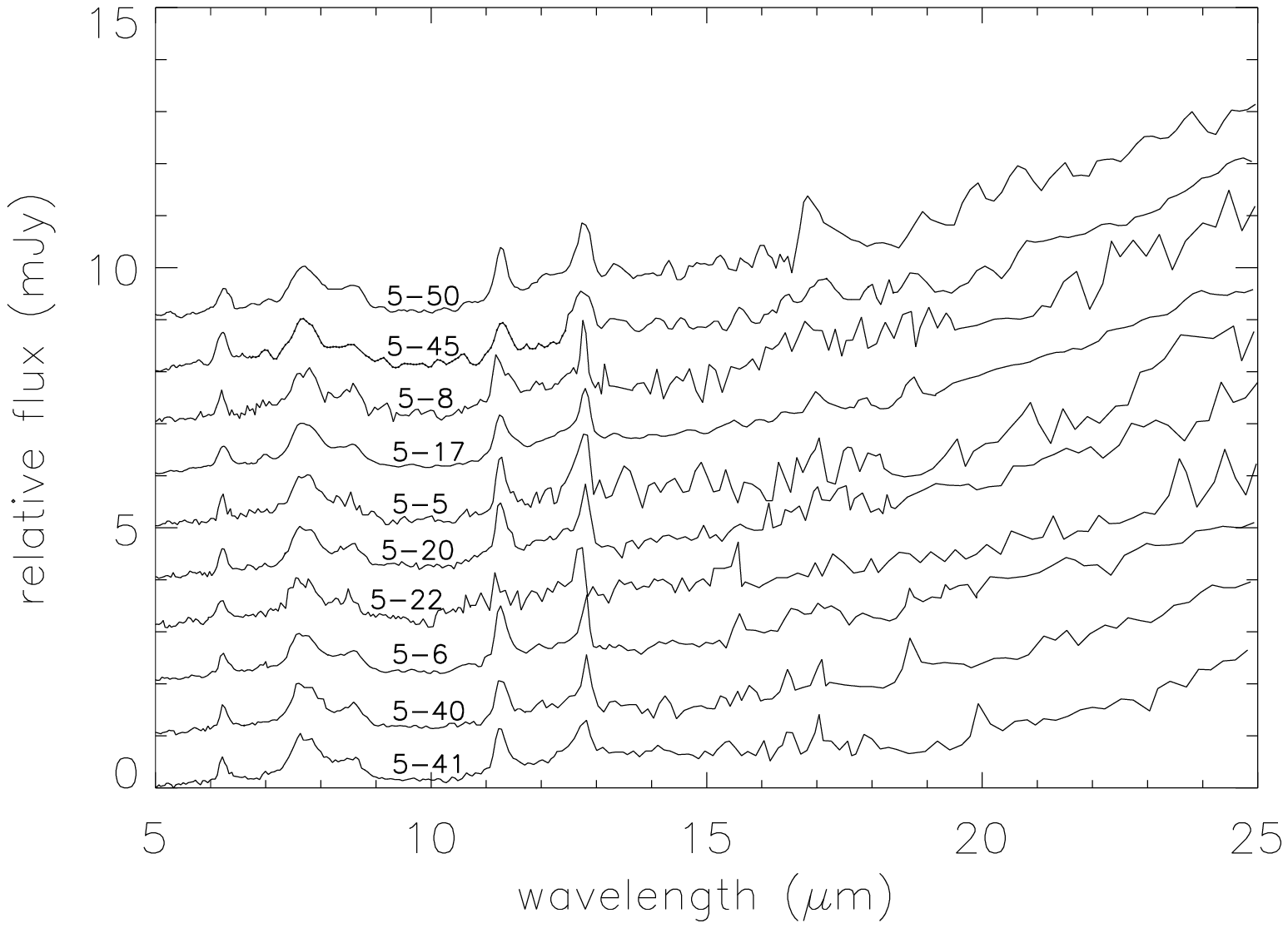}
\caption{Spectra of $Spitzer$ starbursts from Table 5.  All spectra are normalized to f$_{\nu}$(7.7 \ums) = 1.0 mJy, but zero points are displaced by 1 mJy for illustration.  The zero flux level for each spectrum is located 1 mJy below the value at 7.7 \ums. Spectra are labeled by running numbers in Table 5. Spectra are arranged in order of relative continuum strength at 24 \ums, i.e. in order of ratio f$_{\nu}$(24 \ums)/f$_{\nu}$(7.7 \ums) in Table 5. }

\end{figure}

\clearpage

\begin{figure}
\figurenum{4A}
\includegraphics[scale=0.9]{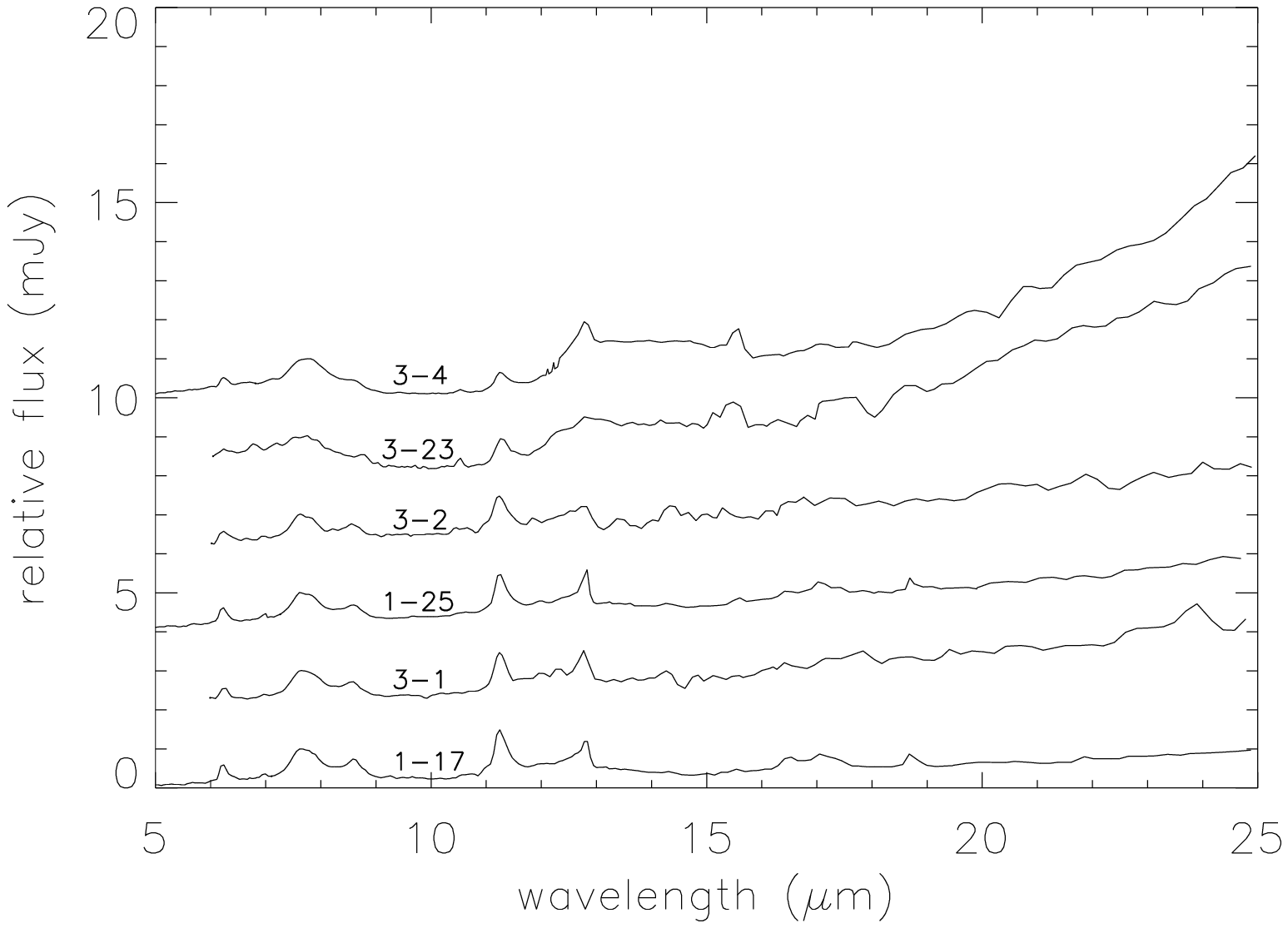}
\caption{Spectra of Markarian or GALEX AGN from Tables 1 and 3.  All spectra are normalized to f$_{\nu}$(7.7 \ums) = 1.0 mJy, but zero points are displaced by 2 mJy for illustration. The zero flux level for each spectrum is located 1 mJy below the value at 7.7 \ums. Spectra are labeled by running numbers in Tables 1 or 3. Spectra are arranged in order of relative continuum strength at 24 \ums, i.e. in order of ratio f$_{\nu}$(24 \ums)/f$_{\nu}$(7.7 \ums) in Tables 1 or 3. Classification of these AGN is discussed in section 4.2. }

\end{figure}

\clearpage
\begin{figure}
\figurenum{4B}
\includegraphics[scale=0.9]{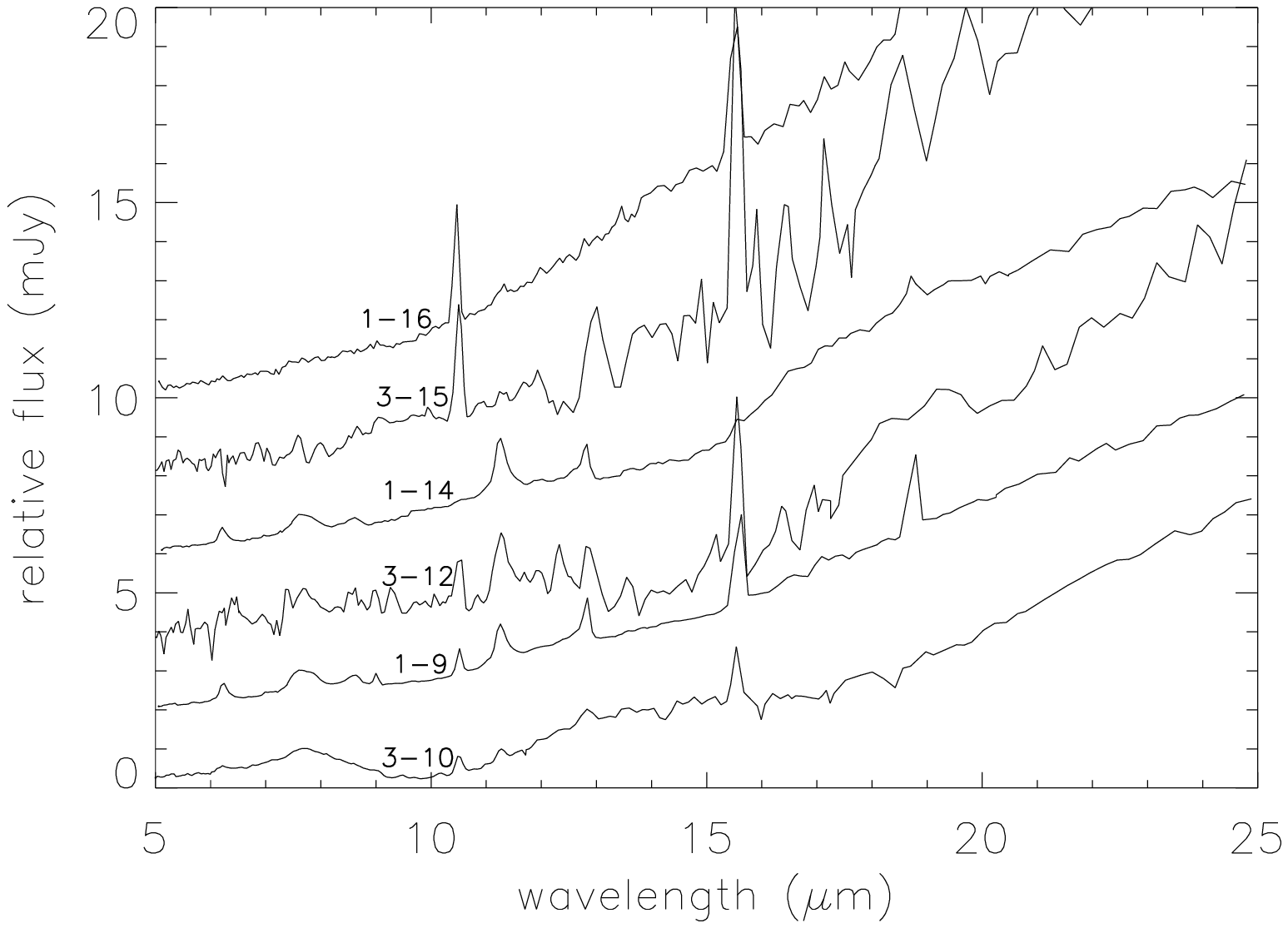}
\caption{Spectra of Markarian or GALEX AGN from Tables 1 and 3.  All spectra are normalized to f$_{\nu}$(7.7 \ums) = 1.0 mJy, but zero points are displaced by 2 mJy for illustration. The zero flux level for each spectrum is located 1 mJy below the value at 7.7 \ums. Spectra are labeled by running numbers in Tables 1 or 3. Spectra are arranged in order of relative continuum strength at 24 \ums, i.e. in order of ratio f$_{\nu}$(24 \ums)/f$_{\nu}$(7.7 \ums) in Tables 1 or 3. Classification of these AGN is discussed in section 4.2. }

\end{figure}

\clearpage

\begin{figure}
\figurenum{5}
\includegraphics[scale=0.9]{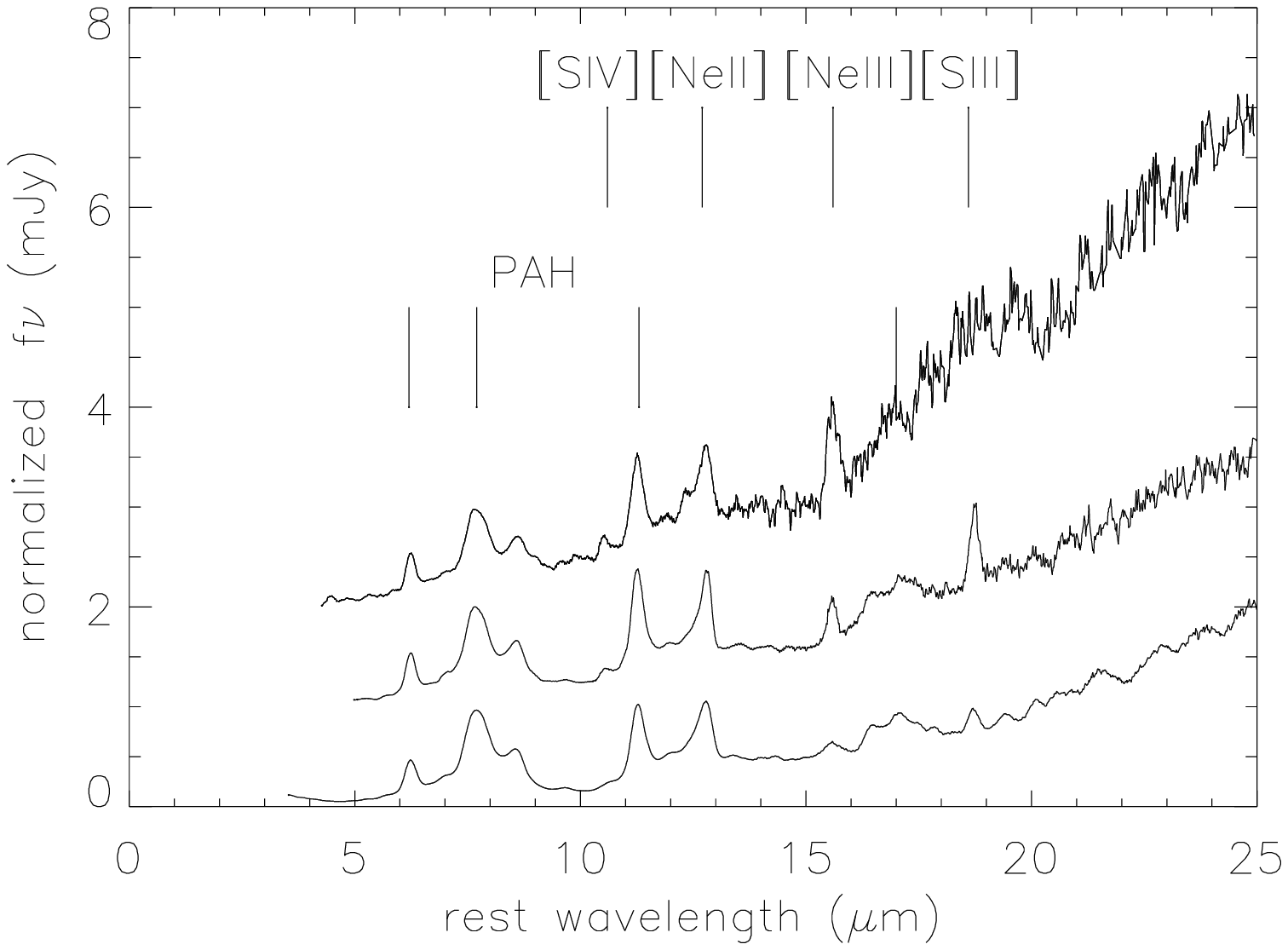}
\caption{Top to bottom are average spectra for starbursts in the Markarian sample (Figure 1), the GALEX sample (Figure 2), and the $Spitzer$ sample (Figure 3).  All spectra are normalized to f$_{\nu}$(7.7 \ums) = 1.0 mJy, but zero points are displaced by 1 mJy for illustration.  The zero flux level for each spectrum is located 1 mJy below the value at 7.7 \ums.  For the Markarian sample, all sources are included in the average except the 5 AGN noted in Tables 1 and 2. For the GALEX sample, all sources are included in the average except the 7 AGN noted in Tables 3 and 4.  For the $Spitzer$ sample, all sources are included because all are classified as starbursts from both SDSS and IRS spectra. } 

\end{figure}

\clearpage

\begin{figure}
\figurenum{6}
\includegraphics[scale=0.9]{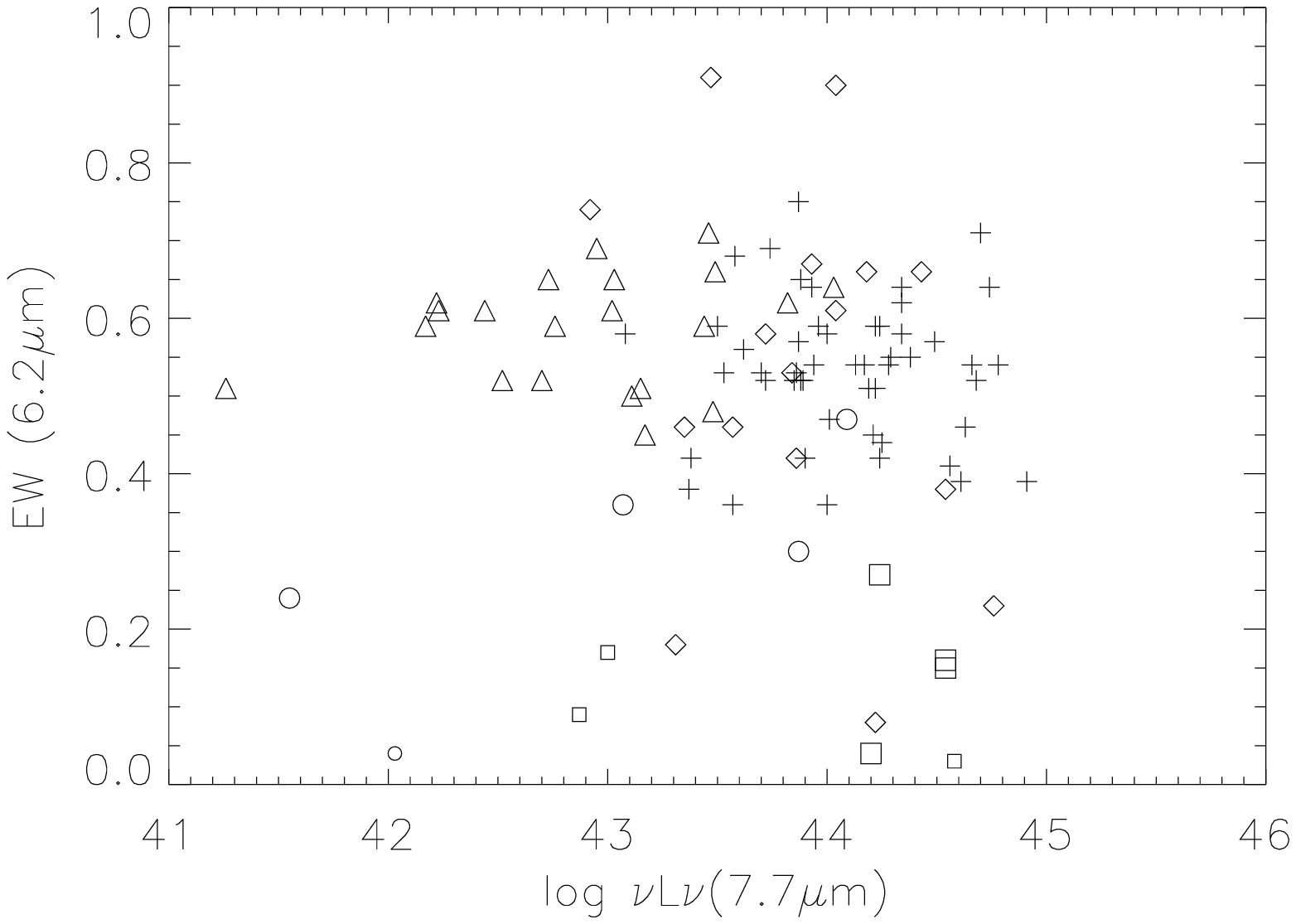}
\caption{PAH 6.2 \um EW (in \ums) compared with luminosity $\nu$L$_{\nu}$(7.7 $\mu$m) in erg s$^{-1}$ (log [$\nu$L$_{\nu}$(\ldot)] = log [$\nu$L$_{\nu}$(erg s$^{-1}$)] - 33.59).  Crosses are $Spitzer$ starbursts from Figure 3.  Triangles are Markarian starbursts from Figure 1, and circles are Markarian AGN from Figure 4.  Diamonds are GALEX starbursts from Figure 2, and squares are GALEX AGN from Figure 4.   Small symbols show limits on undetectable PAH for Markarian and GALEX AGN.  The Markarian starbursts are all found to satisfy the criterion of EW(6.2 \um) $>$ 0.4 \um.  The 3 GALEX sources with starburst classification but EW(6.2 \um) $<$ 0.4 are especially interesting and are discussed in section 4.2.  The $Spitzer$ sample was initially chosen to select only sources with EW(6.2 \um) $\ga$ 0.4 \um because of the well determined correlation between optical classification as starbursts and EW(6.2 \um) $>$ 0.4 \um. The Markarian starburst sample independently confirms this EW criterion, because the Balzano(1983) selection for the Markarian starbursts was defined using only optical criteria for starbursts. } 

\end{figure}

\begin{figure}
\figurenum{7}
\includegraphics[scale=0.9]{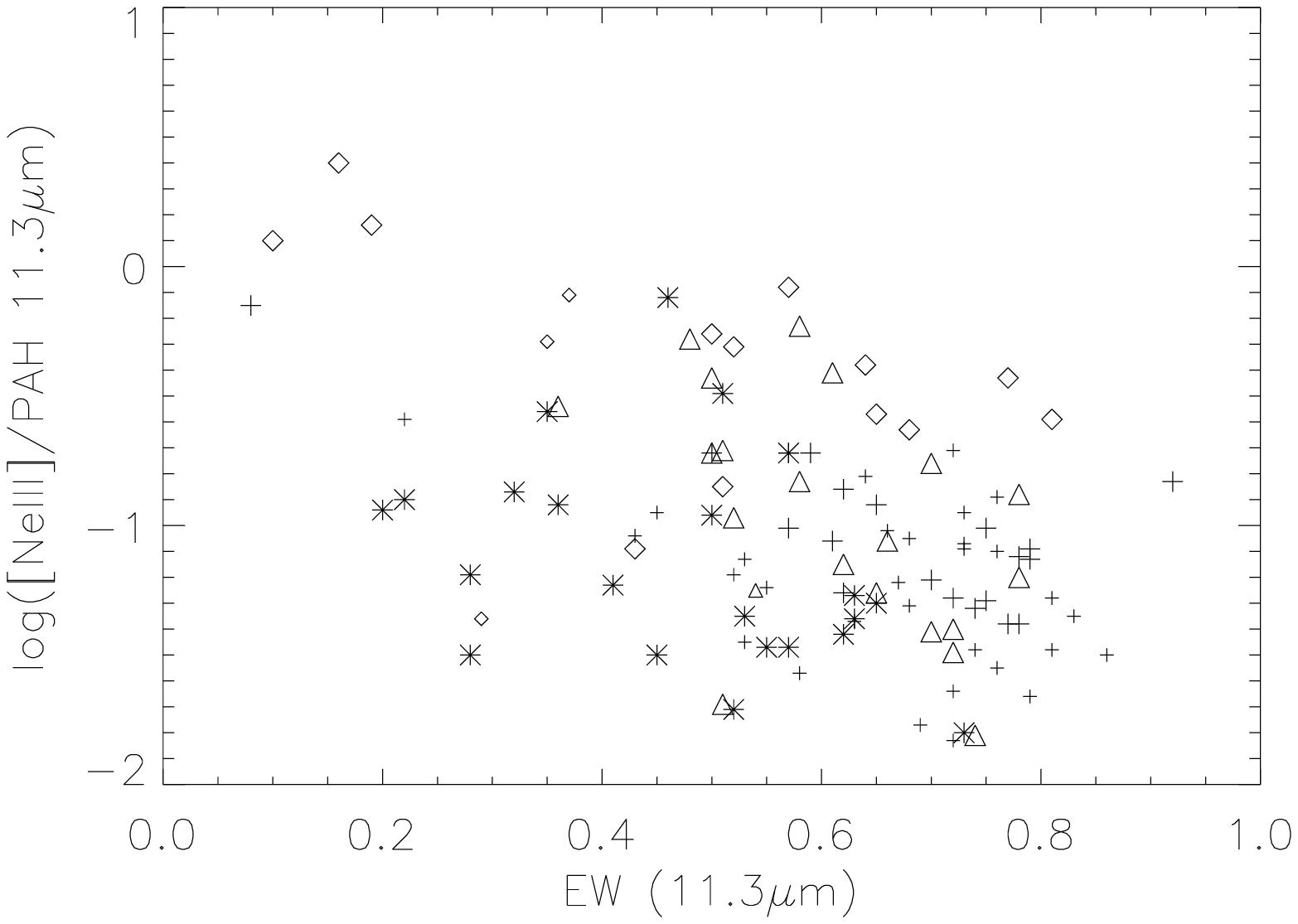}
\caption{Ratio of total flux in [NeIII] 15.5 \um to total flux in PAH 11.3 \um is compared with EW of PAH 11.3 \um (in \ums), using data for starbursts from Tables 2, 4, and 6.  Triangles are Markarian starbursts from Figure 1, diamonds are GALEX starbursts from Figure 2, and crosses are $Spitzer$ starbursts from Figure 3.  Asterisks are nearby starburst galaxies in the \citet{bra06} sample measured with high resolution spectroscopy in \citet{ber09}.   Small symbols show limits on undetectable [NeIII] for all categories. Plot includes only starbursts; the AGN shown with different symbols in Figure 4 are not included. [NeIII]/PAH measures hardness of the ionizing radiation, and EW(11.3 \ums) shows strength of dust continuum relative to strength of PAH.}

\end{figure}

\clearpage

\begin{figure}
\figurenum{8}
\includegraphics[scale=0.9] {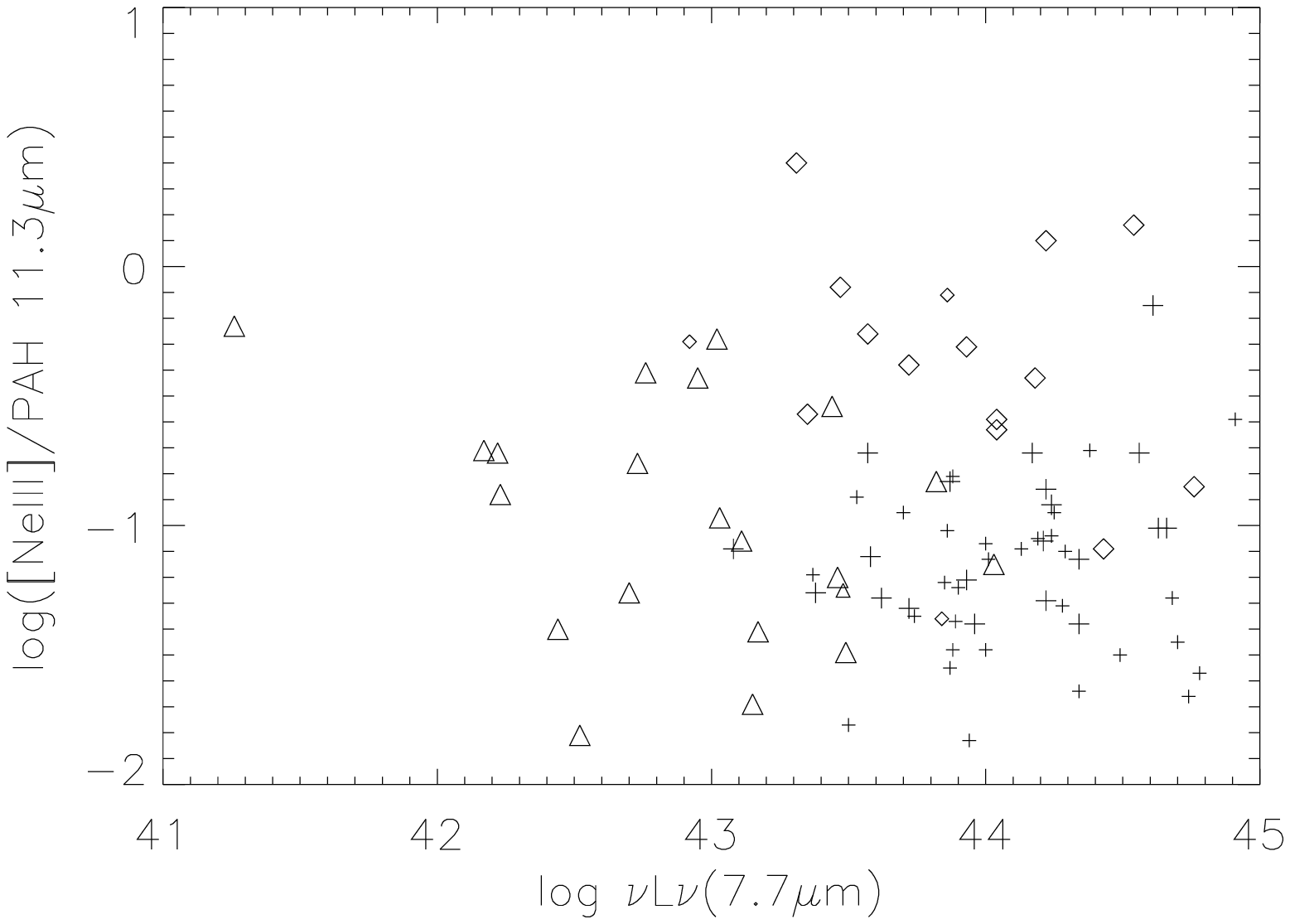}
\caption{Total flux of [NeIII] relative to total flux of PAH 11.3\um compared with luminosity $\nu$L$_{\nu}$(7.7 $\mu$m) in erg s$^{-1}$ (log [$\nu$L$_{\nu}$(\ldot)] = log [$\nu$L$_{\nu}$(erg s$^{-1}$)] - 33.59). Triangles are Markarian starbursts from Figure 1, diamonds are GALEX starbursts from Figure 2, and crosses are $Spitzer$ starbursts from Figure 3.  Small symbols show upper limits for [NeIII] in objects where this could not be measured.  The AGN shown with different symbols in Figure 4 are not included. There is no trend with luminosity for this ratio among the samples, but the values show as in Figure 7 the systematically stronger [NeIII] compared to PAH in the UV-selected Markarian and GALEX samples compared to the $Spitzer$ sample.  This result is also evident in the average spectra of Figure 5.  } 

\end{figure}

\clearpage

\begin{figure}
\figurenum{9}
\includegraphics[scale=0.9] {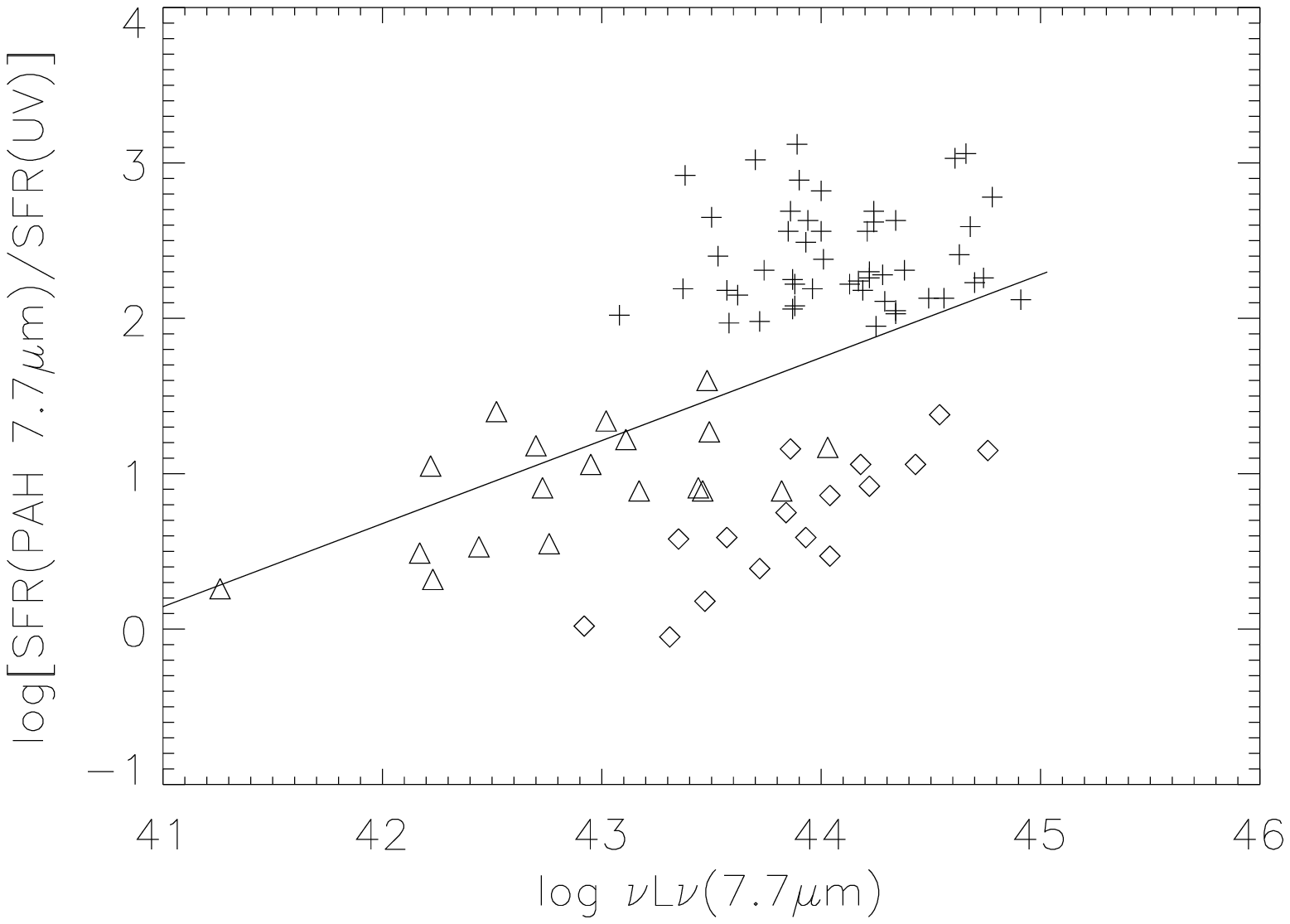}
\caption{Star formation rates from Tables 2, 4, and 6 determined from $\nu$L$_{\nu}$(7.7 $\mu$m) relative to SFRs determined from $\nu$L$_{\nu}$(153 nm), compared with luminosity $\nu$L$_{\nu}$(7.7 $\mu$m) in erg s$^{-1}$ (log [$\nu$L$_{\nu}$(\ldot)] = log [$\nu$L$_{\nu}$(erg s$^{-1}$)] - 33.59).  Triangles are Markarian starbursts from Figure 1, diamonds are GALEX starbursts from Figure 2, and crosses are $Spitzer$ starbursts from Figure 3.  The AGN shown in Figure 4 are not included. Line is the fit to the full $Spitzer$-selected sample of 287 sources in \citet{sar09}; sources in the $Spitzer$ sample from Figure 3 shown with crosses are the most extreme in SFR(PAH)/SFR(UV).  The Markarian and GALEX UV-selected samples show similar values of SFR(PAH)/SFR(UV), with median log SFR(PAH)/SFR(UV) = 0.8.  If the discrepancy between SFR(PAH) and SFR(UV) is caused only by the obscuration of the UV, this result for ultraviolet-selected sources indicates that 15\% of the intrinsic UV luminosity if observed. Much greater obscuration is implied for the $Spitzer$ sample, with median log SFR(PAH)/SFR(UV) = 2.4, indicating that only 0.4\% of the intrinsic ultraviolet luminosity emerges. }

\end{figure}

\clearpage

\begin{figure}
\figurenum{10}
\includegraphics[scale=0.9] {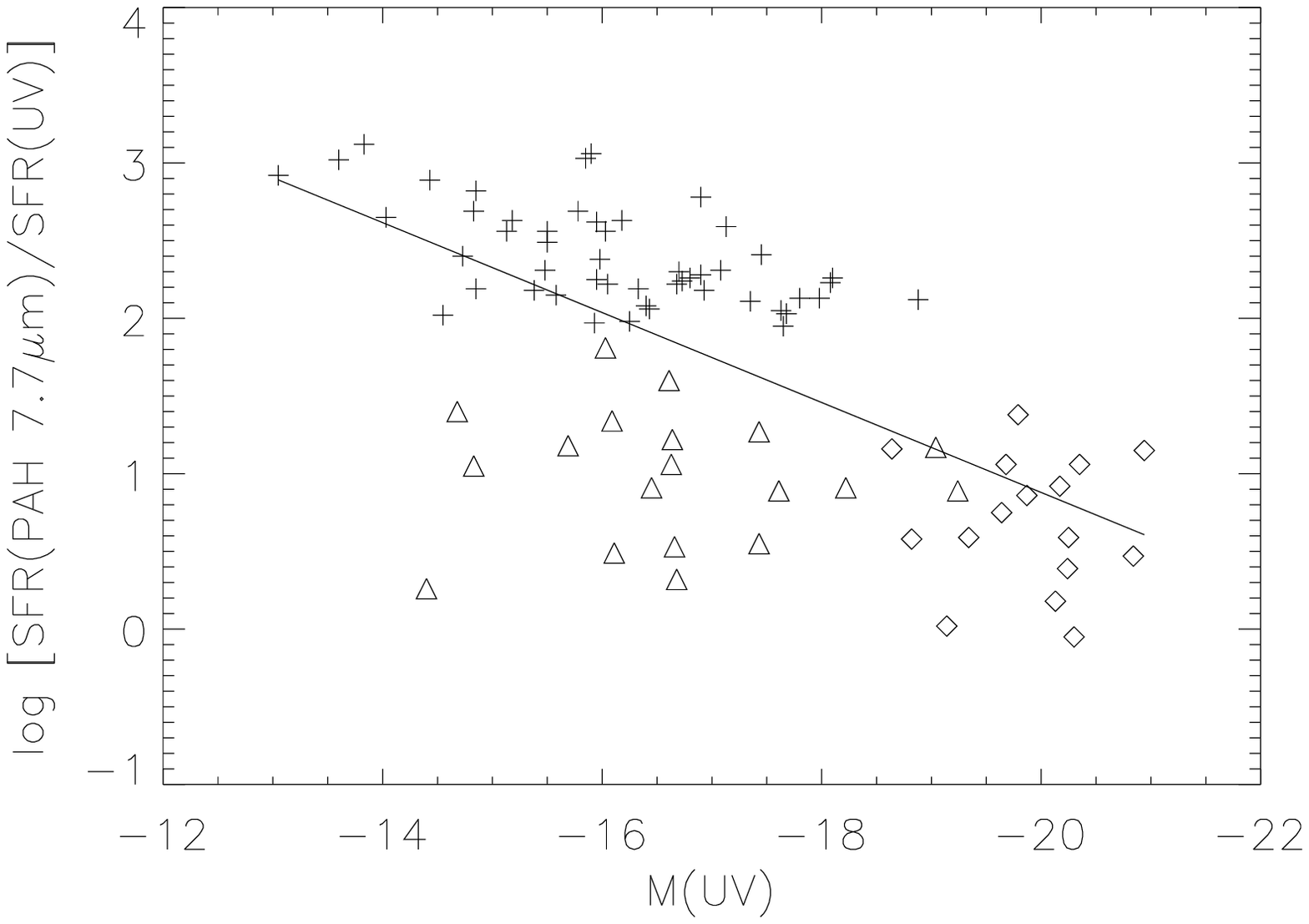}
\caption{Star formation rates from Tables 2, 4, and 6 determined from $\nu$L$_{\nu}$(7.7 $\mu$m) relative to SFRs determined from $\nu$L$_{\nu}$(153 nm), compared with luminosity as measured by M(UV);  M(UV) is an AB magnitude determined at rest frame 153 nm as explained in Tables 2, 4, and 6.  Triangles are Markarian starbursts from Figure 1, diamonds are GALEX starbursts from Figure 2, and crosses are $Spitzer$ starbursts from Figure 3.  The AGN shown in Figure 4 are not included.  Line shown is the formal linear least squares fit to all of the points, of form log[SFR(PAH)/SFR(UV)] = (0.29$\pm$ 0.04)M(UV)+6.67$\pm$0.67.  This line has the opposite slope of the line in Figure 9, which shows that objects selected using UV selection criteria favor sources having less obscuration.  Points show the similarity of Markarian and GALEX sources in terms of obscuration as measured by SFR(PAH)/SFR(UV), but show that the GALEX sample is systematically more luminous.}

\end{figure}

\clearpage

\begin{figure}
\figurenum{11}
\includegraphics[scale=0.9] {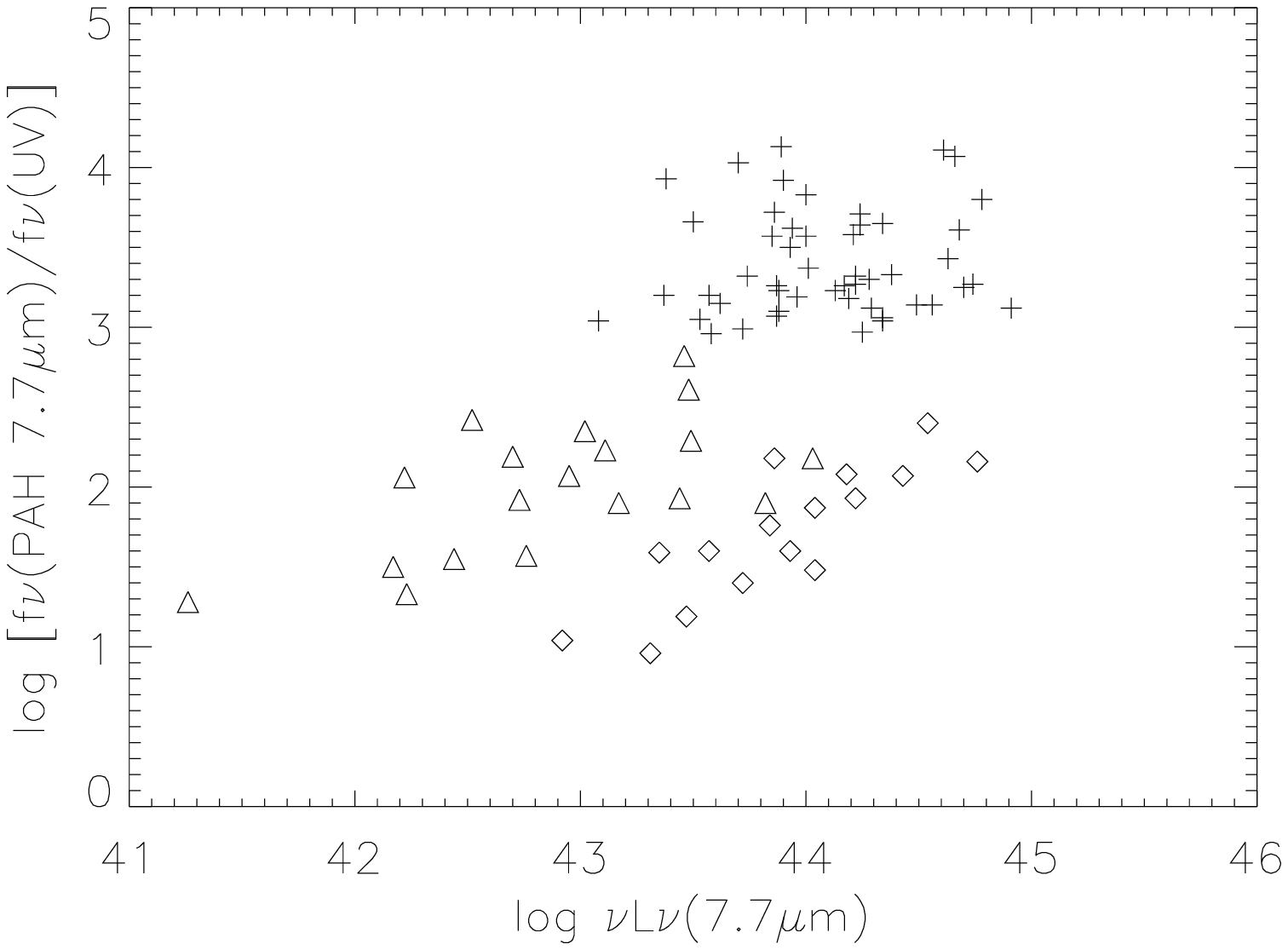}
\caption{Observed flux density of 7.7 \um PAH feature, f$_{\nu}$(7.7 $\mu$m), relative to observed ultraviolet flux density from GALEX, f$_{\nu}$(153 nm), from Tables 1, 3, and 5 compared to luminosity $\nu$L$_{\nu}$(7.7 $\mu$m) in erg s$^{-1}$ (log [$\nu$L$_{\nu}$(\ldot)] = log [$\nu$L$_{\nu}$(erg s$^{-1}$)] - 33.59).  Triangles are Markarian starbursts from Figure 1, diamonds are GALEX starbursts from Figure 2, and crosses are $Spitzer$ starbursts from Figure 3.  The AGN shown in Figure 4 are not included. This illustrates the empirical result that the smallest value of log[f$_{\nu}$(7.7 $\mu$m)/f$_{\nu}$(153 nm)] = 1.   We take this as the empirical value which relates PAH luminosity to ultraviolet luminosity for a starburst with negligible obscuration of the UV. Assuming no obscuration at 7.7 \um, the fraction of UV luminosity intrinsically emitted compared to the luminosity observed, UV(intrinsic)/UV(observed), is then given by log[UV(intrinsic)/UV(observed)] = log[f$_{\nu}$(7.7 $\mu$m)/f$_{\nu}$(153 nm)] - 1.}
  
\end{figure}

\clearpage

\begin{figure}
\figurenum{12}
\includegraphics[scale=0.9] {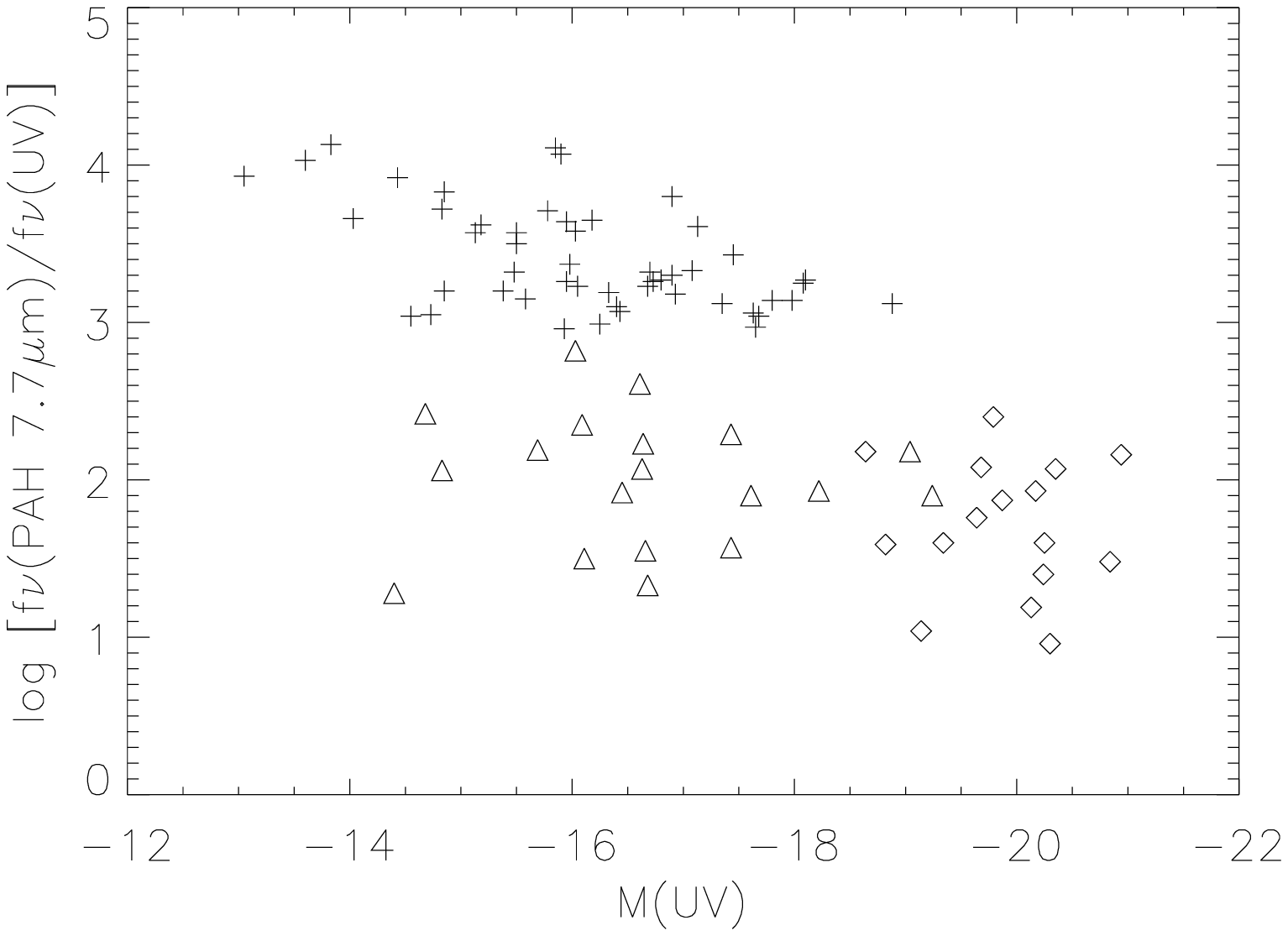}
\caption{Observed flux density of 7.7 \um PAH feature, f$_{\nu}$(7.7 $\mu$m), relative to observed ultraviolet flux density from GALEX, f$_{\nu}$(153 nm), from Tables 1, 3, and 5 compared to luminosity M(UV).  M(UV) is an AB magnitude determined at rest frame 153 nm as explained in Tables 2, 4, and 6.  Triangles are Markarian starbursts from Figure 1, diamonds are GALEX starbursts from Figure 2, and crosses are $Spitzer$ starbursts from Figure 3.  The AGN shown in Figure 4 are not included.  Trend of points is the opposite of trend in Figure 11, which shows that the most luminous sources in M(UV) are also the least obscured. This figure also shows the empirical result that the limiting value of log[f$_{\nu}$(7.7 $\mu$m)/f$_{\nu}$(153 nm)] = 1, which is taken empirically to be the value for an unobscured source. }  

\end{figure}

\clearpage

\begin{figure}
\figurenum{13}
\includegraphics[scale=0.9] {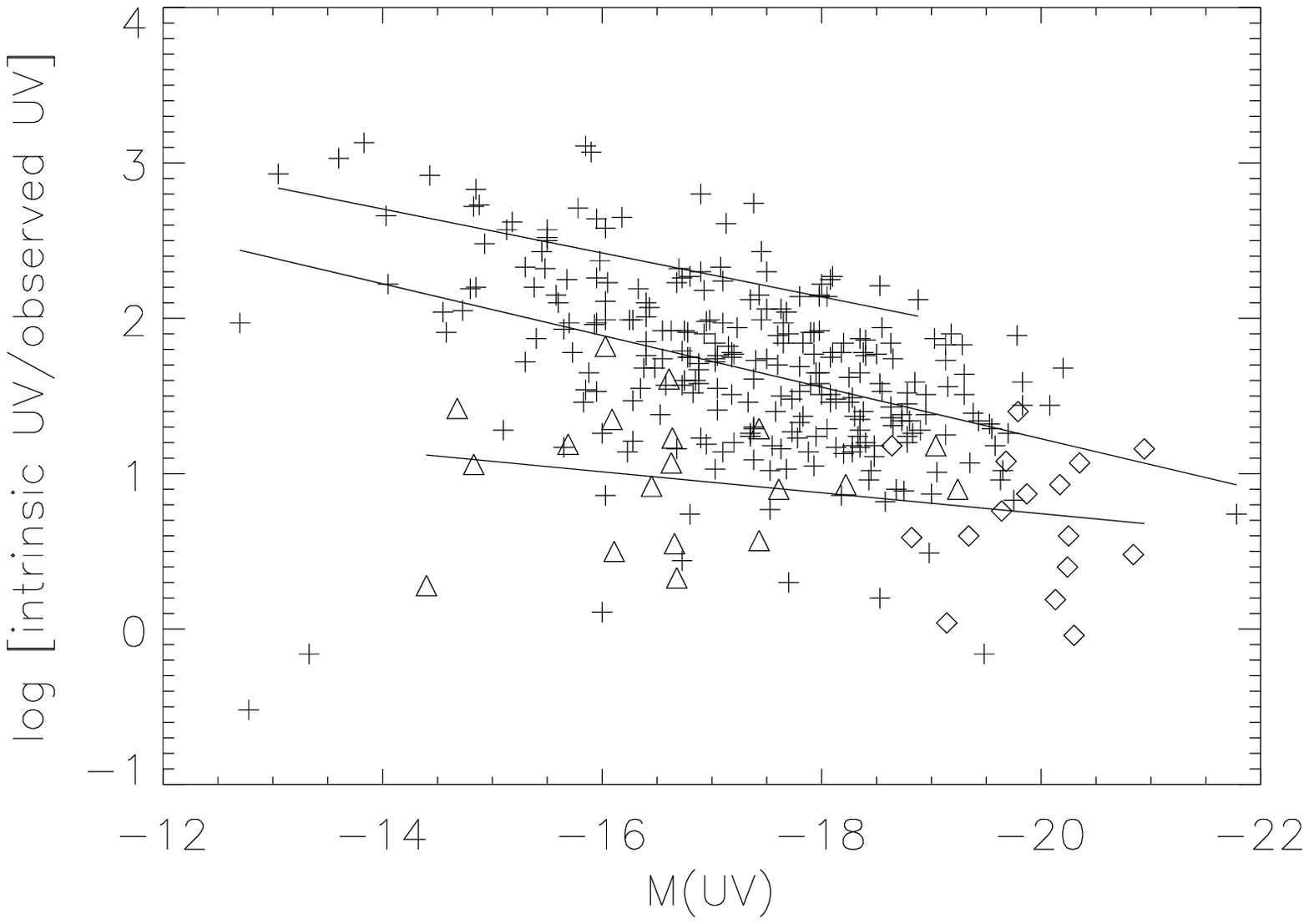}
\caption{Ratio of intrinsic ultraviolet luminosity to emerging luminosity compared to M(UV) for all samples of starbursts.  Triangles are Markarian starbursts from Figure 1, diamonds are GALEX starbursts from Figure 2, and crosses are all $Spitzer$ starbursts from \citet{sar09}.  The AGN shown in Figure 4 are not included.  Intrinsic(UV)/observed(UV) is determined from relation in section 5.2, that log[UV(intrinsic)/UV(observed)] = log[f$_{\nu}$(7.7 $\mu$m)/f$_{\nu}$(153 nm)] - 1.  Lines shown are linear least squares fits to the different samples.  The fit to the ultraviolet-selected Markarian+GALEX sample (lower line) has the form log[UV(intrinsic)/UV(observed)] = 0.07($\pm$0.04)M(UV)+2.09$\pm$0.69; the fit to the full infrared-selected $Spitzer$ sample  (central line) is log[UV(intrinsic)/UV(observed)] = 0.17($\pm$0.02)M(UV)+4.55($\pm$0.4); the fit to the extreme infrared-selected $Spitzer$ sample in Table 5 (upper line) is log [UV(intrinsic)/UV(observed)] = 0.14($\pm$0.03)M(UV)+4.68$\pm$0.50. } 
\end{figure}

\clearpage

\begin{figure}
\figurenum{14}
\includegraphics[scale=0.9] {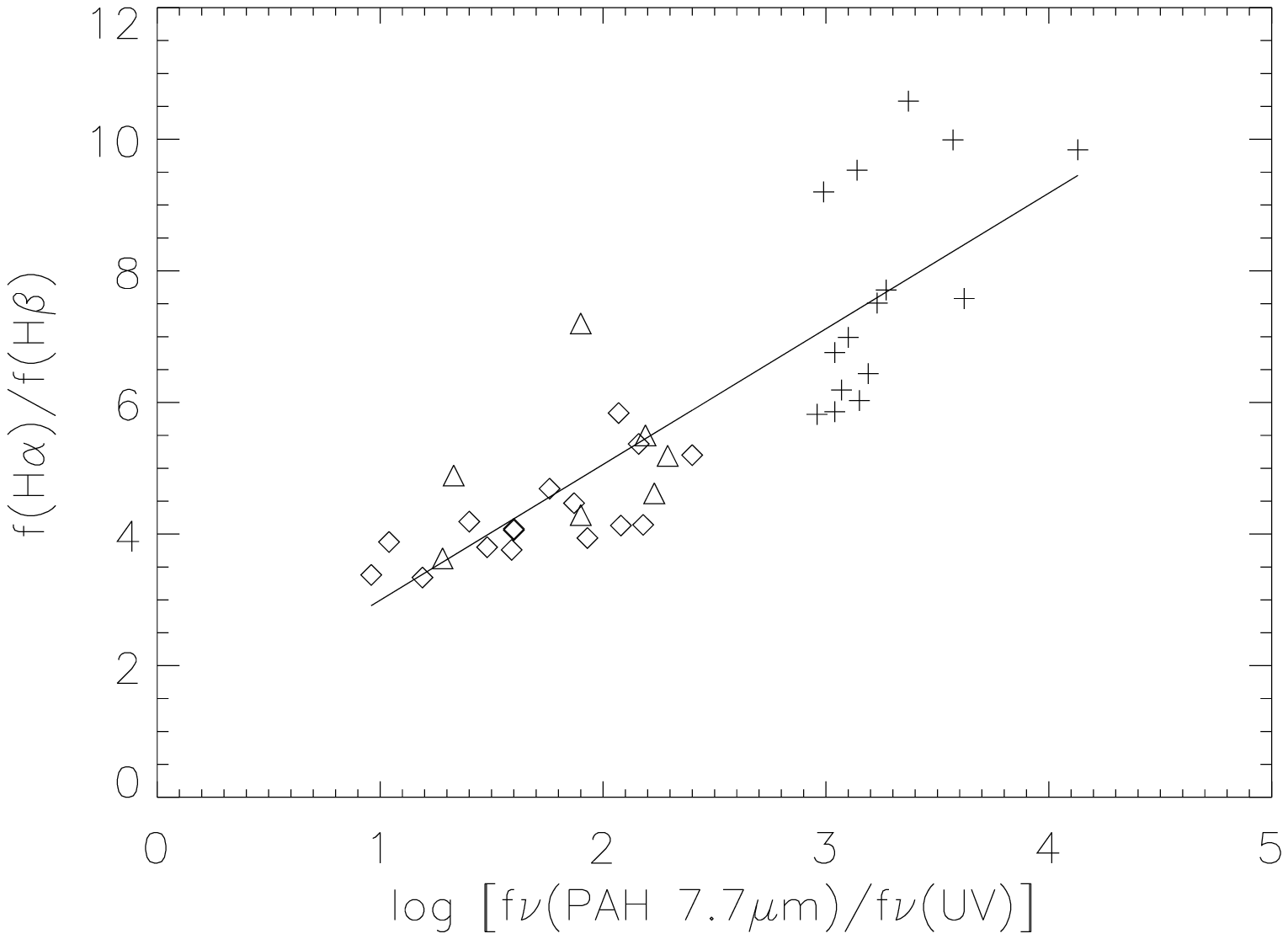}
\caption{Ratio of fluxes H$\alpha$/H$\beta$ from SDSS compared to the f$_{\nu}$(7.7 $\mu$m)/f$_{\nu}$(153 nm), from Tables 1, 3, and 5.   Relative line fluxes are measured using $\lambda$f$_{\lambda}$ taking f$_{\lambda}$ at the peak of the line.  Triangles are Markarian starbursts from Figure 1, diamonds are GALEX starbursts from Figure 2, and crosses are $Spitzer$ starbursts from Figure 3.  The AGN shown in Figure 4 are not included.  Intrinsic ratio in unobscured source should be 2.87, which agrees with lowest value on plot and confirms determination in Figures 11 and 12 that log[f$_{\nu}$(7.7 $\mu$m)/f$_{\nu}$(153 nm)] = 1 for unobscured source.  Line shown is linear least squares fit to all of the points, of form f(H$\alpha$)/f(H$\beta$) = (2.06$\pm$0.21)log[f$_{\nu}$(7.7 $\mu$m)/f$_{\nu}$(153 nm)]+0.93$\pm$0.53.  The slope shows that there is a relation between reddening in the Balmer lines and the obscuration shown by comparing PAH and UV flux densities. }

\end{figure}

\clearpage

\begin{figure}
\figurenum{15}
\includegraphics[scale=0.9] {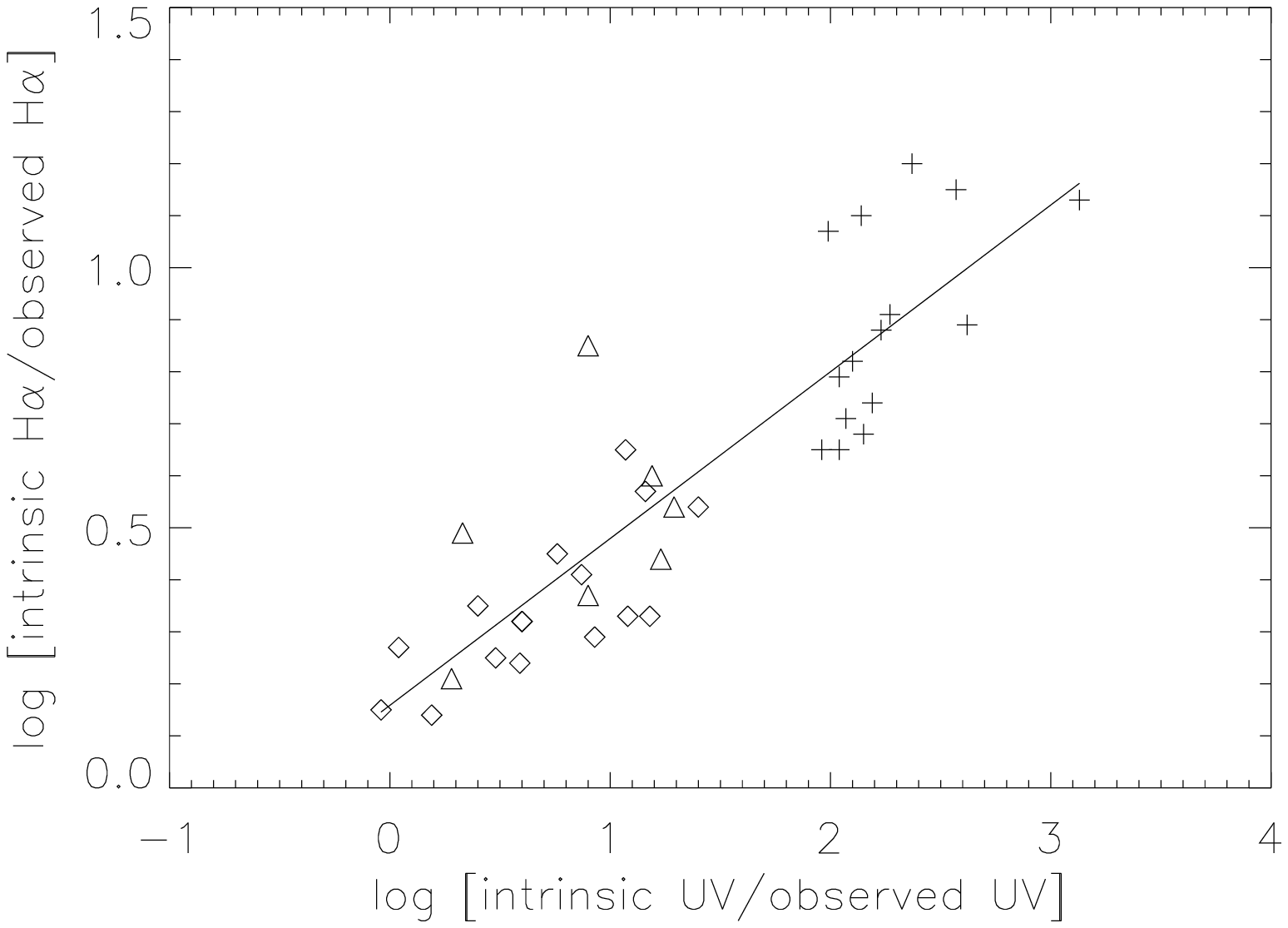}
\caption{ Obscuration at H$\alpha$ determined from reddening of H$\alpha$/H$\beta$ compared to obscuration of ultraviolet determined from f$_{\nu}$(7.7 $\mu$m)/f$_{\nu}$(153 nm).  Triangles are Markarian starbursts from Figure 1, diamonds are GALEX starbursts from Figure 2, and crosses are $Spitzer$ starbursts from Figure 3.  The AGN shown in Figure 4 are not included.  Line is linear least squares fit to all of the points, of form log [intrinsic H$\alpha$/observed H$\alpha$] = 0.32($\pm$0.03)log [intrinsic UV/observed UV] + 0.16$\pm$0.04.  Line shows that much more obscuration is present in the ultraviolet than would be indicated by obscuration at H$\alpha$.  }

\end{figure}

\clearpage

\begin{figure}
\figurenum{16}
\includegraphics[scale=0.9] {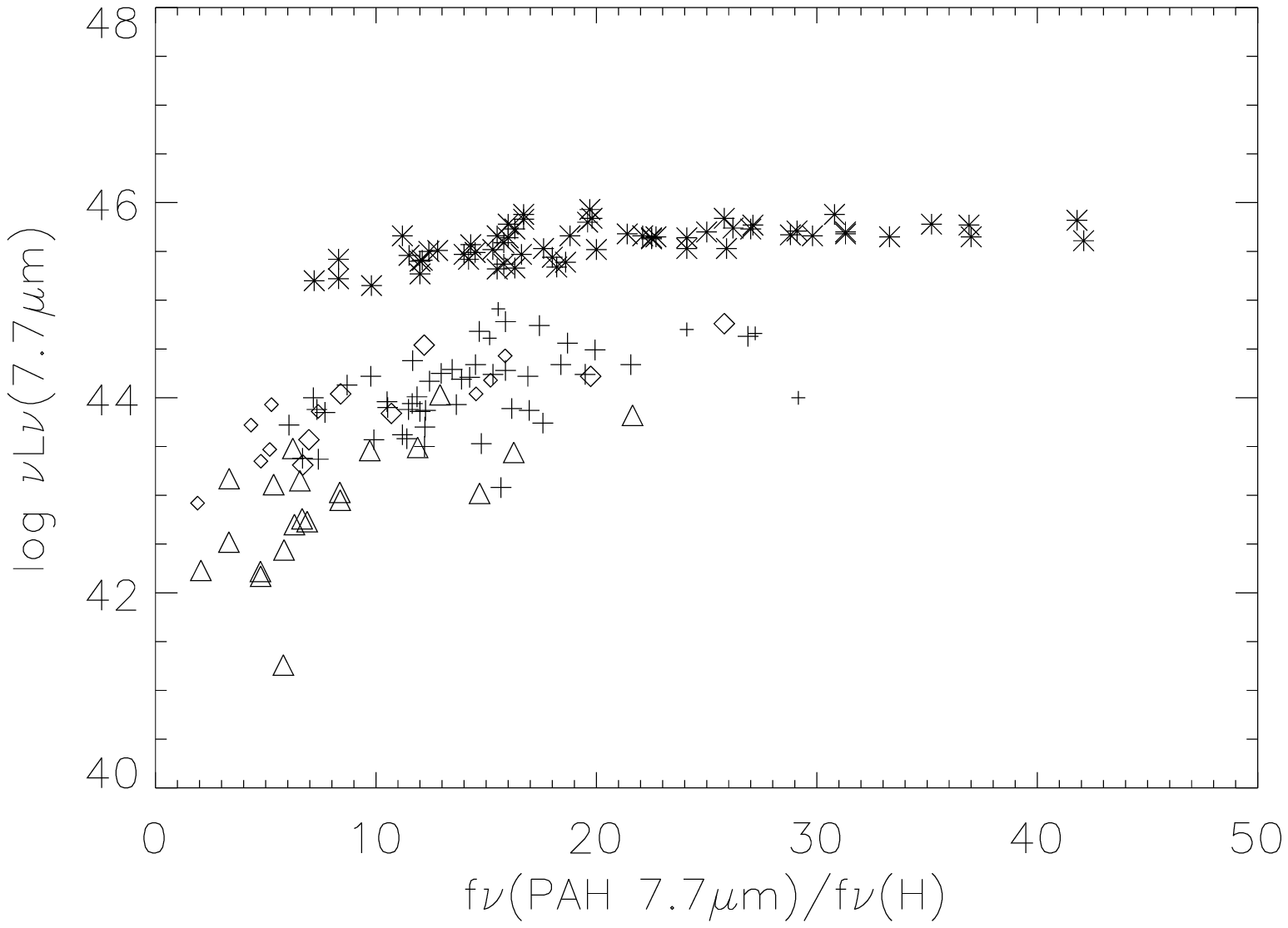}
\caption{Ratio of starburst luminosity to luminosity of stellar continuum shown by ratio f$_{\nu}$(7.7 \ums)/f$_{\nu}$(1.6 \ums) compared with luminosity $\nu$L$_{\nu}$(7.7 $\mu$m) in erg s$^{-1}$ (log [$\nu$L$_{\nu}$(\ldot)] = log [$\nu$L$_{\nu}$(erg s$^{-1}$)] - 33.59).  Triangles are Markarian starbursts from Figure 1, diamonds are GALEX starbursts from Figure 2, and crosses are $Spitzer$ starbursts from Figure 3.  Asterisks are very dusty, luminous starbursts at z $\sim$ 2 discovered by $Spitzer$ with f$_{\nu}$(24 \ums) $\ga$ 0.5 mJy.  The AGN shown in Figure 4 are not included.  Small symbols are sources undetected in 2MASS, for which a limit is shown corresponding to $H$ $>$ 16.5.  The starburst to stellar ratio is similar among all of the samples and not dependent on luminosity. }

\end{figure}

\begin{figure}
\figurenum{17}
\includegraphics[scale=0.9] {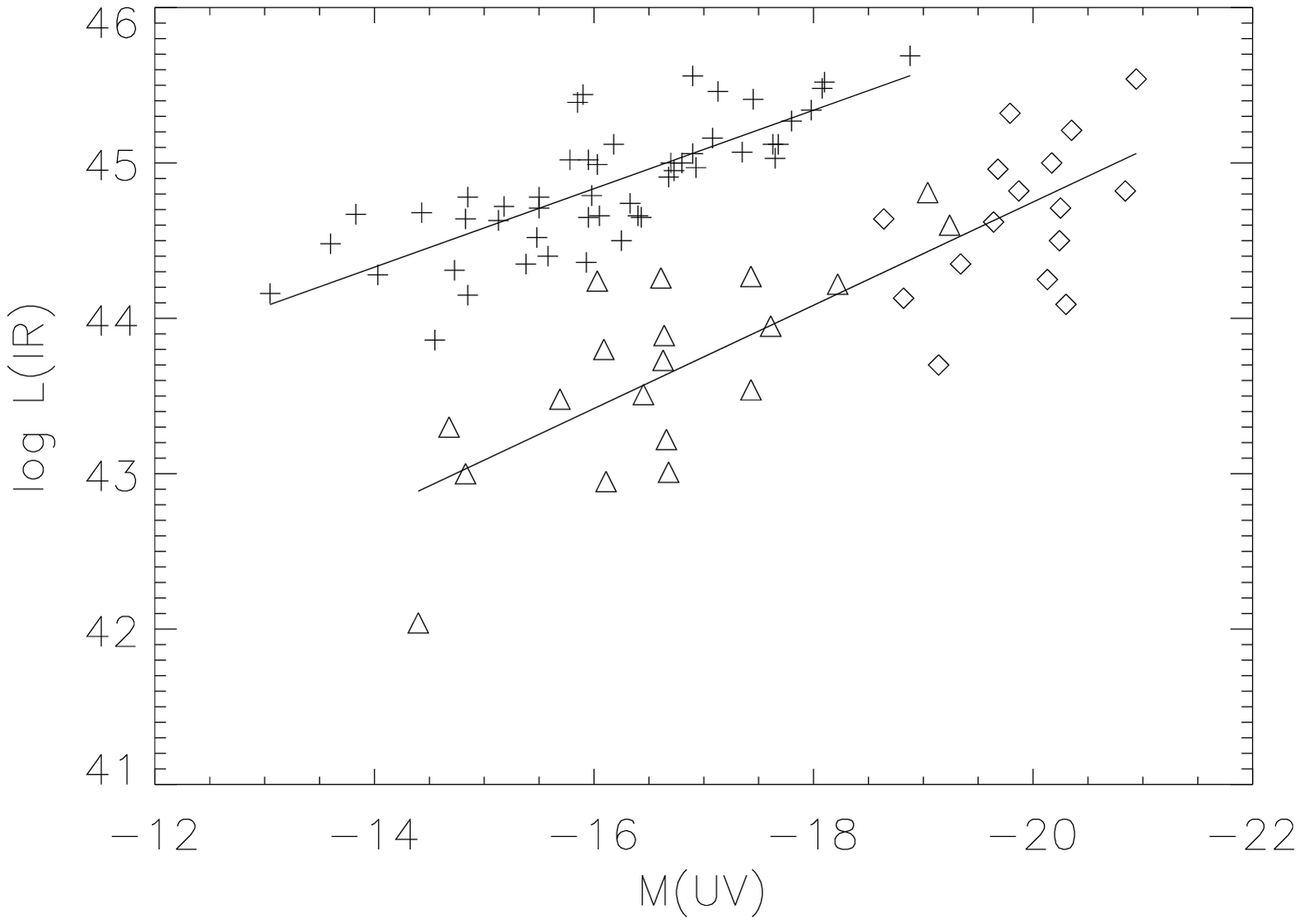}
\caption{Total bolometric luminosity $L_{ir}$ in erg s$^{-1}$ compared to M(UV).  (log [$L_{ir}$(\ldot)] = log [ $L_{ir}$(erg s$^{-1}$)] - 33.59.)  Triangles are Markarian starbursts from Figure 1, diamonds are GALEX starbursts from Figure 2, and crosses are $Spitzer$ starbursts from Figure 3.  Luminosities $L_{ir}$ are determined from the empirical transformation log $L_{ir}$ = log[$\nu$L$_{\nu}$ (7.7$\mu$m)] + 0.78 \citep{hou07} using the $\nu$L$_{\nu}$(7.7$\mu$m) from Tables 2, 4, and 6. Upper line is linear least squares fit to extreme $Spitzer$ sample in Table 5, of form log $L_{ir}$ = -(0.25$\pm$0.03)M(UV)+40.79$\pm$0.48, and to the GALEX+Markarian ultraviolet-selected samples in Tables 1 and 3, of form log $L_{ir}$  = -(0.33$\pm$0.04)M(UV)+38.10$\pm$0.69.  Lines represent the extremes of infrared selection and ultraviolet selection for starbursts and indicate that ultraviolet-selected starbursts of a given M(UV) underestimate bolometric luminosities by about a factor of 30 over all luminosities when compared with the most obscured starbursts. }
\end{figure}

\clearpage

\begin{figure}
\figurenum{18}
\includegraphics[scale=0.9] {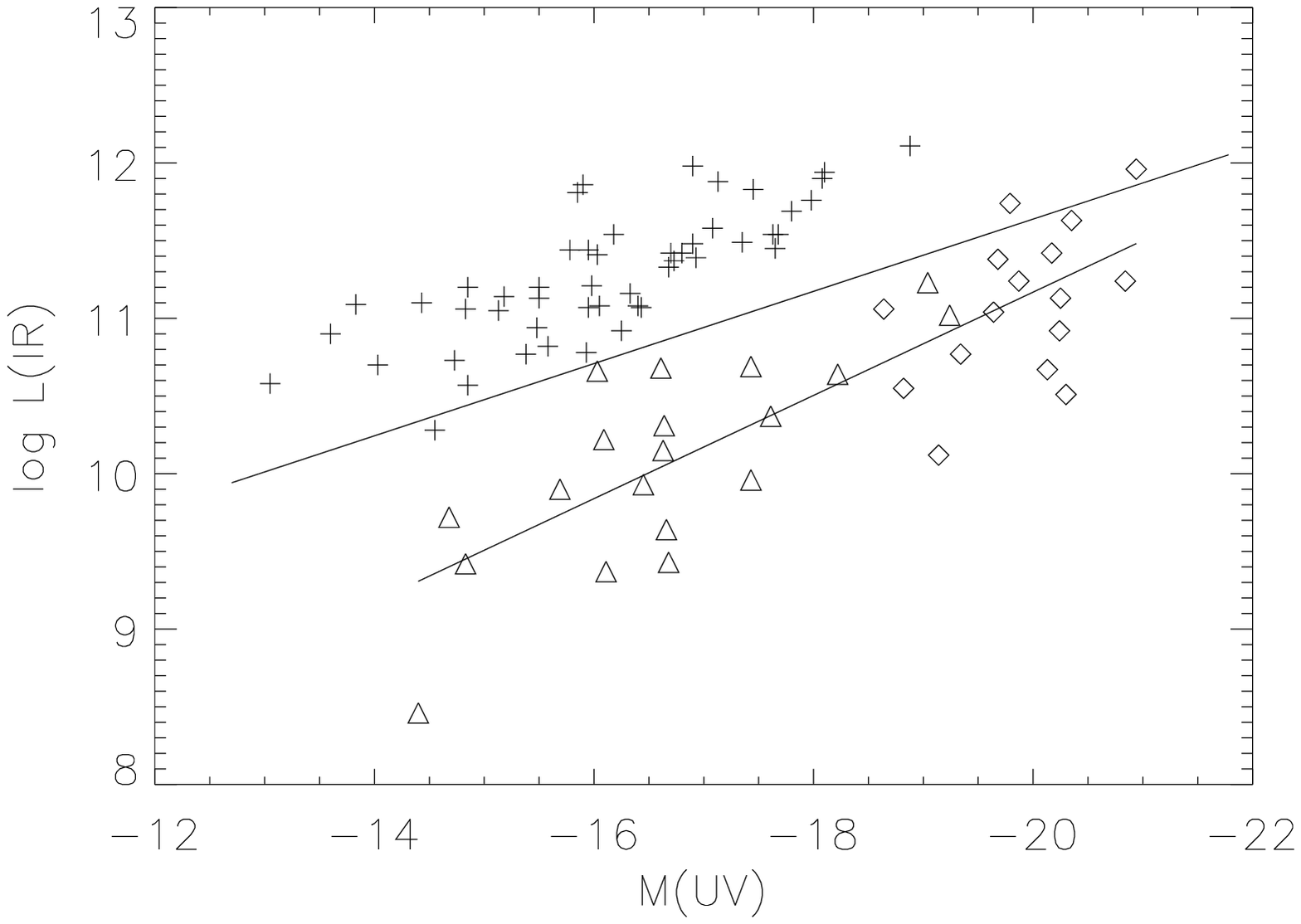}
\caption{Total bolometric luminosity $L_{ir}$ in \ldot~ compared to M(UV).  (log [$L_{ir}$(\ldot)] = log [$L_{ir}$(erg s$^{-1}$)] - 33.59.)  Triangles are Markarian starbursts from Figure 1, diamonds are GALEX starbursts from Figure 2, and crosses are $Spitzer$ starbursts from Figure 3.  Luminosities $L_{ir}$ are determined for all sources from the empirical transformation log $L_{ir}$ = log[$\nu$L$_{\nu}$(7.7$\mu$m)] + 0.78 \citep{hou07} using the $\nu$L$_{\nu}$ (7.7$\mu$m) from Tables 2, 4, and 6.  Lines shown are fit separately to the full $Spitzer$-selected sample of 287 starbursts in \citet{sar09}, of form log $L_{ir}$ = -(0.23$\pm$0.02)M(UV)+6.99$\pm$0.41 (upper line), and to the GALEX+Markarian ultraviolet-selected samples, of form log $L_{ir}$  = -(0.33$\pm$0.04)M(UV)+4.52$\pm$0.69 (lower line).}  
\end{figure}

\clearpage

\begin{figure}
\figurenum{19}
\includegraphics[scale=0.9] {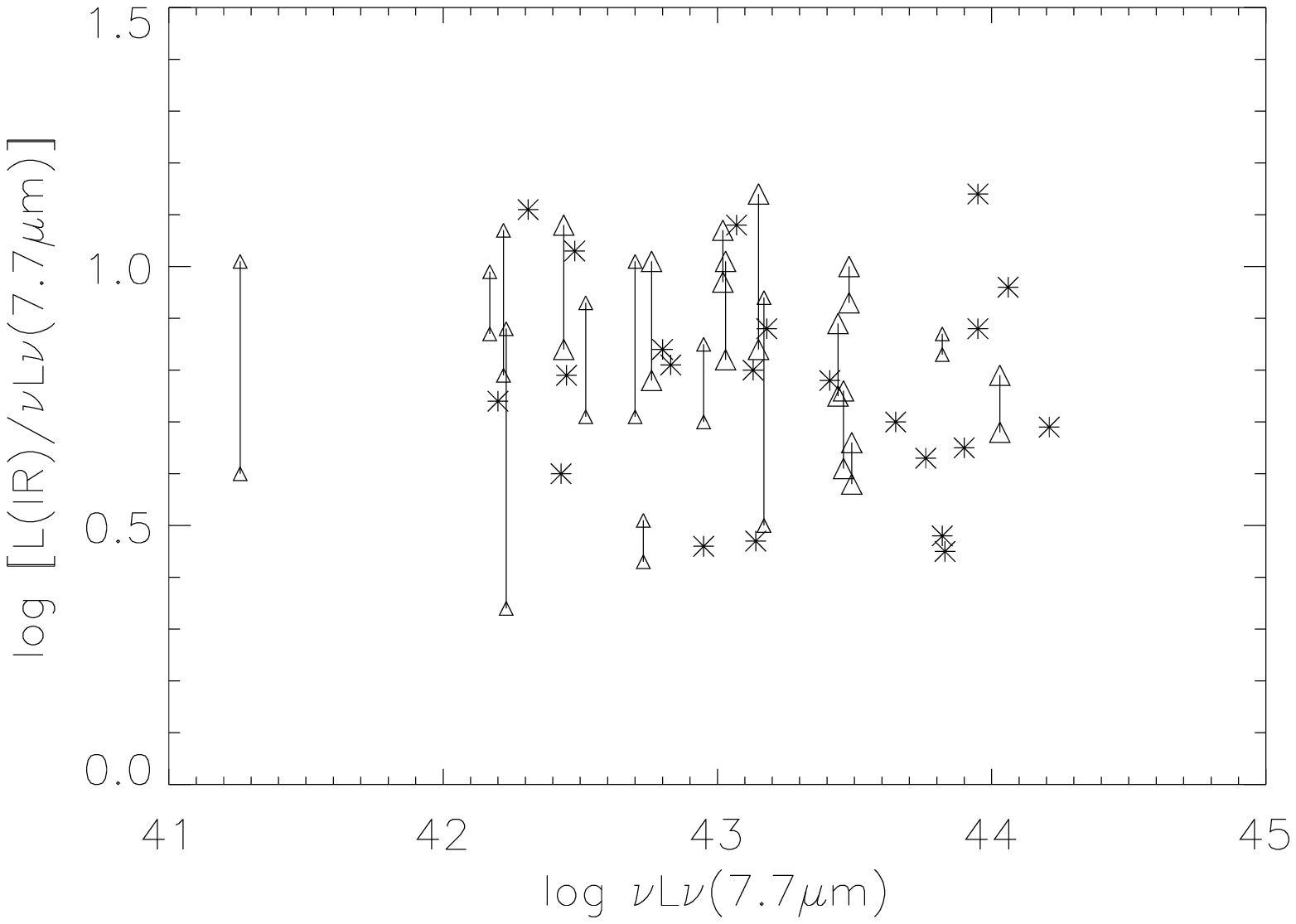}
\caption{Ratio log $L_{ir}$/log[$\nu$L$_{\nu}$(7.7$\mu$m)] compared to luminosity $\nu$L$_{\nu}$(7.7 $\mu$m) in erg s$^{-1}$.  (log [$\nu$L$_{\nu}$(\ldot)] = log [$\nu$L$_{\nu}$(erg s$^{-1}$)] - 33.59.)  Asterisks are starbursts from \citet{bra06} used to determine empirical transformation which we use, log $L_{ir}$ = log[$\nu$L$_{\nu}$ (7.7$\mu$m)] + 0.78.  Vertical lines are results for Markarian starbursts in Table 2.  Results are uncertain because of uncertain aperture corrections when comparing IRS and IRAS flux densities. Upper value for each line is value found if total IRAS flux is compared to observed f$_{\nu}$(7.7 \ums).  Lower value for each line is value found if IRAS fluxes are all decreased by a factor corresponding to the ratio f$_{\nu}$(IRS 25 \ums)/f$_{\nu}$(IRAS 25 \ums) explained in Table 2.  Small triangles represent upper limits for sources without IRAS detections in one or more bands.}

\end{figure}

\clearpage

\begin{figure}
\figurenum{20}
\includegraphics[scale=0.9] {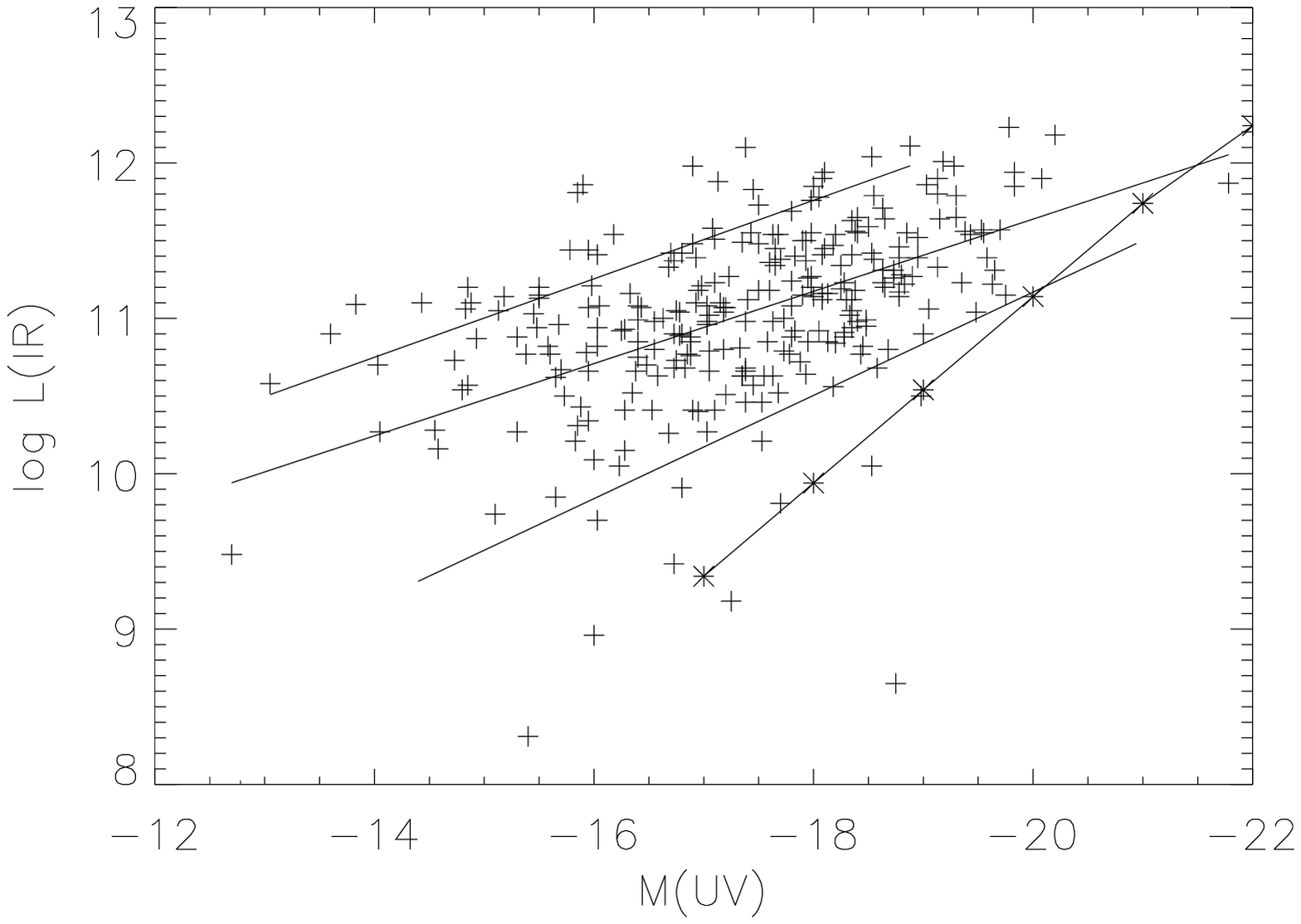}
\caption{Total bolometric luminosity $L_{ir}$ in \ldot~ compared to M(UV) for samples discussed in present paper and shown in Figures 17 and 18 (upper three lines) compared with summary conclusions for LBGs at z = 2.5 in \citet{bou09} (line with asterisks).  Upper line is for the extreme $Spitzer$ starbursts in Table 5; second line is for all of the infrared-selected starbursts in \citet{sar09}, and plotted crosses show this full sample; third line is the combined Markarian and GALEX ultraviolet-selected samples in Tables 1 and 3. Results indicate that the adopted result for LBGs at high redshift (line with asterisks) applies only to the lower envelope for all starbursts.  Differences between this line and other lines show differences that would arise for measures of SFR density.  Adopted results for LBGs agree with the full infrared-selected sample at M(UV) $\sim$ -21, but differ by a factor $>$ 10 at M(UV) = -17.  Differences are greater at fainter M(UV) for reasons discussed in section 5.2, that the faintest sources in the ultraviolet are those with the greatest obscuration by dust. }  
\end{figure}

\clearpage




\end{document}